\documentclass[aps,prb,twocolumn,amsmath,amssymb,superscriptaddress,scrartcl,eqsecnum,longbibliography]{revtex4-1}
\usepackage{extarrows}
\usepackage{slashed}
\usepackage{amsmath,amsfonts,amssymb,graphics,graphicx,epsfig,color,times,indentfirst,layout}
\usepackage{lipsum}
\usepackage{mathrsfs}
\usepackage[unicode=true,pdfusetitle,bookmarks=false,colorlinks=true,citecolor=black,urlcolor=black,linkcolor=black]{hyperref}
\usepackage{tikz}
\usepackage{tikz-cd}
\usetikzlibrary{arrows}
\usetikzlibrary{intersections}
\usetikzlibrary{shapes.geometric}

\RequirePackage[normalem]{ulem} 
\RequirePackage{color}\definecolor{RED}{rgb}{1,0,0}\definecolor{BLUE}{rgb}{0,0,1} 
\providecommand{\DIFaddend}{} 

\begin{document}

\title{Edge theory approach to topological entanglement entropy, mutual information \\
and entanglement negativity in Chern-Simons theories
}

\date{\today}


\author{Xueda Wen}
\affiliation{Department of Physics, University of Illinois at Urbana-Champaign, Urbana, IL 61801, USA}

\author{Shunji Matsuura}
\affiliation{Niels Bohr International Academy and Center for Quantum Devices,
Niels Bohr Institute, Copenhagen University, Blegdamsvej 17, Copenhagen, Denmark}
\affiliation{Yukawa Institute for Theoretical Physics, Kyoto University, Kyoto, Japan}

\author{Shinsei Ryu}
\affiliation{Department of Physics, University of Illinois at Urbana-Champaign, Urbana, IL 61801, USA}

\begin{abstract}
We develop an approach based on edge theories to calculate the entanglement entropy and related quantities in (2+1)-dimensional topologically ordered phases.
Our approach is complementary to, {\it e.g.},
the existing methods using replica trick and Witten's method of surgery,
and applies to a generic spatial manifold of genus $g$, which can be bipartitioned in an arbitrary way.
The effects of fusion and braiding of Wilson lines can be also straightforwardly studied within our framework.
By considering a generic superposition of states
with different Wilson line configurations,
through an interference effect, we can detect,
by the entanglement entropy,
the topological data of Chern-Simons theories, \textit{e.g.}, the
$R$-symbols, monodromy and topological spins of quasiparticles.
Furthermore, by using our method, we calculate other entanglement/correlation measures
such as the mutual information and the entanglement negativity.
In particular, it is found that the entanglement negativity of two adjacent non-contractible regions
on a torus provides a simple way to distinguish Abelian and non-Abelian topological orders.
\end{abstract}
\maketitle

\tableofcontents

\section{Introduction}

Quantum entanglement plays a central role in characterizing and distinguishing various phases realized in quantum many-body systems.
\cite{Kitaev2006,LevinWen2006,Calabrese0905,Eisert0808}
For example,
quantum entanglement as measured by the bipartite entanglement entropy
may be used to distinguish different topological phases,
and to characterize properties of critical points.
\cite{Kitaev2006,LevinWen2006,Calabrese0905,Eisert0808,CC2004,Hsu}
Quantum entanglement has also been
extensively studied in the context of quantum gravity, in particular in the context of the AdS/CFT correspondence.
\cite{RT1,RT2}

In this work, we will mainly focus on the quantum entanglement between spatial regions
in topological quantum field theory (TQFTs) in (2+1) dimensions.
TQFTs were extensively studied after Witten's seminal work on the Chern-Simons gauge theory and its relation to the Jones polynomial.
\cite{Witten1989,Witten1992}
In particular,
in condensed matter physics,
TQFTs are widely used to describe emergent topological phases of matter
in many-body systems,
such as the fractional quantum Hall states,
\cite{FQHe,FQHt,Wen1995}
gapped quantum spin liquids,
\cite{WenSpin}
a $p_x+ip_y$ superconductor,
\cite{Read2000,Stone2006}
and quantum dimer models.
\cite{WenBook,FradkinBook}
Quantum entanglement has been verified to be very useful in characterizing
and extracting the topological data of TQFTs.
For example,
it was found that the quantum entanglement can be used to extract the modular $\mathcal{S}$ and
$\mathcal{T}$ matrices,
which encode the properties of quasi-particles in topological phases.
\cite{YiZhang}

There are different measures of quantum entanglement or correlations, 
which have their own merits depending on the case under study.
Let us start by listing entanglement/correlation measures of our interest in this work.

\subsection{Different entanglement/correlation measures}

First,
when the total system is bipartitioned into two subsystems (regions) $A$ and $B$,
the von Neumann entropy of the region $A$ is defined by
\begin{equation}\label{vN}
S_A^{\text{vN}}=-\text{Tr}\rho_A\ln\rho_A,
\end{equation}
where $\rho_A=\text{Tr}_B\rho$ is the reduced density matrix of the subsystem $A$.
Note that when $\rho$ is a pure state,
$\rho=|\Psi\rangle\langle\Psi|$ where $|\Psi\rangle$ is, {\it e.g.}, the ground state of the total system,
$S_A^{\text{vN}}=S_B^{\text{vN}}$.

An alternative measure of bipartite entanglement is the Renyi entropy
\begin{equation}\label{Renyi}
S_A^{(n)}=\frac{1}{1-n}\ln\text{Tr}\rho_A^n,
\end{equation}
which also satisfies $S_A^{(n)}=S_B^{(n)}$ when $\rho$ is a pure state.
The Renyi entropy can provide more information than the von Neumann entropy,
in that, by knowing the Renyi entropy for arbitrary $n$,
we reconstruct the entanglement spectrum, \textit{i.e.}, all the eigenvalues of $\rho_A$.
The von Neumann entropy and the Renyi entropy are related by $S_A^{\text{vN}}=\lim_{n\to 1}S_A^{(n)}$, or
\begin{equation}\label{vN2}
S_A^{\text{vN}}=-\lim_{n\to1}\partial_n \text{Tr}\rho_A^n.
\end{equation}

For a mixed state, it is found that the quantum and classical correlations cannot be explicitly
separated in these entanglement measures.
As an example, let us consider two subsystems $A_1$ and $A_2$
which are embedded in a larger system.
$A_1$ and $A_2$ are not necessarily complementary to each other, and therefore $\rho_{A_1\cup A_2}$ may correspond to a mixed state.
In this case, a useful quantity that can be constructed based on the entanglement entropy
 is the (Renyi) mutual information
\begin{equation}
\label{def mutual renyi}
I_{A_1A_2}^{(n)}=S_{A_1}^{(n)}+S_{A_2}^{(n)}-S_{A_1\cup A_2}^{(n)},
\end{equation}
which by definition is symmetric in $A_1$ and $A_2$.
By taking the $n\to 1$ limit, we define
the (von Neumann) mutual information:
\begin{equation}
I_{A_1A_2}=\lim_{n\to 1}I_{A_1A_2}^{(n)}.
\end{equation}
However, it is found that the mutual information does not have all the proper features to 
be a quantum entanglement measure. (See, \textit{e.g.}, Ref. \onlinecite{Plenio_2007}, where it is shown 
that the mutual information is finite for most of the separable mixed states.)  It will mix the 
classical and quantum information together, and can only be considered as a correlation measure.

A yet another quantity,
entanglement negativity
has been recently calculated in different many-body systems,
such as
conformal field theories and exactly solvable lattice models.\cite{Calabrese2012a,Calabrese2012b,VidalEN,CastelnovoEN}
The entanglement negativity turned out to be a computable and useful entanglement measure.
\cite{Vidal2002,Plenio2005}
To be concrete, given a reduced density matrix
$\rho_{A_1A_2}$ which describes a mixed state in the Hilbert space $\mathcal{H}_{A_1}\otimes \mathcal{H}_{A_2}$,
we take a partial transposition with respect to the degrees of freedom in region $A_2$
as follows
\begin{equation}\label{partialT}
\langle e_i^{(1)}e_j^{(2)}|\rho_{A_1\cup A_2}^{T_2}|e_k^{(1)}e_l^{(2)}\rangle=
\langle e_i^{(1)}e_l^{(2)}|\rho_{A_1\cup A_2}|e_k^{(1)}e_j^{(2)}\rangle,
\end{equation}
where $T_2$ means the partial transposition on $A_2$, $|e_i^{(1)}\rangle$ and $|e_j^{(2)}\rangle$ are arbitrary bases in $\mathcal{H}_{A_1}$
and $\mathcal{H}_{A_2}$ respectively.
Then the entanglement negativity are defined as
\begin{equation}\label{D1}
\mathcal{E}_{A_1A_2}=\ln\text{Tr}\left|\rho_{A_1\cup A_2}^{T_2}\right|.
\end{equation}
The entanglement negativity in quantum filed theories can be computed by using the replica method as
\cite{Calabrese2012a,Calabrese2012b}
\begin{equation}\label{D2}
\mathcal{E}_{A_1A_2}=\lim_{n_e\to 1}\ln \text{Tr}\left(\rho_{A_1\cup A_2}^{T_2}\right)^{n_e}.
 \end{equation}
The trace of the partial transposition of the density matrix on the replica space $\text{Tr}\left(\rho_{A_1\cup A_2}^{T_2}\right)^{n}$
has different forms depending on whether $n$ is even or odd.
Here we consider the analytic continuation of the
even sequence at $n_e\to 1$.
The formula above has been proved to be of great use in the study of
the entanglement negativity in quantum field theories for both equilibrium cases\cite{Calabrese2012a,Calabrese2012b}
and non-equilibrium cases.
\cite{Coser,Eisler,Hoogeveen,Xueda}

\subsection{Different entanglement/correlation measures for a topological quantum field theory}

These different entanglement/correlation measures have been calculated in TQFTs in (2+1) dimensions.
The topological entanglement entropy (TEE) was first introduced by Kitaev-Preskill and Levin-Wen.
\cite{Kitaev2006,LevinWen2006}
First,
for topologically ordered systems in two spatial dimensions,
it was shown that the von Neumann entanglement entropy for a simply connected region $A$ behaves,
in the limit of zero correlation, as
\begin{equation}\label{Stop}
S_A^{\text{vN}}=\alpha L-\gamma,
\end{equation}
where $\alpha$ is a nonuniversal coefficient,
$L$ is the length of the smooth boundary of $A$,
and $-\gamma$ is a universal negative constant which
is named the `topological entanglement entropy'.
For a general TQFT, $\gamma$ is given by
\begin{equation}
\gamma=\ln \mathcal{D}=\ln\sqrt{\sum_i d_i^2},
\end{equation}
where $d_i$ is the quantum dimension of quasiparticle $i$, and $\mathcal{D}$ is the total quantum dimension
(see Appendix).
%

Dong et al.\ extended the Kitaev-Preskill
results to more general manifolds like torus and a sphere with quasiparticles by using
the replica trick and surgery method.
\cite{Dong}
They found that the entanglement entropy depends on the universal data of a TQFT, \textit{e.g.}, the quantum dimensions and the fusion rules.
In certain cases such as the torus geometry, the entanglement entropy
also depends on the choice of ground state.
Later, Zhang et al.\ studied the entanglement entropy of topological phases on a torus\cite{YiZhang}.
By tuning the ground state and introducing different entanglement cuts,
they found that the modular $\mathcal{S}$ and $\mathcal{T}$ matrices can be extracted from the entanglement entropy.

Besides the entanglement entropy, other entanglement/correlation
measures such as the entanglement negativity and mutual information
which are powerful in the case of mixed states,  turn out to be very useful in characterizing the properties of a TQFT.
Recently, the entanglement negativity was used to study the topological
ordered systems such as the toric code model.
\cite{VidalEN,CastelnovoEN}
It was found that there is a universal topological
entanglement between two adjacent non-contractible regions on a torus.
On the other hand, if the two regions are disjoint,
independent of whether they are contractible or not, there is no
universal topological entanglement between them.
It should be noted that the above results are obtained based on an exactly solvable lattice model.
It is hence desirable to have a more understanding of these results by studying general TQFTs.
The difficulty may be that the operation of `partial transposition',
which is used in the definition of the entanglement negativity [see Eq.\ (\ref{partialT})],
is difficult to realize in practice when one considers a general three dimensional manifold where a TQFT lives.

Most recently, Ref.\ \onlinecite{Qi2015} used mutual information to study the topological ordered phases in (2+1)
dimensions, as well as higher dimensions where topological orders are identified as condensates of membranes.
Therein, the mutual information
can be utilized to define the topological uncertainty principle, which reflects the non-commuting property of non-local order parameters in topological ordered phases \cite{Qi2015}.
Compared to the entanglement entropy of topological ordered phases,
it is noted that the mutual information has the merit of being ultraviolet finite for two disjoint regions.

\subsection{Our motivations}

In this work, our motivations to revisit the topological entanglement entropy and other entanglement/correlation
 measures of a TQFT are mainly as follows.

\begin{enumerate}
\item
 In the calculation of the topological entanglement entropy of
 a Chern-Simons theory on a general manifold,
 one needs to evaluate the Chern-Simons path integral on a 3-manifold.
 In particular, when using the replica trick,
 one needs to consider a $n$-sheeted Riemann surface spacetime and
 glue them together, which may be very complicated.
 In this work, we hope to develop an alternative edge theory approach,
 which may simplify the calculation.
 It should be noted that the edge theory approach to the topological entanglement entropy of a TQFT on a simple manifold such as a sphere, or a cylinder with definite topological flux,
 has been studied in several works \cite{Kitaev2006,Qi1,Das,Oshikawa2013,Taylor2015}.
However, as far as we know,
there are still many open issues to be understood.
For example, how do we use the edge theory approach
to study the topological entanglement entropy of a TQFT
on a general manifold of genus $g$?
How is the effect of fusion and braiding of Wilson lines/quasiparticles reflected in the edge theory approach?
How do we extract topological data of the underlying theory
from the edge theory approach?

\item
Till now, some other entanglement measures such as the entanglement negativity of a TQFT has not yet been studied with the field theory approach.
Although some results have been obtained based on the lattice models \cite{VidalEN,CastelnovoEN},
it is still desirable to understand the general structure of the entanglement negativity for a general TQFT.
Can we use the edge theory approach to fulfill this aim?
Moreover, in Refs.\ \onlinecite{VidalEN,CastelnovoEN}, the lattice model under study is in an Abelian topological ordered phase.
Then it is natural to ask what is the result for a
non-Abelian topological ordered phase?
Is there any qualitative difference between Abelian and non-Abelian theories?
We hope to answer these questions in this work.

\end{enumerate}

\subsection{Summary of main results}

Using the edge theory approach, we found a systematic way to study the topological entanglement entropy,
mutual information and the entanglement negativity for a  (2+1) dimensional Chern-Simons theory on a general manifold.
The effect of braiding and fusion of Wilson lines can be straightforwardly
incorporated in the calculations.
In particular, we have obtained the following results.

\begin{enumerate}
%

\item
 \emph{On topological entanglement entropy.}
By using the edge theory approach, we calculated the entanglement entropy
for given spatial regions
in Chern-Simons theories defined on a general manifold.
Our results agree with the path integral calculations for all the cases considered in
Ref.\ \onlinecite{Dong}.
A technical advantage of our approach,
as compared with the path integral (surgery) calculations,
is that the edge theory approach greatly simplify the calculation in
that we do not have to consider complicated 3-manifolds which may appear
in the surgery method.
The effect of braiding Wilson lines can be also considered,
instead of using skein relation\cite{Dong,Witten1989},
by simply introducing the braiding matrix or $R$-symbols, which makes the calculation more transparent.
We also found that, in the presence of multiple Wilson lines,
By considering a generic superposition of states,
the $R$-symbols, monodromy and topological spins of quasiparticles/anyons
can be detected in the entanglement entropy,
through an interference effect.
Finally, we also applied our edge theory approach to more general manifolds of
$g$-genus,
which may be difficult to handle in the replica trick
due to the complicated 3-manifolds which may arise as a result of surgery.

\item
 \emph{On topological mutual information and entanglement negativity.}
We gave explicit calculations of the topological mutual information and the entanglement negativity in Chern-Simons theories.
In particular, to our knowledge, the results on the entanglement negativity in a Chern-Simons field theory are given for the first time.
Moreover, compared with the previous works on lattice models, we obtained some new results for two adjacent non-contractible regions on a torus.
In Ref.\ \onlinecite{VidalEN}, it was found that the entanglement negativity in this case is independent of the choice of ground state.
Based on our field theory result,
it was found that the entanglement negativity is dependent (independent) on the choice of
ground state if the system is in a non-Abelian (Abelian) topological ordered phase.

\end{enumerate}

Along with these results,
we will also point out that, when using edge theories to calculate entanglement/correlation measures,
the boundary states must be regularized/normalized properly.
In the previous works \cite{Qi1,Das}, the proposed state, which is a superposition of different Ishibshi states,
is regularized as a whole (see next section for details).
We found that this regularization scheme cannot recover the correct topological entanglement entropy for a Chern-Simons
theory on a general manifold.
In this work, we regularized each Ishibashi state separately. Then a general quantum state can be expressed
 as a superposition of different regularized Ishibashi states. With this new regularized state, we can obtain the correct topological entanglement entropy as well as other entanglement/correlation measures for a Chern-Simons theory.

The rest of the paper is organized as follows.
In Sec.\ \ref{LR}, we start by introducing a new regularizd state,
based on which we can calculate the spatial
topological entanglement entropy in a Chern-Simons field theory.
Subsequently in Sec.\ \ref{EE}, we apply our method to
the calculation of the Renyi and von Neumann entanglement entropy for a Chern-Simons theory
defined on various kinds of manifolds.
The effects of braiding and monodromy of  quasiparticles
are also studied in this section.
In Sec.\ \ref{MI}, we study the spatial mutual information in Chern-Simons theories.
We consider different
tripartitions of a torus, and calculate the mutual information accordingly.
In Sec.\ \ref{LREN}, we show how to
calculate the left-right entanglement negativity for a general regularized state.
Then we apply this method to
the calculation of the entanglement negativity on a torus with different tripartitions.
Finally,we conclude our work in Sec.\ \ref{con}.
We also include several appendices containing a brief review of modular tensor categories (Appendix A), the topological
data of $SU(2)_k$ Chern-Simons theories,  and an alternative method of calculating the entanglement negativity for several cases (Appendix B).

\section{Left-right entanglement entropy}
\label{LR}

\subsection{Regularized state at the interface}

In this section, we introduce
basics of boundary states in (1+1) dimensional conformal field theories.
These boundary states will be used later to describe the reduced density matrices
of (2+1) dimensional topologically order phases, but in this section,
we study boundary states and quantum entanglement in isolation.
In particular, we will discuss how we need to regularize these boundary states properly.

In the study of quantum entanglement,
the regularized boundary state was first introduced  in
the quantum quench problem \cite{CC2006,CC2007}.
Later, this concept was used to study the spatial
entanglement entropy of a topological ordered system\cite{Qi1}.
Most recently, the similar idea
was used to study the entanglement entropy between the left and right moving modes
of the regularized boundary state\cite{Das,Zayas}.
To be concrete, in Ref.\ \onlinecite{Qi1,Das}, the regularized boundary state has the expression
\begin{equation}\label{bsOld}
|\mathcal{B}\rangle=\frac{e^{-\epsilon H}}{\sqrt{\mathcal{N}_B}}|B\rangle,
\end{equation}
where $e^{-\epsilon H}$ is a regularization factor,
$H$ is the Hamiltonian,
$\mathcal{N}_B$ is a normalization factor,
and the conformal boundary state $|B\rangle$ can be expressed as a linear combination of Ishibashi states $|h_a\rangle\rangle$,
which are solutions to the conformal boundary condition
\begin{equation}\label{IshibashiC}
L_n|b\rangle=\overline{L}_{-n}|b\rangle, \ \ \forall n\in Z.
\end{equation}
$a$ in $|h_a\rangle\rangle$ is used to label the primary field in a CFT,
or the type of quasiparticles in the corresponding TQFT.
$L_n$ is the generator of chiral conformal transformations which is defined through a Laurent expansion of the stress-energy
tensor, $T(z)=\sum_{n\in Z}z^{-n-2}L_n$,
and $\overline{L}_n$ is the generator of anti-chiral conformal transformations which is defined in a similar way.
Note that the Hilbert space of a CFT can be written in terms of holomorphic and antiholomorphic sectors, i.e.,
\begin{equation}
\mathcal{H}=\bigoplus_{h,\bar{h}}n_{h,\bar{h}}\mathcal{V}_h\otimes \overline{\mathcal{V}}_{\bar{h}},
\end{equation}
where the non-negative integer $n_{h,\bar{h}}$ denotes the number of distinct primary fields with conformal weight $(h,\bar{h})$.
For simplicity, here we only consider the diagonal CFTs with $n_{h,\bar{h}}=\delta_{h,\bar{h}}$. Then the Ishibashi state $|h_a\rangle\rangle$ which
satisfies Eq.\ (\ref{IshibashiC}) can be expressed as a linear combination of states in $\mathcal{V}_{h_a}\otimes \overline{\mathcal{V}}_{\bar{h}_a}$. By using $d_{h_a}(N)$ to label the dimension of subspace for level $N$ of the conformal family, we can denote an orthonormal basis $|h_a,N;j\rangle$ for $\mathcal{V}_{h_a}$, and similarly $\overline{|h_a,N;j}\rangle$
for $\overline{\mathcal{V}}_{\bar{h}_a}$, with $1\le j\le d_{h_a}(N)$.
Then the concrete form of Ishibashi state $|h_a\rangle\rangle$ can be written as
\begin{equation}
|h_a\rangle\rangle\equiv \sum_{N=0}^{\infty}\sum_{j=1}^{d_{h_a}(N)}|h_a,N;j\rangle\otimes \overline{|h_a,N;j}\rangle.
\end{equation}
For a rational CFT (RCFT), in which there are finite number of primary fields, the conformal boundary state may be expressed as
\begin{equation}\label{ConformalBC}
|B_i\rangle=\sum_a\psi_i^a|h_a\rangle\rangle.
\end{equation}
The concrete form of $\psi_i^a$ is related with the modular $\mathcal{S}$-matrix as follows
\begin{equation}\label{Smatrix0}
\psi_i^a=\frac{\mathcal{S}_{ia}}{\sqrt{\mathcal{S}_{0a}}}.
\end{equation}

In Refs.\ \onlinecite{Qi1,Das}, the regularized boundary state in Eq.\ (\ref{bsOld}) was suggested to study the spatial entanglement entropy for a topological ordered system
in (2+1) dimensions.
As will be studied in detail later, it is found that this state can not recover the topological entanglement entropy for
a Chern-Simons theory on a general manifold. There are mainly two reasons as follows:

-- For a conformal boundary state defined in Eq.\ (\ref{ConformalBC}), the amplitude $\psi_i^a$ is fixed through the modular
$\mathcal{S}$ matrix.
However, to study the topological entanglement entropy for a Chern-Simons theory on a general manifold such
as a torus, the ground state can be chosen as an arbitrary superposition of the minimum entangled states (MESs)\cite{YiZhang}.
There is no reason to fix the coefficient $\psi_i^a$ as in Eq.\ (\ref{ConformalBC}).
This indicates that we should choose a state that can be in an arbitrary superposition of Ishibashi states $|h_a\rangle\rangle$.

-- The regularization factor $\frac{e^{-\epsilon H}}{\sqrt{\mathcal{N}_B}}$ in Eq.\ (\ref{bsOld}) acts on the state in a `collective' way (i.e., the regularization factor
is not defined for each Ishibashi state independently, but for the whole superposition
thereof).
This is, however, not the only way to regularize the state.
We may instead regularize each Ishibashi state separately.
This suggests that we may arrange a regularization factor $\frac{e^{-\epsilon H}}{\sqrt{\mathfrak{n}_a}}$ to
each Ishibashi state $|h_a\rangle\rangle$, with the normalization factor $\mathfrak{n}_a$
depending on the primary field $a$.
As will be shown later, this `individual' way of regularization
can correctly recover the spatial topological entanglement entropy
for Chern-Simons theories while the `collective' way of regularization cannot.

Based on the above analysis, we consider an appropriate regularized state as follows
\begin{align}\label{bsNew0}
|\psi\rangle=\sum_a\psi_a|\mathfrak{h}_a\rangle\rangle
\quad
\mbox{where}\quad
|\mathfrak{h}_a\rangle\rangle=\frac{e^{-\epsilon H}}{\sqrt{\mathfrak{n}_a}}|h_a\rangle\rangle,
\end{align}
with $\mathfrak{n}_a$ being a normalization factor so that
\begin{equation}
\langle\langle \mathfrak{h}_a|\mathfrak{h}_b\rangle\rangle=\delta_{ab}.
\end{equation}
Note that $\mathfrak{n}_a$ depends on the type of primary field (or topological sector) $a$.
The amplitude $\psi_a$ in Eq.\ (\ref{bsNew0}) is a complex number which depends on the choice of
ground state of the Chern-Simons field theory on a general manifold.
For the form of the Hamiltonian $H$, following Refs.\ \onlinecite{Qi1,Das}, we consider
\begin{equation}\label{H0}
H=\frac{2\pi}{l}\left(L_0+\overline{L}_0-\frac{c}{12}\right),
\end{equation}
where $l$ is the length of the circle where the state
$|\psi\rangle$ is defined, \textit{e.g.}, the interface
between the subsystems $A$ and $B$ in Fig.\ \ref{sphere} (a).
$c$ is the central charge of the underlying CFT.
The term proportional to $c$ arises from the conformal
transformation from the plane to the cylinder.
It is also instructive to rewrite the Hamiltonian in
Eq.\ (\ref{H0}) as a sum of `chiral Hamiltonian' (or left-moving Hamiltonian) and `anti-chiral Hamiltonian' (or
right-moving Hamiltonian) as $H=H_L+H_R$, where
$
H_L=\frac{2\pi}{l}\left(L_0-\frac{c}{24}\right),
$
and
$
H_R=\frac{2\pi}{l}\left(\overline{L}_0-\frac{c}{24}\right).
$

Now we are ready to calculate the normalization factor $\mathfrak{n}_a$ in $|\mathfrak{h}_a\rangle\rangle$
as follows
\begin{equation}
\begin{split}\label{normalization1}
&\langle\langle \mathfrak{h}_{a'}|\mathfrak{h}_a\rangle\rangle\\
&=
\frac{1}{\sqrt{\mathfrak{n}_{a'}}\sqrt{\mathfrak{n}_{a}}}
\langle\langle h_{a'}|e^{-2\epsilon H}|h_a\rangle\rangle\\
&=\frac{1}{\sqrt{\mathfrak{n}_{a'}}\sqrt{\mathfrak{n}_{a}}}\delta_{a'a}\sum_{N=0}^{\infty}\sum_{j=1}^{d_{h_a}(N)}e^{-\frac{8\pi \epsilon}{l}(h_a+N-\frac{c}{24})}\\
&=:\frac{\delta_{a'a}}{\mathfrak{n}_a}\chi_{h_a}\left(e^{-\frac{8\pi\epsilon}{l}}\right),
\end{split}
\end{equation}
where we have used
\begin{equation}
L_0|h_a,N;j\rangle=(h_a+N)|h_a,N;j\rangle
\end{equation}
and $h_a=\bar{h}_{\bar{a}}$. By requiring that $\langle\langle \mathfrak{h}_a|\mathfrak{h}_b\rangle\rangle=\delta_{ab}$, one can obtain the normalization factor $\mathfrak{n}_a$ as
\begin{equation}\label{na1}
\mathfrak{n}_a=\chi_{h_a}\left(e^{-\frac{8\pi\epsilon}{l}}\right).
\end{equation}
Note that for different primary fields or topological sectors $a$, $\mathfrak{n}_a$ are usually different.

For later use, let us introduce the modular transformation property of the character $\chi$ in CFT, \textit{i.e.},
\begin{equation}\label{na2}
\chi_{h_a}\left(e^{-\frac{8\pi\epsilon}{l}}\right)=\sum_{a'}\mathcal{S}_{aa'}\chi_{h_{a'}}\left(e^{-\frac{\pi l}{2\epsilon}}\right),
\end{equation}
which follows from applying the Poisson summation formula to the explicit expressions
of the character $\chi$ in Eq.\ (\ref{normalization1}), with $\mathcal{S}_{aa'}$ being the matrix elements of the modular $\mathcal{S}$ matrix\cite{BCFT}. In RCFTs, $\mathcal{S}$ is a finite dimensional unitary matrix indexed
by primary fields (or the types of quasiparticles in TQFTs) $\{I, a, b, c\cdots\}$, where $I=0$ labels the identity operator. The anti-quasiparticle of $a$ is denoted by $\bar{a}$, which is the unique quasiparticle that can fuse with $a$ into
$I$ (see appendix for more details).

To avoid confusions, it is helpful to remind ourselves that we will use primary fields, quasiparticles, anyons and topological sectors  back and forth when referring to the label $a$ in $|h_a\rangle\rangle$.

In addition, throughout this work,
we are interested in the spatial entanglement on different closed two-manifold $\mathcal{M}$.
Following Ref. \onlinecite{Dong}, we consider each two dimensional spatial manifold as the boundary of a three-dimensional spacetime manifold $\mathcal{B}$, \textit{i.e.}, $\mathcal{M}=\partial \mathcal{B}$, so that it is convenient to
include the effect of braiding Wilson lines, \textit{etc}. (See Ref. \onlinecite{Dong} for more details.)

\subsection{Left-right entanglement entropy}
\label{LeftRirghtEE}

We now study the reduced density matrix associated to the
(regularized) boundary states,
when we take the partial trace over the right-moving sector.
In particular, we will compute the ``left-right'' entanglement entropy
associated to the reduced density matrix.
This calculation is a necessary exercise for later sections where
we calculate various entanglement/correlation measures in topological quantum liquid.

To see the connection between the left-right entanglement entropy
and the topological entanglement entropy in the simplest setup,
let us consider the geometry in Fig.\ \ref{sphere} (a) for example.
Following Refs.\ \onlinecite{Qi1}, one can use the `cut and glue' strategy.
By cutting the sphere into two semispheres $A$ and $B$,
one has a left-moving chiral CFT (with Hamiltonian $H_L$) and a right-moving antichiral CFT
(with Hamiltonian $H_R$)
on the two physical edges of $A$ and $B$, respectively.
In this case, the left- and right-moving CFTs are the low energy excitations of the subsystems $A$
and $B$, respectively.
Next, by turning on a relevant inter-edge coupling $\lambda H_{LR}$ between the two edges,
the total Hamiltonian for the coupled edge states is $H_L+H_R+\lambda H_{LR}$. For a small
enough $\lambda$, the bulk states in the  subsystems $A$ and $B$ are almost not affected. Therefore,
the entanglement between the subsystem $A$ and subsystem $B$ are reduced to the entanglement between
the left and right moving edge states.

Now let us calculate the left-right entanglement entropy of the regularized state in Eq.\ (\ref{bsNew0}) explicitly.
We start by evaluating the reduced density matrix for the left-moving sector
\begin{equation}
\begin{split}
\rho_L&=\text{Tr}_{R}\left(|\psi\rangle\langle\psi|\right)\\
&=\text{Tr}_{R}\Big(\sum_{a'}\sum_a\psi_{a'}(\psi_a)^{\ast}|\mathfrak{h}_{a'}\rangle\rangle\langle\langle \mathfrak{h}_a|\Big)\\
&=:\sum_a|\psi_a|^2\rho_{L,a},
\end{split}
\end{equation}
where we have defined
\begin{equation}\label{rhoLa}
\begin{split}
\rho_{L,a}&=\text{Tr}_R(|\mathfrak{h}_a\rangle\rangle\langle\langle\mathfrak{h}_a|)\\
&=\sum_{N,j}\frac{1}{\mathfrak{n}_a}e^{-\frac{8\pi \epsilon}{l}(h_a+N-\frac{c}{24})}|h_a,N;j\rangle\langle h_a,N;j|.\\
\end{split}
\end{equation}
(The reduced density matrix for the right-moving part will give the same final result since $S_L=S_R$ for a bipartite system in a pure state.)
To obtain the von Neumann entropy or Renyi entropy, it is convenient to first calculate $\text{Tr}_L\left(\rho_L\right)^n$ as follows
\begin{equation}\label{rhoLn}
\begin{split}
\text{Tr}_L\left(\rho_L\right)^n&=\sum_{a,N,j}|\psi_a|^{2n}\text{Tr}_L(\rho_{L,a})^n\\
&=\sum_{a}\frac{|\psi_a|^{2n}}{\mathfrak{n}_a^n}\chi_{h_a}\left(e^{-\frac{8\pi n \epsilon}{l}}\right)\\
&=\sum_{a}\frac{|\psi_a|^{2n}}{\mathfrak{n}_a^n}\sum_{a'}\mathcal{S}_{aa'}\chi_{h_{a'}}\left(e^{-\frac{\pi l}{2 n\epsilon}}\right),
\end{split}
\end{equation}
where in the last step we have used the modular transformation of the character $\chi_{h_a}$.
By using the explicit form of $\mathfrak{n}_a$ in Eq. (\ref{na1}), $\text{Tr}_L\left(\rho_L\right)^n$ can be further written as
\begin{equation}\label{rho001}
\begin{split}
\text{Tr}_L\left(\rho_L\right)^n
&=\sum_{a}|\psi_a|^{2n}
\frac{\sum_{a'}\mathcal{S}_{aa'}\chi_{h_{a'}}\left(e^{-\frac{\pi l}{2 n\epsilon}}\right)}
{\left[\sum_{a'}\mathcal{S}_{aa'}\chi_{h_{a'}} \left(e^{-\frac{\pi l}{2 \epsilon}}\right)\right]^n  }
\nonumber \\
&\to
e^{\frac{\pi c l}{48 \epsilon}\left(\frac{1}{n}-n\right)} \sum_a |\psi_a|^{2n}(\mathcal{S}_{a0})^{1-n},
\end{split}
\end{equation}
where in the second line, we took the thermodynamic limit $l/\epsilon\to \infty$,
and noted
\begin{equation}
\lim_{l/\epsilon\to \infty}\sum_{a'}\mathcal{S}_{aa'}\chi_{h_{a'}} \left(e^{-\frac{\pi l}{2 n \epsilon}}\right)
=\mathcal{S}_{a0}\times e^{\frac{\pi c l}{48 n \epsilon}},
\end{equation}
i.e., only the identity field $I$,
labeled by ``$0$'' here, survives the limit.
Then based on the definition in Eqs.\ (\ref{Renyi}) and (\ref{vN2}),
one can immediately obtain the Renyi entropy
and the von Neumann entropy as
\begin{align}
\label{Srenyi0}
S^{(n)}_L&=\frac{1}{1-n}\ln \frac{\text{Tr}_L\left(\rho_L\right)^n}{\left(\text{Tr}_L\rho_L\right)^n}
\nonumber \\
&=
\frac{1+n}{n}\cdot \frac{\pi c}{48} \cdot\frac{l}{\epsilon}-\ln \mathcal{D}
+\frac{1}{1-n}\ln \sum_a |\psi_a|^{2n}d_a^{1-n}
\nonumber \\
&\quad
-\frac{n}{1-n}\ln\sum_a|\psi_a|^2,
\nonumber \\
S^{\text{vN}}_L
&=\frac{\pi c}{24}\cdot \frac{l}{\epsilon}-\ln \mathcal{D}
+\frac{\sum_a|\psi_a|^2\ln d_a}{\sum_a|\psi_a|^2}
-\frac{\sum_a |\psi_a|^2\ln |\psi_a|^2}{\sum_a|\psi_a|^2}
\nonumber \\
&\quad
+\ln\sum_a|\psi_a|^2,
\end{align}
where we have used $\mathcal{S}_{a0}=d_a/\mathcal{D}$ (see Eq.\ (\ref{Qdimension})).
The first terms in $S_L^{(n)}$ and $S_L^{\text{vN}}$ in Eq.(\ref{Srenyi0})
are ultraviolet divergent and non-universal, corresponding to the so-called `area law' term in Eq.\ (\ref{Stop}).
The left terms in Eqs.(\ref{Srenyi0}) are independent of the details of
the system. They are determined by the topological property of the system as well as the choice of states, and
therefore are universal.

As a comparison, if one follows the method in Refs.\ \onlinecite{Qi1,Das} to regularize the state in a `collective' way [see Eq.\ (\ref{bsOld})],
then one gets \DIFaddend \cite{Das}
\begin{equation}
\begin{split}\label{DasR}
S^{\text{vN}}_L&=\frac{\pi c}{24}\cdot \frac{l}{\epsilon}
+\ln\sum_a |\psi_a|^2\mathcal{S}_{a0}
-\frac{\sum_a S_{a0}|\psi_a|^2\ln|\psi_a|^2}{\sum_a\mathcal{S}_{a0}|\psi_a|^2}
\end{split}
\end{equation}
which will \emph{not} recover the correct topological entanglement entropy for a Chern-Simons field theory on a general manifold. Nevertheless, it is noted that for
the specific case $|\psi_{a'}|^2=\delta_{aa'}$,
namely the state under consideration is in a definite topological sector $a$, there is no difference between the two methods of regularization.
In this case, both Eq.\ (\ref{Srenyi0}) and (\ref{DasR})  will lead to
$
S^{\text{vN}}_L=\frac{\pi c}{24}\cdot\frac{l}{\epsilon}-\ln \mathcal{D}+\ln d_a.
$

In the rest parts of this work, for most cases we have $\sum_a|\psi_a|^2=1$, and then the Renyi entropy and
the von Neumann entropy for the left moving CFT (or the right moving CFT) can be further simplified as
\begin{align}
\label{Srenyi}
S^{(n)}_L
&=
\frac{1+n}{n}\cdot \frac{\pi c}{48} \cdot\frac{l}{\epsilon}-\ln \mathcal{D}+\frac{1}{1-n}\ln \sum_a |\psi_a|^{2n}d_a^{1-n},
\nonumber \\
S^{\text{vN}}_L
&=\frac{\pi c}{24}\cdot \frac{l}{\epsilon}-\ln \mathcal{D}
+\sum_a|\psi_a|^2\ln d_a
-\sum_a |\psi_a|^2\ln |\psi_a|^2.
\end{align}

Before ending this section,
it is worth mentioning that we will come across in later section
the state of the form
\begin{equation}\label{DirectState}
|\psi\rangle=\bigoplus_a\psi_a|\mathfrak{h}_a\rangle\rangle
\end{equation}
in the study of multi Wilson lines. Then the reduced density matrix $\rho_L$ can be
expressed as $\rho_L=\bigoplus_a|\psi_a|^2\rho_{L,a}$, with the same $\rho_{L,a}$ defined in Eq.\ (\ref{rhoLa}).
It is straightforward to check that $\text{Tr}_L\left(\rho_L\right)^n$ has the same expression as Eq.\ (\ref{rhoLn}).
This indicates that our results in Eqs.\ (\ref{Srenyi0}) and (\ref{Srenyi}) still hold for this case.

\section{Topological entanglement entropy}
\label{EE}

In this section, by using the edge theory approach,
we study the
entanglement entropy associated for a given spatial region in
Chern-Simons theories
defined on different kinds of two spatial manifolds.

\subsection{Sphere}
\label{SphereP}

\begin{figure}[t]
\includegraphics[width=3.25in]{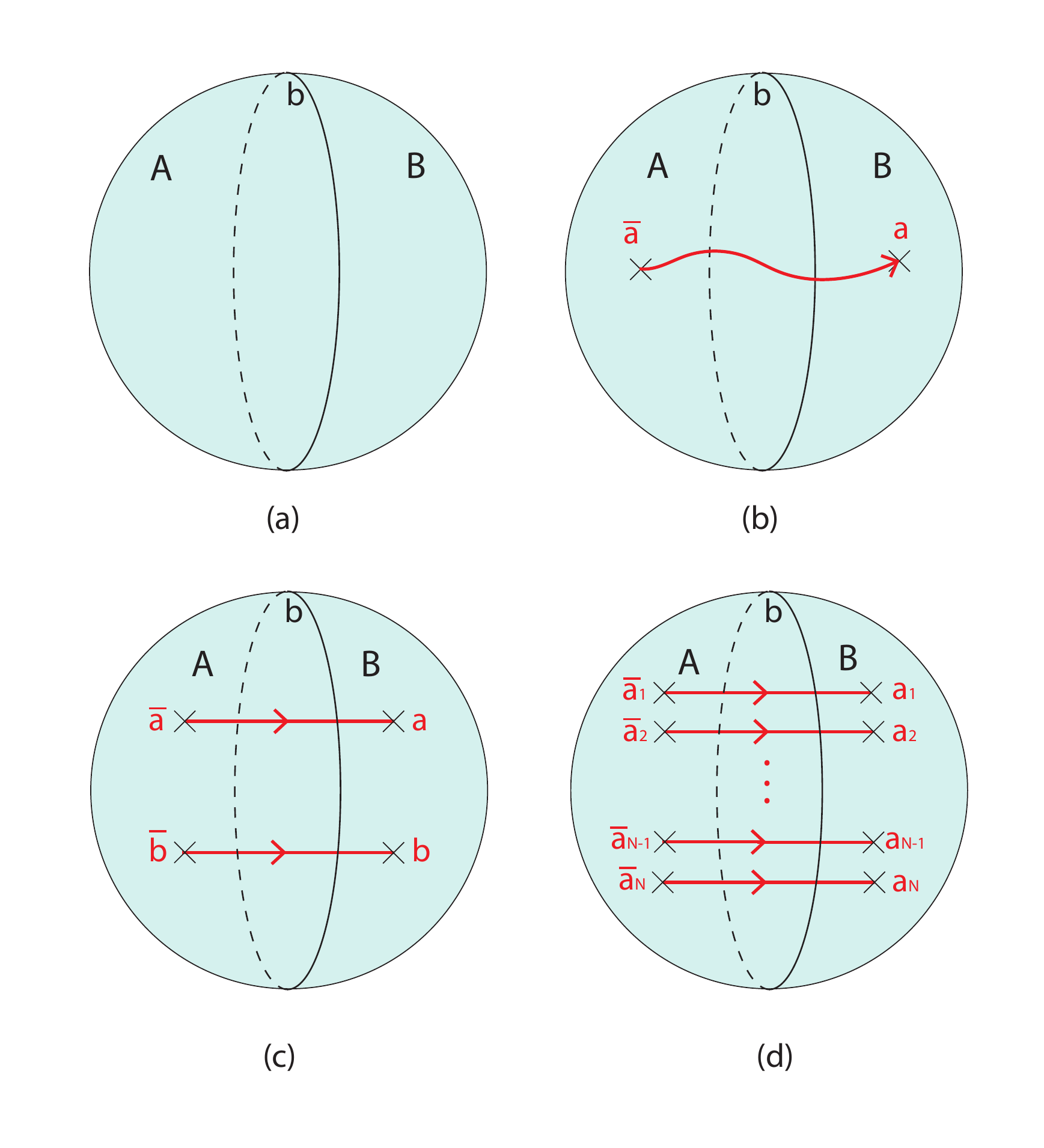}
\caption{
Various setups discussed in Sec.\ \ref{SphereP}.
(a) A $S^2$ is bipartited into two subsystems $A$ and $B$, with the interface labeled by $b$.  (b) A $S^2$ with a quasiparticle $a$ and an anti-quasiparticle $\bar{a}$.  A Wilson line connecting the two quasiparticles
threads through the interface $b$.
The two quasiparticles correspond to two punctures, and therefore the geometry in (b) is equivalent to
a cylinder in topology.
(c) A $S^2$ with two pairs of quasiparticles.
(d) A $S^2$ with $N$ pairs of quasiparticles.
}\label{sphere}
\end{figure}

%

\subsubsection{Sphere}

As shown in Fig.\ \ref{sphere} (a), let us consider a Chern-Simons theory which lives on the simplest closed manifold in two spatial dimensions, \textit{i.e.}, a sphere.
We are interested in the entanglement entropy for the subsystem $A$ ($B$).
For simplicity,
let us first assume that there is no quasiparticle on the sphere,
and therefore no Wilson lines thread through the
interface $b$. In this case, one has  $|\psi_a|^2=\delta_{a0}$ for the regularized state in Eq.\ (\ref{bsNew0}).
Then by using the results in Eqs.\ (\ref{Srenyi}), one can immediately obtain
\begin{align}\label{SphereRenyi}
S^{(n)}_A&=\frac{1+n}{n}\cdot \frac{\pi c}{48} \cdot\frac{l}{\epsilon}-\ln \mathcal{D},\nonumber
\\
S^{\text{vN}}_A&=\frac{\pi c}{24}\cdot\frac{l}{\epsilon}-\ln \mathcal{D}.
\end{align}
Based on the equation above, one can find that
the topological entanglement entropy is independent of the Renyi index $n$,
and only depends on the total quantum dimension $\mathcal{D}$.

The above calculation is based on a $S^2$ with a single interface between $A$ and $B$.
It is straightforward to generalize it to a $S^2$ with
multiple (=$M$) interfaces between $A$ and $B$.
In this case, the wave function under consideration can be expressed as
\begin{equation}
|\psi\rangle=\otimes_{i=1}^M |\mathfrak{h}^i_I\rangle\rangle,
\end{equation}
where $i$ labels the $i$-th component interface, and $I$ refers to the identity primary operator.
By using the method in Sec.\ \ref{LR},
one obtains the Renyi  and the von Neumann entropy as
\begin{align}\label{SphereRenyiM}
S^{(n)}_A&=\frac{1+n}{n}\cdot \frac{\pi c}{48}\cdot \sum_{i=1}^M \frac{ l_i}{\epsilon}-M\ln \mathcal{D},
\nonumber \\
S^{\text{vN}}_A&=\frac{\pi c}{24}\cdot \sum_{i=1}^M\frac{ l_i}{\epsilon}-M\ln \mathcal{D},
\end{align}
where $l_i$ represents the length of the $i$-th component of $AB$ interface.
For the universal part of the entanglement entropy,
one can find that each interface contributes $-\ln \mathcal{D}$.

\subsubsection{Sphere with two quasiparticles = cylinder}

As shown in Fig.\ \ref{sphere} (b),
let us now consider a sphere with two quasiparticles, with $\bar{a}$ in subsystem $A$
and $a$ in subsystem $B$. This configuration corresponds to a $S^2$ with two punctures, which
 is equivalent to cylinderical topology.
In this case, there is a Wilson line corresponding to topological sector $a$ threading through the $AB$ interface. Then
one has $|\psi_{a'}|^2=\delta_{a'a}$
for the regularized state
$|\psi\rangle=\sum_{a'}\psi_{a'}|\mathfrak{h}_{a'}\rangle\rangle$.
Then, based on Eq.(\ref{Srenyi}),
the Renyi and the von Neumann entropy for subsystem $A$ have the expressions as follows
\begin{align}
S^{(n)}_A
&=\frac{1+n}{n}\cdot \frac{\pi c}{48} \cdot\frac{l}{\epsilon}-\ln \mathcal{D}+\ln d_a,
\nonumber \\
S^{\text{vN}}_A&= \frac{\pi c}{24} \cdot\frac{l}{\epsilon}-\ln \mathcal{D}+\ln d_a.
\end{align}
Again, the universal part of entanglement entropy is independent of the Renyi index $n$.
Compared with the results on a sphere with no quasiparticles, the entanglement entropy here is increased
by $\ln d_{a}$.
The physical picture is as follows: For $d_a>1$, the underlying theory is non-Abelian.
The quasiparticle $a$ and antiparticle $\bar{a}$ can fuse into, apart from the identity $I$,
other types of quasiparticles.
This increases the uncertainty that is shared by the two semispheres.
If the underlying theory is Abelian, then $d_{a}=1$ and $\ln d_{a}=0$.
This is because in the Abelian
case, $a$ and $\bar{a}$ can only fuse into $I$, and therefore cannot increase
the uncertainty shared by $A$ and $B$.

\subsubsection{A sphere with $N$ Wilson lines}

As a generalization of the previous part, it is natural to ask what is the entanglement entropy of the subsystem $A$$(B)$
if there are more than one Wilson lines (or more than one pair of quasiparticles) on a sphere, as shown in Fig.\ \ref{sphere} (c) and (d).
The strategy we will use is to fuse the quasiparticles (or anyons) based on the fusion rule:
\begin{equation}
a\otimes b=\bigoplus_cN_{ab}^c c,
\end{equation}
where the fusion coefficients $N_{ab}^c$ are non-negative integers, and $a,b,c$ represent the topological
or anyon charges.
In the following discussions,
for simplicity, we will consider the multiplicity free case, i.e., $N_{ab}^c=0$ or $1$.
For the case with $N_{ab}^c>1$, one needs to include an orthonormal set of bases to
count the number of times that $c$ appears by fusing $a$ and $b$.

As a warm-up, let us first consider the case with two Wilson lines.
As shown in  Fig.\ \ref{sphere} (c), the two
Wilson lines are in topological sectors $a$ and $b$, respectively.
After the fusion, the state at the interface
may be expressed as
\begin{equation}\label{AnyonFusion}
|\psi\rangle=\bigoplus_c \psi_{ab}^c |\mathfrak{h}_{ab\to c}\rangle\rangle.
\end{equation}
For the regularized Ishibashi state $|\mathfrak{h}_{ab\to c}\rangle\rangle$, it has the same
expression as $|\mathfrak{h}_c\rangle\rangle$, as defined in Eq.\ (\ref{bsNew0}). However,
 we use $|\mathfrak{h}_{ab\to c}\rangle\rangle$ instead of $|\mathfrak{h}_c\rangle\rangle$
to emphasize that now
the orthonormal property of $|\mathfrak{h}_{ab\to c}\rangle\rangle$ also depends on the fusion history, \textit{i.e.},
\begin{equation}
\langle\langle \mathfrak{h}_{ab\to c}| \mathfrak{h}_{a'b'\to c'}\rangle\rangle=\delta_{aa'}\delta_{bb'}\delta_{cc'}.
\end{equation}
In the surgery method\cite{Dong},
to obtain this result,
one needs to glue Wilson lines
$a$ and $b$ with Wilson lines $a'$ and $b'$, respectively,
resulting in the factor $\delta_{aa'}\delta_{bb'}$.
From the topological field theory, it can be shown that
$\psi_{ab}^c$ in Eq.\ (\ref{AnyonFusion}) satisfies\cite{Kitaev2006} (see also Appendices)
\begin{equation}\label{Pabc}
|\psi_{ab}^c|^2=P_{ab\to c}=N_{ab}^c\frac{d_c}{d_ad_b},
\end{equation}
where $d_i$ is the quantum dimension of the quasiparticle $i$, and $P_{ab\to c}$ is the probability of fusing $a$ and $b$ into $c$.
 It is required that $\sum_cP_{ab\to c}=1$, and therefore
\begin{equation}\label{P1}
d_ad_b=\sum_cN_{ab}^c d_c.
\end{equation}
The density matrix corresponding to the state (\ref{AnyonFusion})
can be written as
\begin{equation}
\rho=\bigoplus_c|\psi_{ab}^c|^2\rho_c,
\end{equation}
with $\rho_c=|\mathfrak{h}_{ab\to c}\rangle\rangle\langle\langle \mathfrak{h}_{ab\to c}|$.
Based on the discussion around Eq.\ (\ref{DirectState}),
one can directly use the results in Eq.\ (\ref{Srenyi}).
Then the Renyi entropy for subsystem $A$ is expressed as
\begin{equation}
\begin{split}
S^{(n)}_A
&=
\frac{1+n}{n}\cdot \frac{\pi c}{48} \cdot\frac{l}{\epsilon}-\ln \mathcal{D}+\frac{1}{1-n}\ln \sum_c |\psi_{ab}^c|^{2n}d_c^{1-n}.
\end{split}
\end{equation}
By using Eqs.\ (\ref{Pabc}) and (\ref{P1}), one can further obtain
\begin{align}\label{Srenyi2W}
S^{(n)}_A
&=
\frac{1+n}{n}\cdot \frac{\pi c}{48} \cdot\frac{l}{\epsilon}-\ln \mathcal{D}
+
\frac{1}{1-n}\ln \sum_c \frac{N_{ab}^c d_c}{(d_ad_b)^n}
\nonumber \\
&=
\frac{1+n}{n}\cdot \frac{\pi c}{48} \cdot\frac{l}{\epsilon}-\ln \mathcal{D}
+\ln d_a+\ln d_b,
\nonumber \\
S^{\text{vN}}_A&=\frac{\pi c}{24}\cdot \frac{l}{\epsilon}
-\ln \mathcal{D}
+\ln d_a+\ln d_b.
\end{align}

Based on the above example, now we are ready to study the more general case with $N$
Wilson lines threading through the interface, as shown in Fig.\ \ref{sphere} (d).
Suppose that the $N$ Wilson lines
are in topological sectors $a_1, a_2, \cdots, a_N$ respectively, let us fuse them in the
following order.
We first fuse $a_1$ and $a_2$ into $b_1$, and then fuse $b_1$ and $a_3$ into
$b_2$. By repeating this procedure, we finally fuse $b_{N-2}$ and $a_N$ into $c$.
The state we need to consider can be expressed as
\begin{equation}\label{AnyonFusionM}
|\psi\rangle=\bigoplus_{\{b_i\},c} \psi_{a_1,a_2,\cdots,a_N}^c\left( b_1,b_2,\cdots, b_{N-2} \right) |\mathfrak{h}_{a_1\cdots a_N\to c}\rangle\rangle.
\end{equation}
Note that the direct sum is not only over $c$, but also over $\{b_i\}$, which means that
the final fusion result also depends on the fusion channels $\{b_i\}$ in the middle.
For a specific fusion channel in $\{b_i\}$, one has
\begin{equation}
\begin{split}
&\big |\psi_{a_1,a_2,\cdots,a_N}^c\left( b_1,b_2,\cdots, b_{N-2} \right) \big|^2\\
&=
\frac{N_{a_Nb_{N-2}}^c d_{c}}{d_{a_N}d_{b_{N-2}}}\cdots
\frac{N_{a_3b_1}^{b_2}d_{b_2}}{d_{a_3}d_{b_1}} \frac{N_{a_1a_2}^{b_1}d_{b_1}}{d_{a_1}d_{a_2}}\\
&=N_{a_Nb_{N-2}}^c\cdots N_{a_3b_1}^{b_2}N_{a_2a_1}^{b_1}\frac{d_c}{d_{a_N}\cdots d_{a_2}d_{a_1}}.
\end{split}
\end{equation}
Based on the wave function (\ref{AnyonFusionM}), and
relabeling $\psi_{a_1,a_2,\cdots,a_N}^c\left( b_1,b_2,\cdots, b_{N-2} \right)$ as $\psi_a^c(b)$ to simplify notations,
the Renyi entropy of the subsystem $A$ can be expressed as
\begin{equation}
\begin{split}
S^{(n)}_A
&=
\frac{1+n}{n}\cdot \frac{\pi c}{48} \cdot\frac{l}{\epsilon}-\ln \mathcal{D}\\
&+\frac{1}{1-n}\ln \sum_c\sum_{b_1}\sum_{b_2}\cdots\sum_{b_{N-2}} \big|\psi_{a}^c(b)\big|^{2n}d_c^{1-n}.
\end{split}
\end{equation}
After some simple algebra,
one obtains
\begin{align}\label{SRenyimanyW}
S^{(n)}_A
&=
\frac{1+n}{n}\cdot \frac{\pi c}{48} \cdot\frac{l}{\epsilon}-\ln\mathcal{D}+\sum_i \ln {d_{a_i}},
\nonumber \\
S^{\text{vN}}_A&=
\frac{\pi c}{24}\cdot \frac{l}{\epsilon}
-\ln\mathcal{D}
+\sum_i \ln {d_{a_i}}.
\end{align}
This results (\ref{SRenyimanyW}) can be easily understood by considering the
additivity property of entanglement entropy.
Each Wilson line in the topological sector $a_i$ increases the
entanglement entropy by $\ln d_{a_i}$.

\subsection{Torus}

\begin{figure}[t]
\includegraphics[width=3.5in]{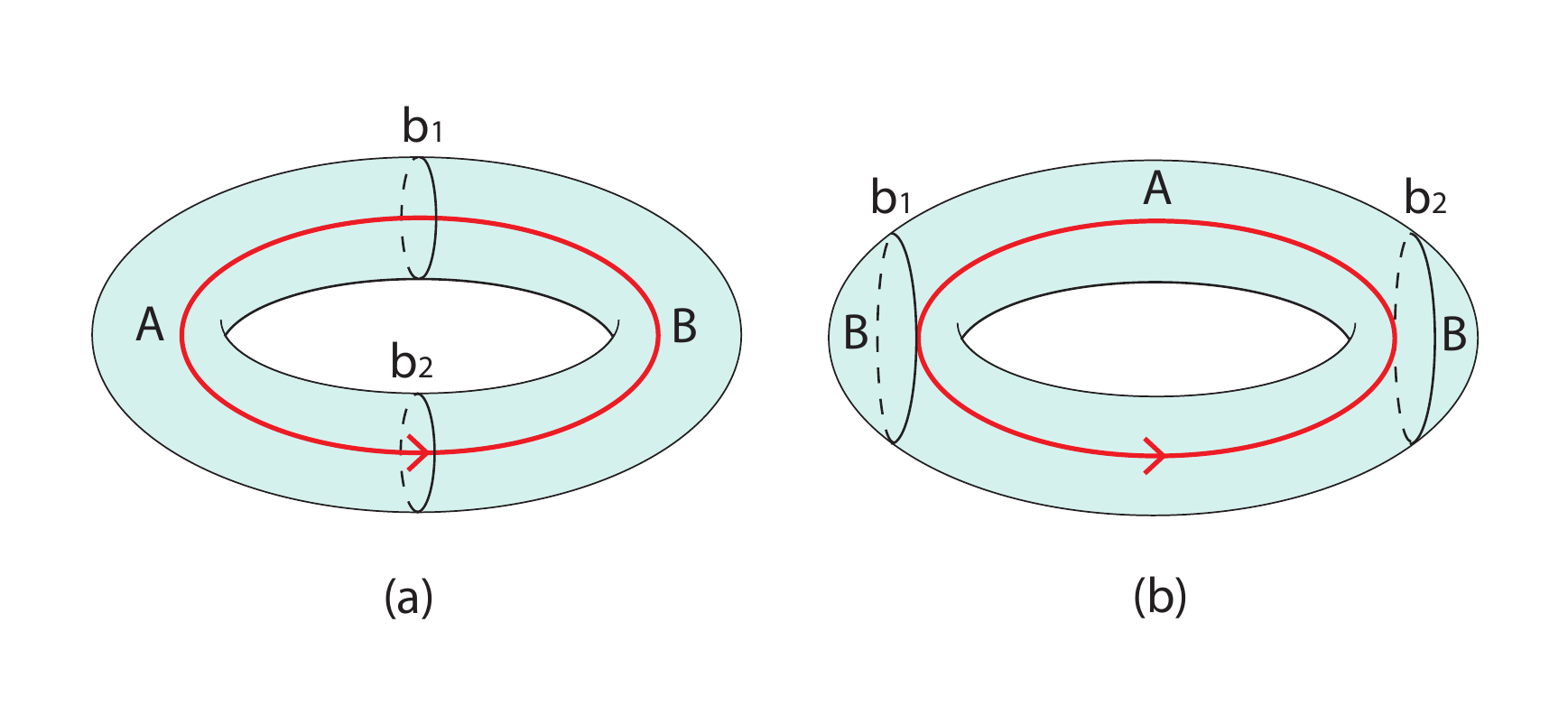}
\caption{
A $T^2$ with a two-component $AB$ interface. The region $B$ is connected in (a) and disconnected in (b).
$b_1$ and $b_2$ denote the interface that separates $A$ from $B$.
The red solid line represents a Wilson loop which may fluctuate among different topological sectors.
}\label{torus}
\end{figure}

In this part, we consider a torus with  a two-component $AB$ interface.
There are many ways to slice the spatial surface,
and here we mainly focus on the two slicing
shown in  Figs.\ \ref{torus} (a) and (b), respectively.

\subsubsection{Connected $B$ region}

As shown in Figs.\ \ref{torus} (a) and (b), for the torus geometry, the Wilson loop can in general fluctuate among different
topological sectors $a$ with probability $|\psi_a|^2$. In this case, the ground state may be written as
\begin{equation}\label{StateTorus}
|\Psi\rangle=\sum_a\psi_a|W_a\rangle,
\end{equation}
where $|W_a\rangle$ represents the state that the Wilson loop is in a definite topological sector $a$.
In Ref.\ \onlinecite{YiZhang}, $|W_a\rangle$ are also called minimal entangled states (MESs).
It is noted that here we use the bulk wavefunction $|\Psi\rangle$ to distinguish it from $|\psi\rangle$ which represents the state at the interface.

For the configuration in Fig.\ \ref{torus} (a), the Wilson loop threads through both
$b_1$ and $b_2$.  Then the wavefunction at the interface may be written as
\begin{align}\label{TwoEdge}
&
|\psi\rangle=\sum_a\psi_a|\mathfrak{h}_a^{b_1}\rangle\rangle\otimes |\mathfrak{h}_a^{b_2}\rangle\rangle
\nonumber \\
&
\mbox{where}\quad
|\mathfrak{h}_a^{b_i}\rangle\rangle=\frac{e^{-\epsilon H_i}}{\sqrt{\mathfrak{n}^{b_i}_a}}|h_a^{b_i}\rangle\rangle,\ \ \ i=1,2
\nonumber \\
&
\hphantom{\mbox{where}}\quad
H_i=\frac{2\pi}{l_i}\left(L_0^i+\overline{L}^i_0-\frac{c}{12}\right),
\nonumber \\
&
\hphantom{\mbox{where}}\quad
\mathfrak{n}_a^{b_i}=\chi_{h_a}\left(e^{-\frac{8\pi\epsilon}{l_i}}\right).
\end{align}
$l_i$ represents the length of the $i$-th component of interface.
Then by following similar procedures in the case of single interface on a cylinder,
we can get the reduced density matrix for the subsystem $A$ as
\begin{equation}
\begin{split}
\rho_A&=\sum_a |\psi_a|^2\rho_{A,a}^{b_1}\otimes \rho_{A,a}^{b_2},
\end{split}
\end{equation}
with
\begin{equation}
\begin{split}
\rho_{A,a}^{b_1}&=\frac{1}{\mathfrak{n}_a^{b_1}}\sum_{N_1,j_1}e^{-\frac{8\pi\epsilon}{l_1}}
|h_a^{b_1},N_1;j_1\rangle\langle h_a^{b_1},N_1;j_1|,\\
\rho_{A,a}^{b_2}&=\frac{1}{\mathfrak{n}_a^{b_2}}\sum_{N_2,j_2}e^{-\frac{8\pi\epsilon}{l_2}}
|\overline{h_a^{b_2},N_2;j_2}\rangle\langle \overline{h_a^{b_2},N_2;j_2}|,\\
\end{split}
\end{equation}
where we have considered that the chirality of edge states at $b_1$ and $b_2$ are opposite to each other, if there is a physical cut.
Then  one can get
\begin{equation}\label{rhoTorus}
\begin{split}
\text{Tr}_A\left(\rho_A^n\right)
=\sum_{a}\big|\psi_a\big|^{2n}\prod_{i=1,2}
\frac{\left[\sum_{a_i}\mathcal{S}_{aa_i}\chi_{h_{a_i}}\left(e^{-\frac{\pi l_i}{2 n\epsilon}}\right)\right]}
{\left[\sum_{a_i}\mathcal{S}_{aa_i}\chi_{h_{a_i}} \left(e^{-\frac{\pi l_i}{2 \epsilon}}\right)\right]^n
 }\\
\end{split}
\end{equation}
where we have used the modular transformation of characters $\chi_{h_{a_i}}$.
In the thermodynamic limit $l/\epsilon\to \infty$, Eq.\ (\ref{rhoTorus}) can be further simplified as
\begin{equation}
\text{Tr}_A\left(\rho_A^n\right)=e^{\frac{\pi c (l_1+l_2)}{48 \epsilon}\left(\frac{1}{n}-n\right)}\sum_a |\psi_a|^{2n}(\mathcal{S}_{a0})^{2-2n}.
\end{equation}
Then by using the definition in Eq.\ (\ref{Renyi}),
one obtains the Renyi and von Neumann entropy for subsystem $A$ as follows
\begin{align}\label{RenyiTorus}
S^{(n)}_A
&=\frac{1+n}{n}\cdot \frac{\pi c}{48} \cdot\frac{l_1+l_2}{\epsilon}
-2\ln \mathcal{D}
\nonumber \\
&\quad
+\frac{1}{1-n}\ln \sum_a |\psi_a|^{2n}d_a^{2-2n},
\nonumber \\
S^{\text{vN}}_A
&=\frac{\pi c}{24} \cdot\frac{l_1+l_2}{\epsilon}-2\ln \mathcal{D}
\nonumber \\
&\quad
+2\sum_a|\psi_a|^2\ln d_a-\sum_a|\psi_a|^2\ln|\psi_a|^2.
\end{align}
The first term above is the area law term. The left terms, which are universal,
are exactly the same as the results obtained with replica trick and surgery method
in Ref.\ \onlinecite{Dong}.
The topological entanglement entropy in this case depends not only on quantum
dimensions but also on the choice of ground state.
On the other hand, it is noted that the formulas in Refs.\ \onlinecite{Qi1,Das} can not recover this result,
because of the inappropriate
regularization scheme.

\subsubsection{ Disconnected $B$ regions}
\label{torusDis}

As shown in Fig.\ \ref{torus} (b), this case is trivial compared with the configuration in Fig.\ \ref{torus} (a),
since there is no Wilson loop threading through the
interface $b_1$ and $b_2$. In this case, we simply make
$|\psi_a|^2=\delta_{a0}$ in Eq.\ (\ref{RenyiTorus}). Then one can obtain
the Renyi entropy and the von Neumann entropy for subsystem $A$ as follows
\begin{align}\label{RenyiTorus2}
S^{(n)}_A
&=\frac{1+n}{n}\cdot\frac{\pi c}{48}\cdot\frac{l_1+l_2}{\epsilon}-2\ln \mathcal{D},
\nonumber \\
S^{\text{vN}}_A
&=\frac{\pi c}{24}\cdot\frac{l_1+l_2}{\epsilon}-2\ln \mathcal{D}.
\end{align}
The universal parts of the entanglement entropy in Eq.\ (\ref{RenyiTorus2}) agree
with the results in Ref.\ \onlinecite{Dong}, as expected. In addition, by comparing with
Eq.\ (\ref{SphereRenyiM}), it is found that the results here are the same
as the entanglement entropy for a $S^2$ with a two-component $AB$ interface.
This is reasonable by considering that the Wilson loop in Fig.\ \ref{torus} (b) does not thread through the
$AB$ interface, and therefore has no effect on the entanglement entropy of the subsystem $A$ $(B)$.

\subsubsection{Effects of the modular $\mathcal{S}$ matrix}

\begin{figure}[t]
\includegraphics[width=2.3in]{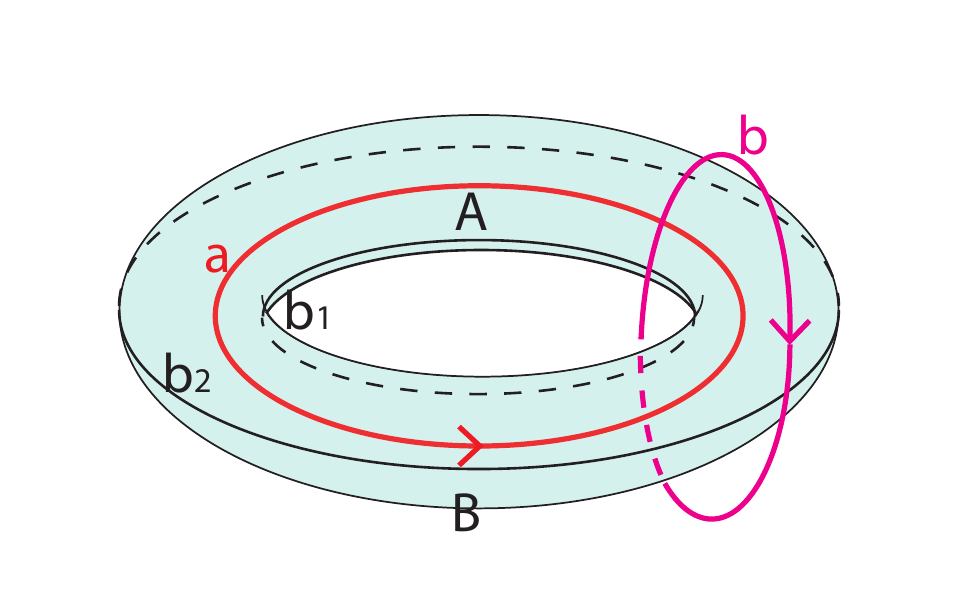}
\caption{
A $T^2$ with a two-component $AB$ interface labeled by $b_1$ and $b_2$.
Compared to Fig.\ \ref{torus}, the bipartition is along the other non-contractible cycle
on $T^2$.
The red (magenta) solid line represents the Wilson loop threading through the interior(exterior) of the torus along the
 longitudinal(meridional) circle.
}\label{TorusS}
\end{figure}

Now we consider the bipartition of a torus as shown in Fig.\ \ref{TorusS}.
In this case, it is convenient to consider the Wilson loop that threads through the entanglement cut, i.e., the Wilson loop
threading through the exterior of the torus around the meridional cycle.
As shown in Fig.\ \ref{TorusS}, by labeling the basis of the degenerate ground state as
$|W_a\rangle_l$ and $|W_b\rangle_m$ respectively ($l$ represents
`longitudinal' and $m$ represents `meridional'), where
$|W_a\rangle_l$ ($|W_b\rangle_m$) represents the state that the Wilson line along the longitudinal(meridional) circle
carries a definite topological flux $a$ ($b$),
 we can express
the state in Eq.\ (\ref{StateTorus}) with either set of bases. In particular, the two sets of bases are related by the modular
$\mathcal{S}$ matrix as follows \cite{YiZhang,Barkeshli}
\begin{equation}
|W_a\rangle_l=\sum_b\mathcal{S}_{ab}|W_b\rangle_m.
\end{equation}
Then the state in Eq.\ (\ref{StateTorus}) may be rewritten as
\begin{equation}\label{StateTorus2}
\begin{split}
|\Psi\rangle&=\sum_a\psi_a|W_a\rangle_l\\
&=\sum_b\left(
\sum_a\psi_aS_{ab}
\right)|W_b\rangle_m\\
&=:\sum_b\phi_b|W_b\rangle_m,
\end{split}
\end{equation}
where we have defined $\phi_a=\sum_i\psi_i\mathcal{S}_{ia}$.
Then the state at the interface can be expressed as
\begin{equation}
|\psi\rangle=\sum_a\phi_a|\mathfrak{h}_a^{b_1}\rangle\rangle\otimes |\mathfrak{h}_a^{b_2}\rangle\rangle.
\end{equation}
By using the formulas (\ref{RenyiTorus}), one can immediately obtain the Renyi and the von Neumann entropy of subsystem $A$ as
\begin{align}
S^{(n)}_A
&=\frac{1+n}{n}\cdot \frac{\pi c}{48} \cdot\frac{l_1+l_2}{\epsilon}-2\ln \mathcal{D}
\nonumber \\
&\quad
+\frac{1}{1-n}\ln \sum_a |\phi_a|^{2n}d_a^{2-2n},
\nonumber \\
S^{\text{vN}}_A&=\frac{\pi c}{24}\cdot \frac{l_1+l_2}{\epsilon}-2\ln \mathcal{D}
\nonumber \\
&\quad
+2\sum_a|\phi_a|^2\ln d_a
-\sum_a |\phi_a|^2\ln |\phi_a|^2.
\end{align}
As an example, let us consider the specific case $\psi_a=\delta_{a0}$ in Eq.\ (\ref{StateTorus}),
i.e., the Wilson loop $a$  in the longitudinal circle
is in the identity topological sector $I$. For the entanglement cut in Fig.\ \ref{torus} (a), the universal parts of $S_A^{(n)}$ and $S_A^{\text{vN}}$ are both $-2\ln \mathcal{D}$, which is in the minimal value.
On the other hand, for the entanglement cut in Fig.\ \ref{TorusS}, we have $\phi_a=\mathcal{S}_{0a}=d_a/\mathcal{D}$, and then
it is straightforward to check that the universal parts of $S_A^{(n)}$ and $S_A^{\text{vN}}$ are both $0$,
which is in the maximal value. This is as expected by considering that the Wilson loop operators corresponding to
 the longitudinal and meridional circles do not commute with each other.

\subsection{Manifolds of genus $g$}

In this part, we consider general manifolds of genus $g$.
As a warm-up, we will first consider a simple case with
$g=2$, and then move on to the general case with arbitrary $g$.

\subsubsection{Double torus}

\begin{figure}[t]
\includegraphics[width=3.2in]{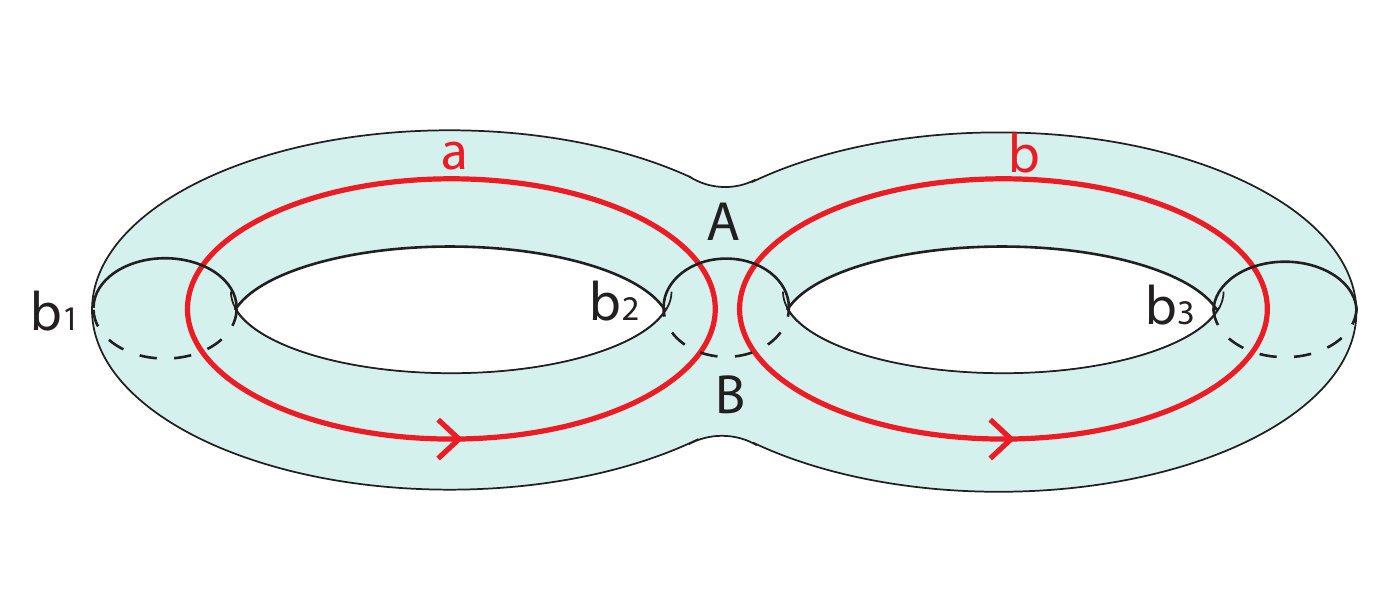}
\caption{A manifold of genus $g=2$. We have three components of $AB$ interfaces
labeled by $b_1$, $b_2$ and $b_3$, respectively.
The red solid lines $a$ and $b$ represent two independent Wilson loops
threading through the interior of the double torus along the longitudinal circles.
}\label{Torus2g}
\end{figure}

\begin{figure*}
\includegraphics[width=6.0in]{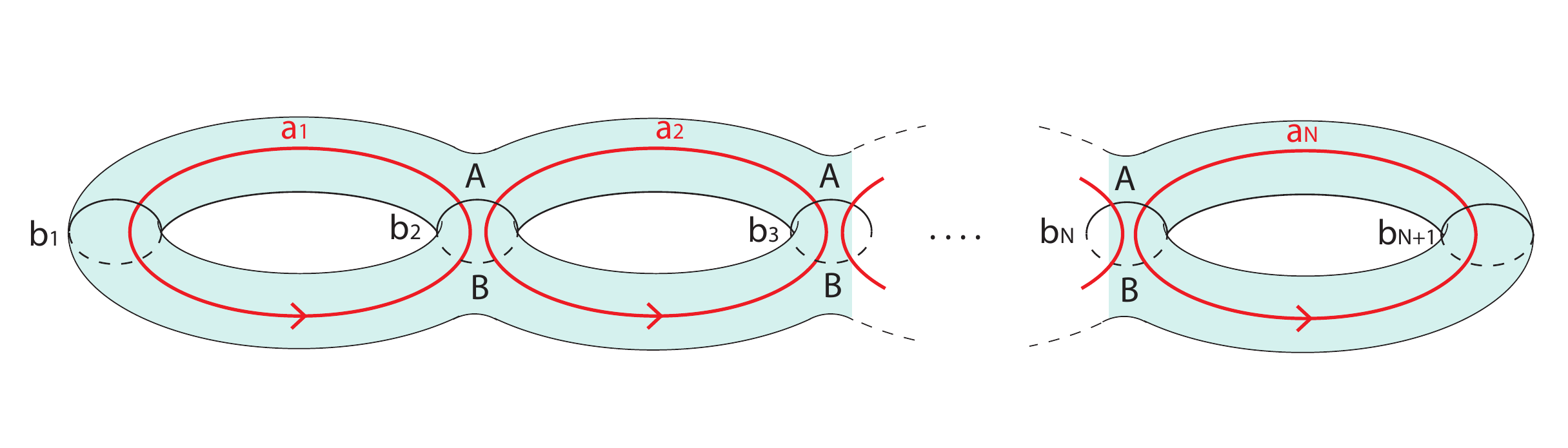}
\caption{A manifold of genus $g$ with $g=N$.
We have a $(N+1)$-component interface labeled by $b_1, b_2, \cdots, b_{N+1}$, respectively.
We consider $N$ independent Wilson loops that thread through the interior of the manifold along the longitudinal circles. Each Wilson loop (red solid lines) can fluctuate among different topological sectors independently.
}\label{Gtorus}
\end{figure*}

Let us consider a double torus with three components of $AB$ interfaces
as shown in Fig.\ \ref{Torus2g}.
We consider two independent Wilson loops that thread through the $AB$ interface along the longitudinal circles
\footnote{Generally, we allow a third Wilson $c$ to connect Wilson loops $a$ and $b$,
where the topological sector $c$ satisfies the constrain that both $N_{a\bar{a}}^c$ and $N_{b\bar{b}}^c$ are non-vanishing.
Here, for simplicity, we consider the case that $c=I$, \textit{i.e.}, the third Wilson line
is in the identity sector, so that the ground state can be expressed in terms 
of two independent Wilson loops. For the general case with $c\neq I$, one 
can use the $F$-move to change the basis (See, \textit{e.g.}, the operation in Fig.4 of a recent paper Ref. \onlinecite{MB}.)
Then there is only one Wilson line threading through the interface $b_2$ in Fig. \ref{Torus2g}, and one can
immediately write down the corresponding Ishibashi state. The same procedures apply to a manilfold of genus $g$.
}.
For the configuration in Fig.\ \ref{Torus2g}, where the Wilson loops $a$ and $b$ fluctuate independently, the bulk wave function may be written as
\begin{equation}
|\Psi\rangle=\left(\sum_a\psi_a|W_a\rangle\right)\bigotimes\left(\sum_b\psi_b|W_b\rangle\right).
\end{equation}
Focusing on the $AB$ interface $b_1$, $b_2$ and $b_3$,
the wave function may be expressed as
\begin{equation}\label{2genusState}
|\psi\rangle=\sum_{ab} \psi_a\psi_b
|\mathfrak{h}_a^{b_1}\rangle\rangle
\otimes \left(\oplus_c \psi_{ab}^c |\mathfrak{h}_{ab\to c}^{b_2}\rangle\rangle\right)
\otimes |\mathfrak{h}_b^{b_3}\rangle\rangle,
\end{equation}
where we have used $b_i$ with $i=1,2,3$ to label the $i$-th component of $AB$ interface. The fusion
probability at interface $b_2$ has the form $|\psi_{ab}^c|^2=N_{ab}^cd_c/d_ad_b$.
Then the reduced density matrix for the subsystem $A$  may be written as
\begin{align}\label{rhoA2torus}
\rho_A&=\text{Tr}_B|\psi\rangle\langle \psi|
\nonumber \\
&=\sum_{a,b}|\psi_a|^2|\psi_b|^2\frac{1}{\mathfrak{n}_a\mathfrak{n}_b}
\nonumber \\
&\quad \sum_{N_1,j_1}e^{-\frac{8\pi \epsilon}{l_1}(h_a+N_1-\frac{c}{24})}|h_a^{b_1},N_1;j_1\rangle\langle h_a^{b_1},N_1;j_1|
\nonumber \\
& \quad \otimes\sum_{N_3,j_3}e^{-\frac{8\pi \epsilon}{l_3}(h_b+N_3-\frac{c}{24})}|h_b^{b_3},N_3;j_3\rangle\langle h_b^{b_3},N_3;j_3|
\nonumber \\
&\quad\otimes\Big(\oplus_{c}\frac{N_{ab}^c}{\mathfrak{n}_c}\frac{d_c}{d_ad_b}\cdot
\sum_{N_2,j_2}e^{-\frac{8\pi\epsilon(h_c+N_2-\frac{c}{24})}{l_2}}
\nonumber \\
&\qquad
\times |h_c^{b_2},N_2;j_2\rangle\langle h_c^{b_2},N_2;j_2|\Big).
\end{align}
Note that for the configuration in Fig.\ \ref{Torus2g},
imagining a physical cut along $b_1$, $b_2$ and $b_3$, then
there may be an ambiguity in defining the chirality of edge states
for the subsystem $A$ ($B$). Here, for simplicity, we choose all the
edge states to be left-moving.
In fact, it can be checked that the freedom of choosing the chirality of
edge states has no effect on the entanglement entropy.
In the rest of this work, once there is an ambiguity
in defining the chirality of edge states, without affecting the results,
we may choose it to be left-moving.

Based on $\rho_A$ in Eq.\ (\ref{rhoA2torus}), one can obtain
\begin{equation}\label{rhoN3g}
\begin{split}
\text{Tr}\left(\rho_A^n\right)
&=
\sum_{a,b,c}|\psi_a|^{2n}|\psi_b|^{2n}\left(\frac{1}{\mathfrak{n}_a
\mathfrak{n}_b\mathfrak{n}_c
}\right)^n
\left(\frac{d_c}{d_ad_b}\right)^nN_{ab}^{c}\\
&\quad
\times \chi_{h_a}\left(e^{-\frac{8\pi n\epsilon}{l_1}}\right)
\chi_{h_b}\left(e^{-\frac{8\pi n\epsilon}{l_3}}\right)
\chi_{h_c}\left(e^{-\frac{8\pi n\epsilon}{l_2}}\right).
\end{split}
\end{equation}
In particular, for $n=1$, this reduces to
\begin{equation}
\begin{split}
\text{Tr}(\rho_A)&=\sum_{a,b,c}|\psi_a|^2|\psi_b|^2\frac{d_c}{d_ad_b}N_{ab}^{c}\\
&=\sum_{a,b}|\psi_a|^2|\psi_b|^2\\
&=\Big(\sum_a|\psi_a|^2\Big)\Big(\sum_b|\psi_b|^2\Big),
\end{split}
\end{equation}
as expected.
Here we consider the normalization condition $\sum_a|\psi_a|^2=\sum_b|\psi_b|^2=1$, and therefore
$\text{Tr}(\rho_A)=1$.
By using the modular transformation property of the character $\chi_{h_i}$, $\text{Tr}(\rho_A^n)$ in Eq.\ (\ref{rhoN3g})
may be rewritten as
\begin{equation}
\begin{split}
\text{Tr}\left(\rho_A^n\right)
&=
\sum_{a,b,c}|\psi_a|^{2n}|\psi_b|^{2n}
\left(\frac{d_c}{d_ad_b}\right)^nN_{ab}^{c}\\
&\quad\prod_{i=a,b,c}\frac{\sum_{a'}\mathcal{S}_{ia'}\chi\left(e^{-\frac{\pi l_i}{2n\epsilon}}\right)}{\left[
\sum_{a'}\mathcal{S}_{ia'}\chi\left(e^{-\frac{\pi l_i}{2\epsilon}}\right)
\right]^n}.
\end{split}
\end{equation}
Taking the thermodynamic limit $l_i/\epsilon\to \infty$, we obtain
\begin{equation}
\begin{split}
\text{Tr}\left(\rho_A^n\right)
&=
\sum_{a,b,c}|\psi_a|^{2n}|\psi_b|^{2n}
\left(\frac{d_c}{d_ad_b}\right)^nN_{ab}^{c}\cdot\\
&\quad
\times
\left(
\frac{d_a}{\mathcal{D}}\cdot
\frac{d_b}{\mathcal{D}}\cdot
\frac{d_c}{\mathcal{D}}
\right)^{1-n}e^{\frac{\pi c}{48}\cdot \frac{l_1+l_2+l_3}{\epsilon}\left(\frac{1}{n}-n\right)},
\end{split}
\end{equation}
which, after some simple algebra, can be further simplified as
\begin{equation}
\begin{split}
\text{Tr}\left(\rho_A^n\right)
&=\left(\sum_{a}|\psi_a|^{2n}d_a^{2-2n}\right)\left(\sum_{b}|\psi_b|^{2n}d_b^{2-2n}\right)\\
&\quad
\times
\frac{1}{\mathcal{D}^{3-3n}}e^{\frac{\pi c}{48}\cdot \frac{l_1+l_2+l_3}{\epsilon}\left(\frac{1}{n}-n\right)}.\\
\end{split}
\end{equation}
Then the Renyi and the von Neumann entropy of subsystem $A$ can be obtained as
\begin{align}
\label{SrenyiG2}
S^{(n)}_A
&=
\frac{1+n}{n}\cdot \frac{\pi c}{48} \cdot\frac{l_1+l_2+l_3}{\epsilon}-3\ln\mathcal{D}
\nonumber \\
&\quad
+\frac{1}{1-n}\ln \sum_a |\psi_a|^{2n}d_a^{2-2n}
\nonumber \\
&\quad
+\frac{1}{1-n}\ln \sum_b |\psi_b|^{2n}d_b^{2-2n},
\nonumber \\
S^{\text{vN}}_A
&=\frac{\pi c}{24} \cdot\frac{l_1+l_2+l_3}{\epsilon}-3\ln \mathcal{D}
\nonumber \\
&\quad
+2\sum_a|\psi_a|^2\ln d_a   -\sum_a |\psi_a|^2\ln |\psi_a|^2
\nonumber \\
&\quad
+2\sum_b|\psi_b|^2\ln d_b   -\sum_b |\psi_b|^2\ln |\psi_b|^2.
\end{align}
Compared with Eq.\ (\ref{RenyiTorus}) for a torus with $g=1$,
the above result is easy to understand by considering the additivity property of the entanglement entropy.
Take the Renyi entropy for example, each component of interface contributes to $-\ln \mathcal{D}$; and each
Wilson loop contributes to $\frac{1}{1-n}\ln \sum_i |\psi_i|^{2n}d_i^{2-2n}$.

\subsubsection{Manifolds of genus $g$}

Now we study the case of a manifold of genus $g$ with $g=N$.
As shown in Fig.\ \ref{Gtorus}, we consider $N$ independent Wilson loops labeled by $a_1,a_2,\cdots, a_N$ threading through the interior of the manifold along the longitudinal circles.
Each Wilson loop can fluctuate among different topological sectors independently.
Then the bulk wave function may be written as
\begin{equation}
|\Psi\rangle=\bigotimes_{i=1}^N \left(\sum_{a_i}\psi_{a_i}|W_{a_i}\rangle\right).
\end{equation}
Now we choose the entanglement cut as shown in Fig.\ \ref{Gtorus}, so that
we have a $(N+1)$-component interface.
Then the state at the interface can be expressed as
\begin{equation}
\begin{split}
|\psi\rangle&=\sum_{a_1a_2\cdots a_N} \psi_{a_1}\psi_{a_2}\cdots\psi_{a_{N}}
|\mathfrak{h}_{a_1}^{b_1}\rangle\rangle\\
&\quad \otimes\Big( \oplus_{c_2} \psi_{a_1a_2}^{c_2}|\mathfrak{h}_{a_1a_2\to c_2}^{b_2}\rangle\rangle\Big)\otimes\cdots\\
&\quad\otimes \Big(\oplus_{c_{N}}\psi_{a_{N-1}a_N}^{c_{N}}|\mathfrak{h}_{a_{N-1}a_N\to c_{N}}^{b_N}\rangle\rangle\Big)
\otimes |\mathfrak{h}_{a_N}^{b_{N+1}}\rangle\rangle,
\end{split}
\end{equation}
where the probability of fusing quasiparticles $a_{i-1}$ and $a_i$ into $c_i$ is
$|\psi_{a_ia_{i+1}}^{c_{i+1}}|^2=N_{a_ia_{i+1}}^{c_{i+1}}d_{c_{i+1}}/d_{a_i}d_{a_{i+1}}$.
Following similar procedures in the previous part, one can get
\begin{equation}\label{rhoNgN}
\begin{split}
&
\text{Tr}\left(\rho_A^n\right)
\nonumber \\
&=
\sum_{a_{1,2,\ldots,N}}
|\psi_{a_1}|^{2n}|\psi_{a_2}|^{2n}\cdots |\psi_{a_N}|^{2n}
\left(\frac{1}{\mathfrak{n}_{a_1}^{b_1}
\mathfrak{n}_{a_N}^{b_{N+1}}
}\right)^n\\
&\quad\times
\sum_{c_{2,3,\ldots, N}}
\frac{|\psi_{a_1a_2}^{c_2}|^{2n}}{(\mathfrak{n}_{c_2}^{b_2})^n}\frac{|\psi_{a_2a_3}^{c_3}|^{2n}}{(\mathfrak{n}_{c_3}^{b_3})^n}\cdots
\frac{|\psi_{a_{N-1}a_N}^{c_N}|^{2n}}{(\mathfrak{n}_{c_N}^{b_N})^n}\\
&\quad
\times
\chi_{h_{a_1}}\left(e^{-\frac{8\pi n\epsilon}{l_1}}\right)
\chi_{h_{a_N}}\left(e^{-\frac{8\pi n\epsilon}{l_{N+1}}}\right)
\prod_{i=2}^N\chi_{h_{c_i}}\left(e^{-\frac{8\pi n\epsilon}{l_i}}\right).
\end{split}
\end{equation}
By using the modular transformation property of the character $\chi_{h_i}$,
and taking the thermodynamic limit $l_i/\epsilon\to \infty$,
$\text{Tr}(\rho_A^n)$ can be simplified as
\begin{equation}\label{rhoNgN2}
\begin{split}
&\text{Tr}\left(\rho_A^n\right)\\
&=
\sum_{a_{1,2,\cdots,N}}|\psi_{a_1}|^{2n}|\psi_{a_2}|^{2n}\cdots |\psi_{a_N}|^{2n}\\
& \quad
\times
\sum_{c_2,\cdots,c_N}\Big(\frac{d_{c_2}}{d_{a_1}d_{a_2}}\Big)^nN_{a_1a_2}^{c_2}
\cdots \Big(\frac{d_{c_N}}{d_{a_{N-1}}d_{a_N}}\Big)^nN_{a_{N-1}a_N}^{c_N}
\\
&\quad \times
\Big(
\frac{d_{a_1}}{\mathcal{D}}\cdot\frac{d_{a_N}}{\mathcal{D}}\cdot\prod_{i=2}^N\frac{d_{c_i}}{\mathcal{D}}
\Big)^{1-n}e^{\frac{\pi c}{48}\cdot \frac{\sum_{i=1}^{N+1}l_i}{\epsilon}\left(\frac{1}{n}-n\right)}.
\end{split}
\end{equation}
The sum $\sum_{c_2,\cdots,c_N}$ can be easily done by considering that $\sum_{c_i}N_{a_{i-1}a_i}^{c_i}d_{c_i}/d_{a_{i-1}a_i}=1$.
Then Eq.\ (\ref{rhoNgN2}) can be further simplified as
\begin{equation}\label{rhoNgN3}
\begin{split}
&
\text{Tr}\left(\rho_A^n\right)
\nonumber \\
&=
\sum_{a_{1,2,\cdots,N}}|\psi_{a_1}|^{2n}|\psi_{a_2}|^{2n}\cdots |\psi_{a_N}|^{2n}\\
&
\quad
\times\Big(\prod_{i=1}^Nd_{a_i}\Big)^{2-2n}
\Big(\frac{1}{\mathcal{D}^{1+N}}\Big)^{1-n}
e^{\frac{\pi c}{48}\cdot \frac{\sum_{i=1}^{N+1}l_i}{\epsilon}\left(\frac{1}{n}-n\right)}
\\
&=\prod_{i=1}^N\left(\sum_{a_i}|\psi_{a_i}|^{2n}d_{a_i}^{2-2n}\right)
\Big(\frac{1}{\mathcal{D}^{1+N}}\Big)^{1-n}
e^{\frac{\pi c}{48}\cdot \frac{\sum_{i=1}^{N+1}l_i}{\epsilon}\left(\frac{1}{n}-n\right)},\\
\end{split}
\end{equation}
based on which we can immediately obtain the Renyi entropy and the von Neumann entropy of subsystem $A$
as follows
\begin{align}\label{RenyiNgenus}
S^{(n)}_A
&=\frac{1+n}{n}\cdot\frac{\pi c}{48}\cdot\frac{l_1+\cdots+l_{N+1}}{\epsilon}-(N+1)\ln\mathcal{D}
\nonumber \\
&\quad
+\frac{1}{1-n}\sum_{i=1}^N\ln\left[\sum_{a_i}|\psi_{a_i}|^{2n}\left(d_{a_i}\right)^{2-2n}\right],
\nonumber \\
S^{\text{vN}}_A
&=\frac{\pi c}{24} \cdot\frac{l_1+\cdots+l_{N+1}}{\epsilon}-(N+1)\ln \mathcal{D}
\nonumber \\
&\quad
+\sum_{i=1}^N\Big(2\sum_{a_i}|\psi_{a_i}|^2\ln d_{a_i}   -\sum_{a_i} |\psi_{a_i}|^2\ln|\psi_{a_i}|^2\Big).
\end{align}
For $N=1$ and $2$, we recover the results (\ref{RenyiTorus})
and (\ref{SrenyiG2}), respectively. It is found that the coefficient
in front of $-\ln \mathcal{D}$ equals the number of components of the
$AB$ interface.
For each Wilson loop $a_i$ that threads through the entanglement
cut with probability $|\psi_{a_i}|^2$, it contributes to
the Renyi and von Neumann entropy as
\begin{align}
\Delta S^{(n)}_{A,a_i} &=\frac{1}{1-n}\sum_{i=1}^N\ln\left[\sum_{a_i}|\psi_{a_i}|^{2n}\left(d_{a_i}\right)^{2-2n}\right],
\nonumber \\
\Delta S^{\text{vN}}_{A,a_i}&=\sum_{a_i}|\psi_{a_i}|^2\ln \frac{d_{a_i}^2}{|\psi_{a_i}|^2}.
\end{align}

In fact, we also checked the Renyi entropy and the von Neumann entropy for a
$g$-genus manifold with replica and surgery methods.
The results we obtained are exactly the same as
the universal parts in Eq.\ (\ref{RenyiNgenus}).

\subsection{A sphere with four quasiparticles }

Although we have studied the entanglement entropy for several examples
in the presence of quasiparticles,
it is still interesting to ask if we can extract more topological data of Chern-Simons theories,
such as the braiding property of Wilson lines and so on.
In this part, we demonstrate that our edge theory approach is powerful enough to
study these more complicated cases.

Following Ref.\ \onlinecite{Dong}, we consider a $S^2$ with four quasiparticles,
with two quasiparticles carrying anyon charge $a$, and the other two carrying anyon charge $\bar{a}$.
According to different distributions of the four quasiparticles, we need to study the entanglement
entropy case by case, as discussed in the following.

\subsubsection{$A$ with $a$ and $\bar{a}$}

\begin{figure}[t]
\includegraphics[width=3.20in]{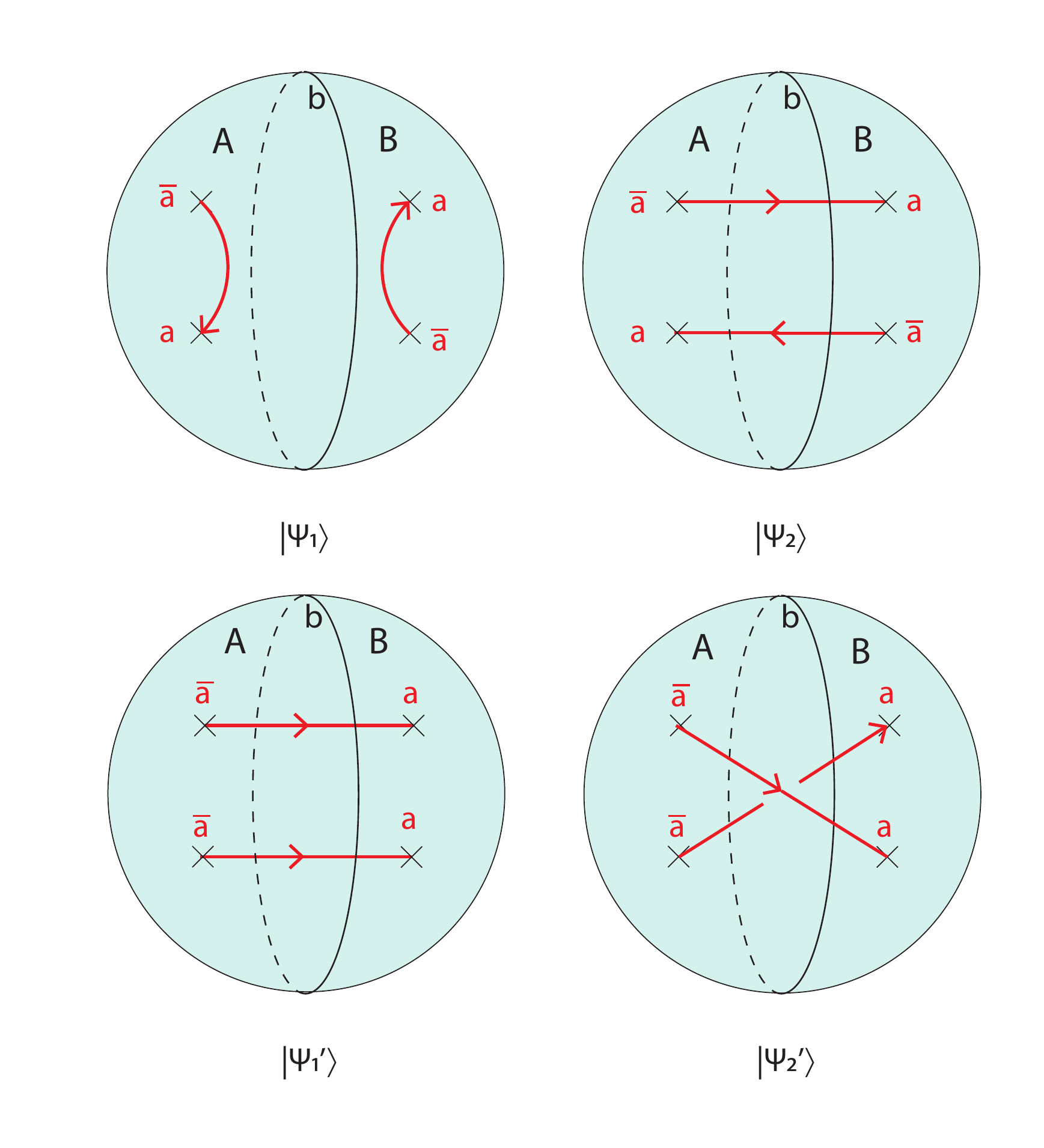}
\caption{
Top row:
A $S^2$ with four quasi-particles, with the subsystem $A$ containing
two quasiparticles $a$ and $\bar{a}$.
Each red solid line represents a Wilson line connecting quasiparticles $\bar{a}$ and $a$.
The two configurations represent two states $|\Psi_1\rangle$ and $|\Psi_2\rangle$, respectively.
Bottom row:
A $S^2$ with four quasi-particles, with two quasiparticles $\bar{a}$ and $\bar{a}$ in the  subsystem $A$,
and the other two quasiparticles $a$ and $a$ in the subsystem $B$.
Each red solid line represent a Wilson line that connects $\bar{a}$ and $a$.
The two configurations represent two states $|\Psi'_1\rangle$ and $|\Psi'_2\rangle$, respectively.}
\label{4puncture}
\end{figure}

Let us consider the case where
there are two quasiparticles $a$ and $\bar{a}$ in the subsystem $A$, with the other two quasiparticles
$\bar{a}$ and $a$ in subsystem $B$.
As shown in Fig.\ \ref{4puncture} (top row)
there are two configurations
which correspond to states $|\Psi_1\rangle$ and $|\Psi_2\rangle$, respectively.
We want to calculate the entanglement entropy of the subsystem $A$ for a general state
\begin{equation}\label{WF4puncture}
|\Psi\rangle=a|\Psi_1\rangle+b|\Psi_2\rangle.
\end{equation}
For $|\Psi_1\rangle$, there is no Wilson line threading through the $AB$ interface,
and therefore the corresponding
state at the interface is $|\mathfrak{h}_{a\bar{a}\to I}\rangle\rangle$,
with $I$ being the identity topological sector.
For $|\Psi_2\rangle$, among different fusion channels, there is a fusion channel $a\otimes\bar{a}\to I$.
Then the state at the interface may be expressed as
\begin{equation}\label{State4puncture}
|\psi\rangle=\bigoplus_c\psi_{ab}^c|\mathfrak{h}_{a\bar{a}\to c}\rangle\rangle,
\end{equation}
where
\begin{equation}
|\psi_{ab}^c|^2=
\left\{
\begin{split}
&\left|a+b\frac{1}{\sqrt{d_{a}d_{\bar{a}}}}\right|^2, \ \ \ c=I,\\
&N_{a\bar{a}}^c\frac{d_c}{d_ad_{\bar{a}}}|b|^2, \ \ \ \ \ \ \ \ c\neq I.
\end{split}
\right.
\end{equation}
Note that for a general TQFT, one always has $d_{\bar{a}}=d_{a}$.
It is also noted that for the state in Eq.\ (\ref{State4puncture}), $\sum_c|\psi_{ab}^c|^2\neq 1$, but has the following expression
\begin{equation}
\sum_{c}|\psi_c|^2=\Big |a+\frac{b}{d_{a}}\Big |^2+|b|^2\left(1-\frac{1}{d_a^2}\right).
\end{equation}
In this case, to obtain the Renyi and the von Neumann entropy,
we can use the results  (\ref{Srenyi0}) directly.
Let us check the von Neumann entropy first. For convenience, we rewrite $S^{\text{vN}}$ in Eq.\ (\ref{Srenyi0})
in the following form
\begin{equation}\label{SvN4punture}
\begin{split}
S^{\text{vN}}
&=\frac{\pi c}{24}\cdot \frac{l}{\epsilon}-\ln\mathcal{D}
+\frac{\sum_i|\psi_i|^2\ln \left(d_i/|\psi_i|^2\right)}{\sum_i|\psi_i|^2}
+\ln\sum_i|\psi_i|^2.\\
\end{split}
\end{equation}
It is found that
\begin{equation}
\frac{d_i}{|\psi_i|^2}=\left\{
\begin{split}
&\frac{1}{|a+b/d_a|^2}, \ \ \  i=I,\\
&\frac{d_a^2}{|b|^2}N_{a\bar{a}}^i, \ \ \ \ \ \ \ \ \   i\neq I,
\end{split}
\right.
\end{equation}
is independent of $d_i$. Then the von Neumann entropy for the subsystem $A$, after some straightforward algebra,
can be obtained as follows
\begin{equation}\label{4pSvon}
S^{\text{vN}}_A= \frac{\pi c}{24}\cdot \frac{l}{\epsilon}-\ln \mathcal{D}-\lambda_1\ln \lambda_1-(d_a^2-1)\lambda_2\ln\lambda_2,
\end{equation}
where $\lambda_1$ and $\lambda_2$ are defined as
\begin{equation}
\left\{
\begin{split}
\lambda_1=\frac{|ad_a+b|^2}{|ad_a+b|^2+(d_a^2-1)|b|^2},\\
\lambda_2=\frac{|b|^2}{|ad_a+b|^2+(d_a^2-1)|b|^2}.
\end{split}
\right.
\end{equation}
One can find that the universal parts of the entanglement entropy in Eq.\ (\ref{4pSvon}) are exactly the same as the results obtained with the method of replica trick and surgery in Ref.\ \onlinecite{Dong}.

Similarly, we can obtain the Renyi entropy as follows
\begin{equation}
\begin{split}
S^{(n)}_A&=\frac{1+n}{n}\cdot \frac{\pi c}{48} \cdot\frac{l}{\epsilon}
-\ln\mathcal{D}\\
&\quad +
\frac{1}{1-n}\ln\left[
\Big|a+\frac{b}{d_{\alpha}}\Big|^{2n}+\left(\frac{|b|}{d_{\alpha}}\right)^{2n}(d_{\alpha}^2-1)
\right]\\
&\quad -\frac{n}{1-n}\ln\left[
\Big |a+\frac{b}{d_{\alpha}}\Big |^2+\frac{|b|^2}{d_a^2}\left(d_a^2-1\right)
\right].
\end{split}
\end{equation}

\subsubsection{ Effect of braiding and $R$-symbols}

In this part, we will study how the braiding of Wilson lines can show up in the entanglement entropy.
We consider a generic superposition of two states
\begin{equation}\label{WF4punctureR}
|\Psi'\rangle=a|\Psi'_1\rangle+b|\Psi'_2\rangle,
\end{equation}
where $|\Psi'_1\rangle$ and $|\Psi'_2\rangle$ are shown in Fig.\ \ref{4puncture} (bottom row).
In this case, the two quasiparticles in subsystem $A$ are both
 in topological sector $\bar{a}$.
Compared to the configuration in $|\Psi'_1\rangle$,
one can find that there is braiding of Wilson lines in $|\Psi'_2\rangle$.

At the interface, the states corresponding to $|\Psi'_1\rangle$ and $|\Psi'_2\rangle$ may be expressed as
\begin{equation}
\left\{
\begin{split}
|\psi'_1\rangle&=\oplus_c\psi_{aa}^c|\mathfrak{h}_{aa\to c}\rangle\rangle,\\
|\psi'_2\rangle&=\oplus_c\psi_{aa}^cR^{aa}_c|\mathfrak{h}_{aa\to c}\rangle\rangle.\\
\end{split}
\right.
\end{equation}
where $|\psi_{aa}^c|^2=N_{aa}^cd_c/d_{a}^2$,
and $R_c^{aa}$ are the so-called $R$-symbols,
which describe the effects of braiding of anyons/Wilson lines (see Appendix A for details).
The $R$-symbol is in general a unitary matrix,
but reduces to a collection of phases in a fusion multiplicity free theory.
In particular, $R^{ab}_c$ represents the phase picked up by exchanging anyons $a$ and $b$ which fuse into channel $c$.
Then the state at the interface may be written as
\begin{equation}\label{WFRsymbol}
\begin{split}
|\psi'\rangle&=a|\psi'_1\rangle+b|\psi'_2\rangle,\\
&=\oplus_c\left(a+bR_c^{aa}\right)\psi_{aa}^c|\mathfrak{h}_{aa\to c}\rangle\rangle\\
&=:\oplus_c\phi_c|\mathfrak{h}_{aa\to c}\rangle\rangle.
\end{split}
\end{equation}
Based on the wave function above, we can obtain the Renyi
entropy as well as the von Neumann entropy of the subsystem $A$($B$)
by using Eq.\ (\ref{Srenyi0}) directly.

In the following, we are mainly interested in the $SU(2)_k$ theory,
in which the $R$-symbol has an explicit expression
\begin{equation}
R_j^{j_1,j_2}=(-1)^{j-j_1-j_2}q^{\frac{1}{2}\left[j_1(j_1+1)+j_2(j_2+1)-j(j+1)\right]},
\end{equation}
where $q=e^{-2\pi i/(2+k)}$,
and $j$ represents the anyonic charge of $SU(2)_k$ theory,
which is labeled by integers and half-integers as
$\mathcal{C}=\{0,\frac{1}{2},1,\cdots, \frac{k}{2}\}$.
(Here, for the definition of $q$, we follow the convention in Ref.\ \onlinecite{Dong}.
It is noted that in some literatures $q=e^{2\pi i/(2+k)}$ is used, and therefore the expression of $R$-symbols are slightly modified
accordingly.)
In addition, the fusion rule in the $SU(2)_k$ theory is
\begin{equation}\label{fusionSU2a}
\begin{split}
j_1\times j_2&=\sum_{|j_1-j_2|}^{\text{min}\{j_1+j_2,k-j_1-j_2\}} j\\
&=|j_1-j_2|+\left(|j_1-j_2|+1\right)+\cdots\\
&\quad+\text{min}\{j_1+j_2,k-j_1-j_2\}.
\end{split}
\end{equation}

Relabeling $a=\bar{a}=j$ and using Eq.\ (\ref{Srenyi0}),
we can immediately write down the Renyi entropy and the von Neumann entropy for the subsystem $A$  as follows
\begin{align}
\label{SrenyiRsymbol}
S^{(n)}_A
&=\frac{1+n}{n}\cdot \frac{\pi c}{48} \cdot\frac{l}{\epsilon}-\ln\mathcal{D}
\nonumber \\
&\quad
+\frac{1}{1-n}\ln\sum_{i=0}^{\text{min}\{2j,k-2j\}}\Big|
\frac{a+bR_i^{jj}}{d_j}
\Big|^{2n}d_i
\nonumber \\
&\quad
-\frac{n}{1-n}\ln\sum_{i=0}^{\text{min}\{2j,k-2j\}}\Big|
\frac{a+bR_i^{jj}}{d_j}
\Big|^{2}d_i,
\nonumber \\
S^{\text{vN}}_A
&=\frac{\pi c}{24}\cdot \frac{l}{\epsilon}-\ln\mathcal{D}
\nonumber \\
&\quad
+\frac{\sum_{i=0}^{\text{min}\{2j,k-2j\}}\Big|
a+bR_i^{jj}
\Big|^{2}d_i\ln\left|\frac{d_j}{a+bR^{jj}_i}\right|^2}{\sum_{i=0}^{\text{min}\{2j,k-2j\}}\Big|
a+bR_i^{jj}
\Big|^{2}d_i}
\nonumber \\
&\quad
+\ln\sum_{i=0}^{\text{min}\{2j,k-2j\}}\Big|
\frac{a+bR_i^{jj}}{d_j}
\Big|^{2}d_i,
\end{align}
where the quantum dimension $d_j$ is defined as
\begin{equation}\label{qdSU2}
d_j=\frac{\sin\left(\frac{(2j+1)\pi}{k+2}\right)}{\sin\left(\frac{\pi}{k+2}\right)}.
\end{equation}

Before we end this part, it is emphasized that
the $R$-symbols usually depend on the
choice of bases in the topological Hilbert space, which indicates that $R$-symbols are usually gauge dependent.
An exception is $R^{aa}_b$, which is gauge invariant (see Appendix \ref{gauge}).
That is to say, our results on the entanglement entropy
in Eq.\ (\ref{SrenyiRsymbol}) are gauge invariant, as it should be.

\begin{center}
\emph{
(a) Specific case $a=\bar{a}=\frac{1}{2}$
}
\end{center}

In Ref.\ \onlinecite{Dong}, the specific case of $a=\bar{a}=\frac{1}{2}$ is studied based on the replica trick and surgery method.
In this part, based on our general formula in Eq.\ (\ref{SrenyiRsymbol}),
we make a comparison with the results in Ref.\ \onlinecite{Dong}.

For $a=\bar{a}=\frac{1}{2}$ in a $SU(2)_k$ theory, the fusion rule of two anyons $a$ is simply
\begin{equation}
\frac{1}{2}\otimes \frac{1}{2}=0\oplus 1.
\end{equation}
For convenience, we label the quasiparticles with $j=0,\frac{1}{2},1$ as $\omega, \alpha$ and $\sigma$, respectively.
Based on Eq.\ (\ref{qdSU2}), it can be checked that
$d_{\omega}=1$, $d_{\alpha}=2\cos\frac{\pi}{k+2}$ and $d_{\sigma}=2\cos\frac{2\pi}{2+k}+1=\frac{\sin\frac{3\pi}{k+2}}{\sin\frac{\pi}{k+2}}$. From Eq.\ (\ref{SrenyiRsymbol}), the universal parts of the von Neumann entropy may be expressed as follows
\begin{widetext}
\begin{equation}
\begin{split}
S^{\text{vN}}_{A,\text{top}}
&=-\ln\mathcal{D}
+\frac{\sum_{i=0}^{\text{min}\{2j,k-2j\}}\Big|
a+bR_i^{jj}
\Big|^{2}d_i\ln\left|\frac{d_j}{a+bR^{jj}_i}\right|^2}{\sum_{i=0}^{\text{min}\{2j,k-2j\}}\Big|
a+bR_i^{jj}
\Big|^{2}d_i}
+\ln\sum_{i=0}^{\text{min}\{2j,k-2j\}}\Big|
\frac{a+bR_i^{jj}}{d_j}
\Big|^{2}d_i\\
&=:-\ln \mathcal{D}-d_{\omega}\lambda_1\ln\lambda_1-d_{\sigma}\lambda_2\ln\lambda_2,
\end{split}
\end{equation}
\end{widetext}
where $\lambda_1$ and $\lambda_2$ are defined as
\begin{equation}\label{lambda12a}
\left\{
\begin{split}
\lambda_1&=\frac{|a+bR^{\alpha\alpha}_{\omega}|^2}{d_{\omega}|a+bR_{\omega}^{\alpha\alpha}|^2+d_{\sigma}|a+bR_{\sigma}^{\alpha\alpha}|^2},\\
\lambda_2&=\frac{|a+bR^{\alpha\alpha}_{\sigma}|^2}{d_{\omega}|a+bR_{\omega}^{\alpha\alpha}|^2+d_{\sigma}|a+bR_{\sigma}^{\alpha\alpha}|^2}.
\end{split}
\right.
\end{equation}
For a $SU(2)_k$ theory, one has
\begin{equation}
\left\{
\begin{split}
R_{\omega}^{\alpha\alpha}&=-q^{\frac{3}{4}},\\
R_{\sigma}^{\alpha\alpha}&=q^{-\frac{1}{4}}.
\end{split}
\right.
\end{equation}
Therefore, $\lambda_1$ and $\lambda_2$ in Eq.\ (\ref{lambda12a})
can be rewritten as
\begin{equation}\label{lambda12b}
\left\{
\begin{split}
\lambda_1&=\frac{|a-bq^{\frac{3}{4}}|^2}{d_{\omega}|a-bq^{\frac{3}{4}}|^2+d_{\sigma}|a+bq^{-\frac{1}{4}}|^2},\\
\lambda_2&=\frac{|a+bq^{-\frac{1}{4}}|^2}{d_{\omega}|a-bq^{\frac{3}{4}}|^2+d_{\sigma}|a+bq^{-\frac{1}{4}}|^2},
\end{split}
\right.
\end{equation}
which agrees with the result in Ref.\ \onlinecite{Dong}.
It is noted that if we focus on either $|\Psi'_1\rangle$  or $|\Psi'_2\rangle$ separately, the universal
parts of the Renyi entropy or von Neumann entropy are simply $S_{A,\text{top}}^{\text{vN}}=-\ln\mathcal{D}+2\ln d_{\alpha}$.
Hence, the $R$-symbols cannot be detected.
In other words, the effects of braiding or $R$-symbols can be detected only
through the interference effect in the entanglement entropy.

\subsubsection{Effects of monodromy and topological spin}

\begin{figure}[t]
\includegraphics[width=3.20in]{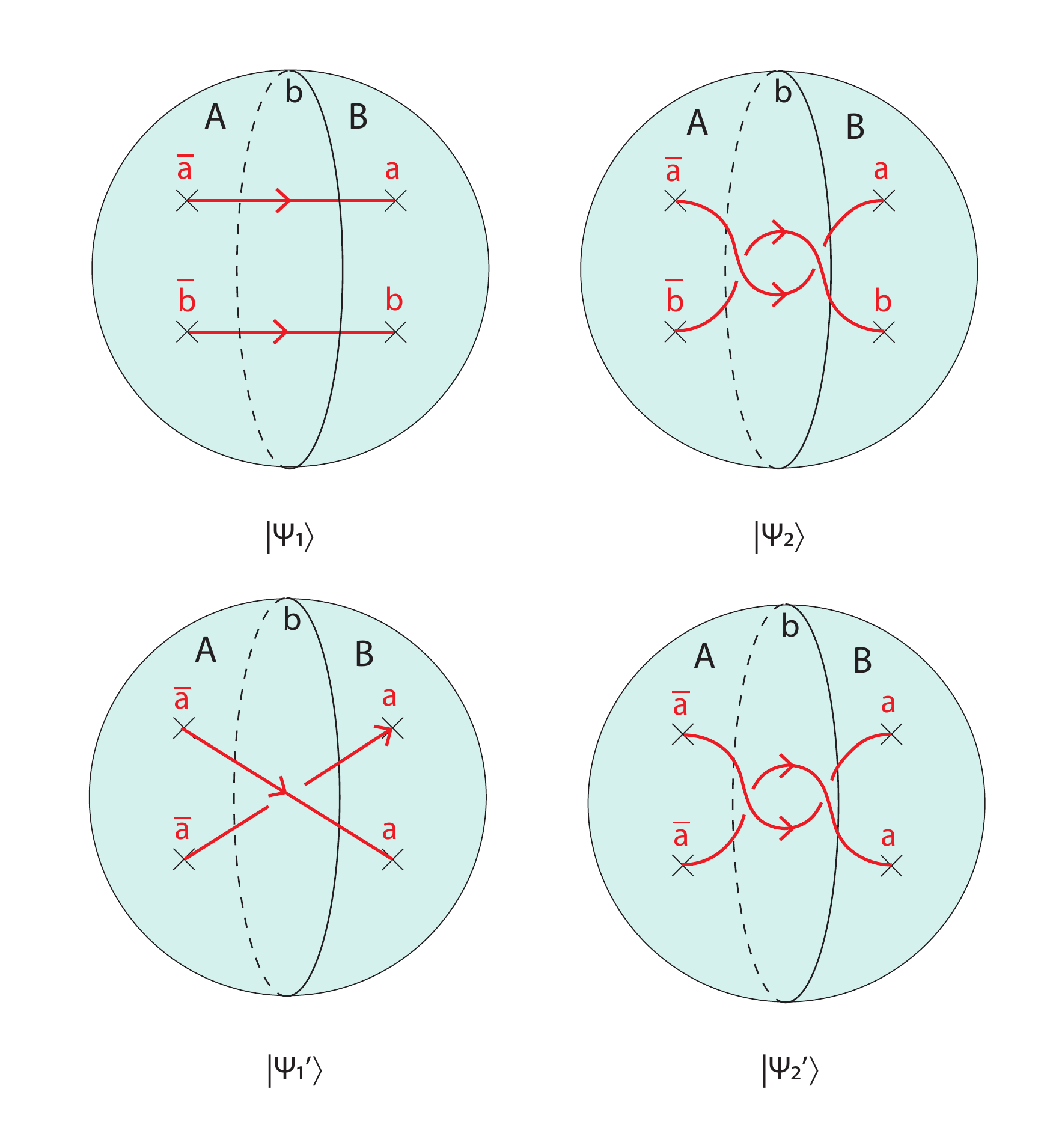}
\caption{ Top row: A $S^2$ with four quasi-particles, with two quasiparticles $\bar{a}$ and $\bar{b}$ in subsystem $A$,
and the other two quasiparticles $a$ and $b$ in subsystem $B$.
The red solid lines are Wilson lines which connect $\bar{a}(\bar{b})$ and $a(b)$.
The two configurations represent two states $|\Psi_1\rangle$ and $|\Psi_2\rangle$, respectively.
Bottom row: A $S^2$ with four quasi-particles, with two quasiparticles $\bar{a}$ and $\bar{a}$ in subsystem $A$,
and the other two quasiparticles $a$ and $a$ in subsystem $B$.
The two configurations correspond to the two states $|\Psi'_1\rangle$ and $|\Psi'_2\rangle$, respectively.
}\label{Monodromy}
\end{figure}

%
%

The effect of monodromy, or double braiding,  of two quasiparticles/Wilson lines $a$ and $b$ is
governed by the monodromy equation or
ribbon equation as follows
\begin{equation}\label{doubleR}
\sum_{\lambda}\left[R^{ab}_c\right]_{\mu\lambda}\left[R^{ba}_c\right]_{\lambda\nu}
=\frac{\theta_c}{\theta_a\theta_b}\delta_{\mu\nu},
\end{equation}
which is associated with the mutual statistics of $a$ and $b$ fused into channel $c$.
For the multiplicity free case we are interested in here, Eq.\ (\ref{doubleR}) reduces to
\begin{equation}\label{MonodromyDef}
R^{ab}_cR^{ba}_c=\frac{\theta_c}{\theta_a\theta_b}=:M^{ab}_c.
\end{equation}
The topological spin $\theta_a$,
also known as twist factor, is related to the spin or conformal scaling dimension $h_a$ of $a$ as
\begin{equation}
\theta_a=e^{i2\pi h_a}.
\end{equation}
Therefore, $M^{ab}_c$in Eq.\ (\ref{MonodromyDef}) can be rewritten as $M_c^{ab}=e^{i2\pi(h_c-h_a-h_b)}$.
To see the effect of the monodromy on the entanglement entropy, we consider
a general state $|\Psi\rangle=a|\Psi_1\rangle+b|\Psi_2\rangle$, where $|\Psi_1\rangle$ and $|\Psi_2\rangle$ are
shown in Fig.\ \ref{Monodromy} (top row).
It is noted that for the configuration in $|\Psi_2\rangle$,
the two Wilson lines braid for two times. Compared to the configuration in Fig.\ \ref{4puncture},
this double braiding of two Wilson lines allows us to study the case $a\neq b$.

At the interface, the states corresponding to $|\Psi_1\rangle$ and $|\Psi_2\rangle$ may be written as
\begin{equation}
\left\{
\begin{split}
|\psi_1\rangle&=\oplus_c\psi_{ab}^c|\mathfrak{h}_{ab\to c}\rangle\rangle,\\
|\psi_2\rangle&=\oplus_c\psi_{ab}^cM^{ab}_c|\mathfrak{h}_{ab\to c}\rangle\rangle,\\
\end{split}
\right.
\end{equation}
based on which one can write down the state corresponding to $|\Psi\rangle$ as
\begin{equation}\label{WFMonodromy}
\begin{split}
|\psi\rangle&=a|\psi_1\rangle+b|\psi_2\rangle,\\
&=\oplus_c\left(a+bM_c^{ab}\right)\psi_{ab}^c|\mathfrak{h}_{ab\to c}\rangle\rangle\\
&=\oplus_c\left(a+b\frac{\theta_c}{\theta_a\theta_b}\right)\psi_{ab}^c|\mathfrak{h}_c\rangle\rangle\\
&=:\oplus_c\phi_c|\mathfrak{h}_{ab\to c}\rangle\rangle.
\end{split}
\end{equation}
where $|\psi_{ab}^c|^2=N_{ab}^cd_c/d_ad_b$. Then one can immediately obtain the Renyi entropy
and the von Neumann entropy of the subsystem $A$($B$)
by using the results in Eq.\ (\ref{Srenyi0}).

Now we are interested in the $SU(2)_k$ theories, where the topological spins are expressed as
\begin{equation}
\theta_j=e^{i2\pi\frac{j(j+1)}{k+2}}.
\end{equation}
Relabeling the anyonic charges as $a=j_1$ and $b=j_2$, then we have
\begin{equation}
\begin{split}\label{SrenyMonodromy}
S^{(n)}_A
&=\frac{1+n}{n}\cdot \frac{\pi c}{48} \cdot\frac{l}{\epsilon}-\ln\mathcal{D}\\
&\quad+\frac{1}{1-n}\ln\sum_{j=|j_1-j_2|}^{\text{min}\{j_1+j_2,k-j_1-j_2\}}\Big|
a+b\frac{\theta_j}{\theta_{j_1}\theta_{j_2}}
\Big|^{2n}\frac{d_j}{d^n_{j_1}d^n_{j_2}}\\
&\quad-\frac{n}{1-n}\ln\sum_{j=|j_1-j_2|}^{\text{min}\{j_1+j_2,k-j_1-j_2\}}\Big|
a+b\frac{\theta_j}{\theta_{j_1}\theta_{j_2}}
\Big|^{2}\frac{d_j}{d_{j_1}d_{j_2}},\\
\end{split}
\end{equation}
and
\begin{equation}\label{SvNMonodromy}
\begin{split}
S^{\text{vN}}_A
&=\frac{\pi c}{24}\cdot \frac{l}{\epsilon}-\ln\mathcal{D}\\
&\quad+\frac{\sum\limits_{j=|j_1-j_2|}^{\text{min}\{j_1+j_2,k-j_1-j_2\}}\Big|
a+b\frac{\theta_j}{\theta_{j_1}\theta_{j_2}}
\Big|^{2}d_j\ln\frac{d_{j_1}d_{j_2}}{\left|a+b\theta_j/\theta_{j_1}\theta_{j_2}\right|^2}}
{\sum_{j=|j_1-j_2|}^{\text{min}\{j_1+j_2,k-j_1-j_2\}}\Big|
a+b\frac{\theta_j}{\theta_{j_1}\theta_{j_2}}
\Big|^{2}d_j}\\
&\quad +\ln\sum_{j=|j_1-j_2|}^{\text{min}\{j_1+j_2,k-j_1-j_2\}}\Big|
a+b\frac{\theta_j}{\theta_{j_1}\theta_{j_2}}
\Big|^{2}\frac{d_j}{d_{j_1}d_{j_2}}.
\end{split}
\end{equation}

Similar with the previous calculation involving the $R$-symbols,
the effects of monodromy can be detected only through the
interference effect. One can check that for either $|\Psi_1\rangle$ or $|\Psi_2\rangle$ separately, the universal
parts of the Renyi entropy and the von Neumann entropy are simply
$S_{A,\text{top}}^{(n)}=S_{A,\text{top}}^{\text{vN}}=-\ln\mathcal{D}+\ln d_{j_1}+\ln d_{j_2}$.

As a specific example, it is interesting to check the case with anyonic charges $j_1=j_2=\frac{1}{2}$.
As before, we label the anyons with $j=0,\frac{1}{2}, 1$  as $\omega, \alpha$ and $\sigma$, respectively.
Then based on Eq.\ (\ref{SvNMonodromy}), one can obtain
\begin{equation}
S_{A,\text{top}}^{\text{vN}}
=-\ln\mathcal{D}-d_{\omega}\lambda_1\ln\lambda_1-d_{\sigma}\lambda_2\ln\lambda_2,
\end{equation}
where $\lambda_1$ and $\lambda_2$ are defined as
\begin{equation}
\left\{
\begin{split}
\lambda_1&=\frac{\left|a+b\frac{\theta_{\omega}}{\theta_{\alpha}\theta_{\alpha}}\right|^2}{d_{\omega}\left|a+b\frac{\theta_{\omega}}{\theta_{\alpha}\theta_{\alpha}}\right|^2+d_{\sigma}\left|a+b\frac{\theta_{\sigma}}{\theta_{\alpha}\theta_{\alpha}}\right|^2},\\
\lambda_2&=\frac{\left|a+b\frac{\theta_{\sigma}}{\theta_{\alpha}\theta_{\alpha}}\right|^2}{d_{\omega}\left|a+b\frac{\theta_{\omega}}{\theta_{\alpha}\theta_{\alpha}}\right|^2+d_{\sigma}\left|a+b\frac{\theta_{\sigma}}{\theta_{\alpha}\theta_{\alpha}}\right|^2},
\end{split}
\right.
\end{equation}
which may be further rewritten as
\begin{equation}
\left\{
\begin{split}
\lambda_1&=\frac{|a+bq^{\frac{3}{2}}|^2}{d_{\omega}|a+bq^{\frac{3}{2}}|^2+d_{\sigma}|a+bq^{-\frac{1}{2}}|^2},\\
\lambda_2&=\frac{|a+bq^{-\frac{1}{2}}|^2}{d_{\omega}|a+bq^{\frac{3}{2}}|^2+d_{\sigma}|a+bq^{-\frac{1}{2}}|^2},
\end{split}
\right.
\end{equation}
where $q=e^{-2\pi i/(2+k)}$.
Before we end this part, it is noted that  $M^{ab}_c=R^{ab}_cR^{ba}_c$
in Eqs.\ (\ref{SrenyMonodromy}) and (\ref{SvNMonodromy}) is a gauge invariant quantity,
although $R^{ab}_c$ for $a\neq b$ is not gauge invariant itself
(see Appendix \ref{gauge}).
This is expected since
that the entanglement entropy should be gauge independent.

\subsubsection{Discussion: Relative phase in interference effect}

From the discussions above, it is found that both the $R$-symbols and the monodromy can be detected
through the interference effect, in which the $R$-symbols and the monodromy appear as relative phases between
two sets of bases in $|\psi_1\rangle$ and $|\psi_2\rangle$.
To understand this interference effect better, let us consider another state
\begin{equation}
|\Psi'\rangle=a|\Psi'_1\rangle+b|\Psi'_2\rangle,
\end{equation}
where
$|\Psi'_1\rangle$ and $|\Psi'_2\rangle$ are shown in Fig.\ \ref{Monodromy} (bottom row).
In particular, the two Wilson lines are braided once in
$|\Psi'_1\rangle$ and twice in $|\Psi'_2\rangle$.
Then the corresponding states at the interface can be written as
\begin{equation}
\left\{
\begin{split}
|\psi'_1\rangle&=\oplus_c\psi_{aa}^cR^{aa}_c|\mathfrak{h}_{aa\to c}\rangle\rangle,\\
|\psi'_2\rangle&=\oplus_c\psi_{aa}^cM^{aa}_c|\mathfrak{h}_{aa\to c}\rangle\rangle.\\
\end{split}
\right.
\end{equation}
where $|\psi^c_{aa}|^2=N_{aa}^cd_c/d_a^2$, and $M^{aa}_c$ is defined through Eq.\ (\ref{MonodromyDef}), i.e., $M^{aa}_c=R^{aa}_cR^{aa}_c$. Note that for the multiplicity free case we consider here, both $R^{ab}_c$ and
$M^{ab}_c$ are simply complex phases.
Then the state at the interface can be written as
\begin{equation}\label{WFRelative}
\begin{split}
|\psi'\rangle&=a|\psi'_1\rangle+b|\psi'_2\rangle,\\
&=\oplus_cR^{aa}_c\left(a+bR_c^{aa}\right)\psi_{aa}^c|\mathfrak{h}_{aa\to c}\rangle\rangle\\
&=:\oplus_c\phi_c|\mathfrak{h}_{aa\to c}\rangle\rangle,
\end{split}
\end{equation}
By comparing the states in Eq.\ (\ref{WFRelative}) and Eq.\ (\ref{WFRsymbol}), it is straightforward
to check that $S_A^{(n)}$ and $S_A^{\text{vN}}$ corresponding to the state in Eq.\ (\ref{WFRelative})
have the same  expressions as those in Eq.\ (\ref{SrenyiRsymbol}).
This is as expected because what we detect in the interference is the relative phase.

\section{Topological mutual information}
\label{MI}

As mentioned in the introduction,
the Renyi and the von Neumann entropy are good measures for bipartite entanglement.
For a tripartite system, or more generally a mixed state, it is convenient to introduce other entanglement/correlation measures such as the mutual information and the entanglement negativity. Since the mutual information is expressed in terms of the entanglement entropy,
 one can directly use the results in the previous section.
 In the following, we will give several examples on the mutual information
 between two spatial regions on a torus for Chern-Simons theories.

\begin{figure}[t]
\includegraphics[width=3.40in]{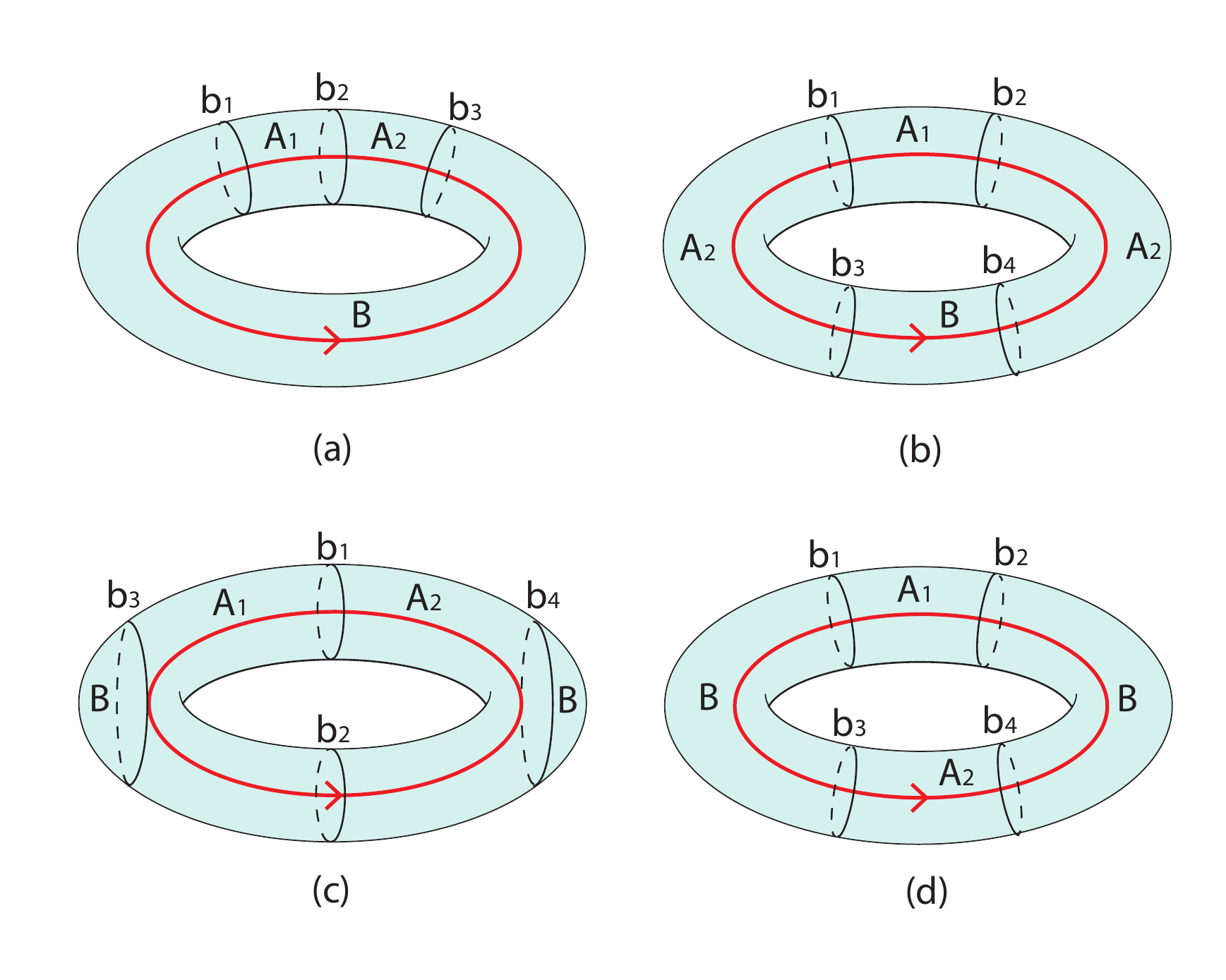}
\caption{
Four setups in calculating the mutual information and the entanglement negativity.
Two adjacent non-contractible regions $A_1$ and $A_2$ on a torus with non-contractible ((a) and (b)) and contractible (c) $B$.
(d)
Two disjoint non-contractible regions $A_1$ and $A_2$ on a torus with non-contractible region $B$.
The red solid line represents a Wilson loop threading through the interior of the torus.
}\label{MIC}
\end{figure}

\subsection{Two adjacent non-contractible regions on a torus with non-contractible $B$}

Let us consider two adjacent non-contractible regions $A_1$ and $A_2$ on a torus
with their compliment $B$ which is also non-contractible.
Here we mainly consider two nontrivial cases, shown in
Figs.\ \ref{MIC} (a) and (b).
The two regions $A_1$ and $A_2$ share a one-component $A_1A_2$ interface in Fig.\ \ref{MIC} (a)
and a two-component $A_1A_2$ interface in  Fig.\ \ref{MIC} (b). In the following, we will calculate the mutual information between $A_1$ and $A_2$ for these two cases respectively.

\subsubsection{One component interface}

As shown in Fig.\ \ref{MIC} (a), the two adjacent non-contractible regions $A_1$ and $A_2$ share a one-component $A_1A_2$ interface.
This case can be easily studied based on our
previous results on the bipartite entanglement of a torus. To be concrete, let us consider the Renyi mutual information defined
in Eq.\ \eqref{def mutual renyi}.
For the two adjacent regions $A_1$ and $A_2$ as shown in Fig.\ \ref{MIC} (a),
the subsystem $A=A_1\cup A_2$ has the same topology as $A_1$ ($A_2$), which
is simply a cylinder. Therefore, for a general state in Eq.\ (\ref{StateTorus}), \textit{i.e.},
\begin{equation}\label{state001}
|\Psi\rangle=\sum_a\psi_a|W_a\rangle,
\end{equation}
one can directly use the result in Eq.\ (\ref{RenyiTorus}), and the Renyi mutual information can be obtained as
\begin{equation}\label{RMIadjacent}
\begin{split}
I^{(n)}_{A_1A_2}
&=\frac{1+n}{n}\cdot \frac{2\pi c}{48}\cdot \frac{l_2}{\epsilon}
-2\ln\mathcal{D}\\
&\quad+\frac{1}{1-n}\ln \sum_a|\psi_a|^{2n}d_a^{2-2n},
\end{split}
\end{equation}
where $l_2$ is the length of the interface $b_2$.
And the von Neumann mutual information has the following expression
\begin{equation}\label{vNMIadjacent}
\begin{split}
I_{A_1A_2}&=
\lim_{n\to 1}I^{(n)}_{A_1A_2}\\
&=\frac{2\pi c}{24}\cdot \frac{l_2}{\epsilon}-2\ln \mathcal{D}+2\sum_a|\psi_a|^2\ln d_a\\
&\quad -\sum_a|\psi_a|^2\ln |\psi_a|^2.
\end{split}
\end{equation}
For a later comparison with the entanglement negativity, it should be noted that $I_{A_1A_2}^{(n)}$ and $I_{A_1A_2}$
depend on the choice of ground state for both Abelian and non-Abelian Chern-Simons theories.

\subsubsection{Two component interface}

As shown in Fig.\ \ref{MIC} (b),
let us consider the two adjacent non-contractible regions $A_1$ and $A_2$
which share a two-component $A_1A_2$ interface.
In this case, the subsystem $A_2$ itself is composed of two disjoint regions. To obtain the mutual information
between $A_1$ and $A_2$, we need to calculate the entanglement entropy of the subsystem $A_2$ first.

 For the general ground state in Eq.\ (\ref{state001}),
the state at the interface (including the components $b_1$, $b_2$, $b_3$ and $b_4$) has the following expression
\begin{equation}\label{TwoDisjointState}
|\psi\rangle=\sum_a\psi_a \bigotimes_{i=1,\cdots,4}|\mathfrak{h}_a^{b_i}\rangle\rangle.
\end{equation}
Following similar procedures in the previous sections, one can obtain the reduced density matrix for the subsystem
$A_2$ as follows
\begin{equation}
\begin{split}
\rho_{A_2}&=\sum_a |\psi_a|^2\frac{1}{\prod_{i=1,\cdots, 4}\mathfrak{n}_a^{b_i}}\\
&\quad\bigotimes_{i=1,\cdots,4}
\sum_{N_i,j_i}e^{-\frac{8\pi\epsilon}{l_i}}
|h_a^{b_i},N_i;j_i\rangle\langle h_a^{b_i},N_i;j_i|,\\
\end{split}
\end{equation}
based on which one can get
\begin{equation}\label{rhoTorus4}
\begin{split}
&\text{Tr}\left(\rho_{A_2}^n\right)
=\sum_{a}|\psi_a|^{2n}\prod_{i=1,\cdots,4}
\frac{\sum_{a_i}\mathcal{S}_{aa_i}\chi_{h_{a_i}}\left(e^{-\frac{\pi l_i}{2 n\epsilon}}\right)}
{\left[\sum_{a_i}\mathcal{S}_{aa_i}\chi_{h_{a_i}} \left(e^{-\frac{\pi l_i}{2 \epsilon}}\right)\right]^n
 },
\end{split}
\end{equation}
where we have used the modular transformation property of the character $\chi_{h_i}$.
In the thermodynamic limit $l_i/\epsilon\to \infty$, Eq.\ (\ref{rhoTorus4}) can be further simplified as
\begin{equation}\label{rhoA1A2}
\text{Tr}\left(\rho_{A_2}^n\right)
=e^{\frac{\pi c(l_1+l_2+l_3+l_4)}{48 \epsilon}\left(\frac{1}{n}-n\right)}\sum_a |\psi_a|^{2n}(\mathcal{S}_{a0})^{4-4n}.
\end{equation}
Then one can obtain the Renyi and the von Neumann entropy of $A_2$ as
\begin{align}\label{AdjacentBr}
S_{A_2}^{(n)}&=
\frac{1+n}{n}\cdot\frac{\pi c}{48}\cdot\frac{l_1+l_2+l_3+l_4}{\epsilon}-4\ln\mathcal{D}
\nonumber \\
&\quad
+\frac{1}{1-n}\ln \sum_a|\psi_a|^{2n}d_a^{4-4n},
\nonumber \\
S_{A_2}^{\text{vN}}
&=\frac{2\pi c}{48}\cdot\frac{l_1+l_2+l_3+l_4}{\epsilon}-4\ln \mathcal{D}
\nonumber \\
&\quad
+4\sum_a|\psi_a|^2\ln d_a
-\sum_a|\psi_a|^2\ln |\psi_a|^2.
\end{align}
Based on the results in Eqs.\ (\ref{RenyiTorus}) and (\ref{AdjacentBr}), one can
obtain the mutual information between $A_1$ and $A_2$ as follows
\begin{align}\label{RMIadjacentBA1}
I^{(n)}_{A_1A_2}
&=\frac{1+n}{n}\cdot \frac{2\pi c}{48}\cdot \frac{l_1+l_2}{\epsilon}
-4\ln\mathcal{D}
\nonumber \\
&\quad
+\frac{1}{1-n}\ln \sum_a|\psi_a|^{2n}d_a^{4-4n},
\nonumber \\
I_{A_1A_2}
&=\frac{2\pi c}{24}\cdot \frac{l_1+l_2}{\epsilon}-4\ln \mathcal{D}+4\sum_a|\psi_a|^2\ln d_a
\nonumber \\
&\quad
-\sum_a|\psi_a|^2\ln|\psi_a|^2.
\end{align}
Similar with the one-component $A_1A_2$ interface case, the mutual information in Eq.\ (\ref{RMIadjacentBA1})
depends on the choice of ground state for both Abelian and non-Abelian Chern-Simons theories.

\subsection{Two adjacent non-contractible regions on a torus with contractible $B$}

In this part, as shown in Fig.\ \ref{MIC} (c), we will calculate the mutual information of two adjacent non-contractible regions
$A_1$ and $A_2$ with a contractible region $B$.
In section \ref{EE}, the entanglement entropy of $A=A_1\cup A_2$ has already been calculated [see Eq.\ (\ref{RenyiTorus2})].
To calculate the mutual information between $A_1$ and $A_2$, one only needs to further calculate $S_{A_1(A_2)}$ as follows.

Given the ground state in Eq.\ (\ref{state001}), the state at the interface (including the components
$b_1$, $b_2$ and $b_3$) can be written as
\begin{equation}
|\psi\rangle=|\mathfrak{h}_I^{b_3}\rangle\rangle \otimes \sum_a\psi_a |\mathfrak{h}_a^{b_1}\rangle\rangle\otimes
|\mathfrak{h}_a^{b_2}\rangle\rangle.
\end{equation}
Then it is straightforward to check that the reduced density matrix for $A_1$ has the expression
\begin{equation}
\begin{split}
\rho_{A_1}&=\text{tr}_{A_2\cup B}|\psi\rangle\langle \psi|\\
&=\rho^{b_3}_{A_1,I}\otimes \sum_{a} |\psi_a|^2\rho_{A_1,a}^{b_1}\otimes\rho_{A_1,a}^{b_2},\\
\end{split}
\end{equation}
where $\rho_{A_i,a}^{b_i}$ has the form
\begin{equation}
\rho_{A1,a}^{b_i}
=\sum_{N_i,j_i}\frac{1}{\mathfrak{n}_a^{b_i}}e^{-\frac{8\pi \epsilon}{l_i}(h_a+N-\frac{c}{24})}
|h_a^{b_i},N_i;j_i\rangle\langle h_a^{b_i},N_i;j_i|.
\end{equation}
Then one can obtain
\begin{equation}\label{rhoNbb}
\begin{split}
\text{Tr}(\rho_{A_1}^n)
&=e^{\frac{\pi c (l_1+l_2+l_3)}{48 \epsilon}\left(\frac{1}{n}-n\right)}
(\mathcal{S}_{00})^{1-n}\sum_a |\psi_a|^{2n}(\mathcal{S}_{a0})^{2-2n},
\end{split}
\end{equation}
where we have used modular transformation of the character $\chi_{h_i}$ and taken the thermodynamic limit $l_i/\epsilon\to \infty$.
Based on $\text{Tr}(\rho_{A_1}^n)$ in Eq.\ (\ref{rhoNbb}),
we can obtain the Renyi entropy and the von Neumann entropy of subsystem $A_1$ as follows
\begin{align}\label{TorusAdConB1}
S^{(n)}_{A_1}&=\frac{1+n}{n}\cdot\frac{\pi c}{48}\cdot\frac{l_1+l_2+l_3}{\epsilon}-3\ln\mathcal{D}
\nonumber \\
&\quad
+\frac{1}{1-n}\ln  \sum_a|\psi_a|^{2n}d_a^{2-2n},
\nonumber \\
S^{\text{vN}}_{A_1}&=
\frac{2 \pi c}{48}\cdot\frac{l_1+l_2+l_3}{\epsilon}-3\ln \mathcal{D}
\nonumber \\
&\quad
+2\sum_a|\psi_a|^2\ln d_a-\sum_a|\psi_a|^2\ln |\psi_a|^2.
\end{align}
The same results can be obtained for $S_{A_2}^{(n)}$ and $S_{A_2}^{\text{vN}}$ by simply replacing $l_3$ with $l_4$.
Then based on Eqs.\ (\ref{RenyiTorus2}) and (\ref{TorusAdConB1}), one can obtain the mutual information between $A_1$ and $A_2$ as follows
\begin{align}\label{MIA1A2cB0}
I^{(n)}_{A_1A_2}
&=\frac{1+n}{n}\cdot \frac{2\pi c}{48}\cdot \frac{l_1+l_2}{\epsilon}
-4\ln\mathcal{D}
\nonumber \\
&\quad
+\frac{2}{1-n}\ln \sum_a |\psi_a|^{2n}d_a^{2-2n},
\nonumber \\
I_{A_1A_2}
&= \frac{4\pi c}{48}\cdot \frac{l_1+l_2}{\epsilon}
-4\ln\mathcal{D}
\nonumber \\
&\quad
+4\sum_a|\psi_a|^2\ln d_a
-2\sum_a|\psi_a|^2\ln|\psi_a|^2.
\end{align}
It is found that the mutual information in Eq.\ (\ref{MIA1A2cB0})
does not change if we take $B\to \varnothing$, which corresponds to the bipartition of a torus (see Fig.\ \ref{torus}).

\subsection{Two disjoint non-contractible regions on a torus}

%
%

In this part, we consider two disjoint non-contractible regions $A_1$ and $A_2$ on a torus, as shown in Fig.\ \ref{MIC} (d).
For this case, the mutual information between $A_1$ and $A_2$ can be easily calculated based on our previous results.
First, it is straightforward to check that $S_{A}^{(n)}=S_{B}^{(n)}$, with $A=A_1\cup A_2$. This can
be understood based on
the fact that the torus is bipartited into $A=A_1\cup A_2$ and $B$. Then, based on  Eqs.\ (\ref{RenyiTorus})
and (\ref{AdjacentBr}), one can immediately get the Renyi and von Neumann mutual information
between $A_1$ and $A_2$ as follows
\begin{align}\label{RMIdisjoint}
I^{(n)}_{A_1A_2}
&=\frac{2}{1-n}\ln \sum_a|\psi_a|^{2n}\left(\frac{d_a}{\mathcal{D}}\right)^{2-2n}
\nonumber \\
&\quad
-\frac{1}{1-n}\ln \sum_a|\psi_a|^{2n}\left(\frac{d_a}{\mathcal{D}}\right)^{4-4n}
\nonumber \\
&=\frac{1}{1-n}\ln \frac{\left(\sum_a|\psi_a|^{2n}d_a^{2-2n}\right)^2}
{\sum_a|\psi_a|^{2n}d_a^{4-4n}},
\nonumber \\
I_{A_1A_2}
&=-\sum_a|\psi_a|^2\ln|\psi_a|^2.
\end{align}

Some remarks on the results of mutual information
in Eq.\ (\ref{RMIdisjoint}) are in order:

$\bullet$
For both $I^{(n)}_{A_1A_2}$ and $I_{A_1A_2}$, the area law term disappears.
That is to say, short-scale degrees of freedom cancel in the mutual information of two disjoint regions. This is
very helpful for numerical calculations, because one needs not to calculate the entanglement entropy for different
lengths of interface. It is noted that for the mutual information of two adjacent regions in Eqs.\ (\ref{RMIadjacent})
and (\ref{vNMIadjacent}), the short-scale degrees of freedom does not cancel.

$\bullet$
The universal parts of $I^{(n)}_{A_1A_2}$ and $I_{A_1A_2}$ result from the fluctuations of the Wilson loop. If we set $\psi_{a'}=\delta_{aa'}$, i.e., the Wilson loop stays in a definite topological sector $a$, then both $I^{(n)}_{A_1A_2}$ and $I_{A_1A_2}$ vanish.

$\bullet$
The result of mutual information $I_{A_1A_2}$ in Eq.\ (\ref{RMIdisjoint}) was also obtained in Ref.\ \onlinecite{Qi2015} by using the surgery method.
In that work, the mutual information $I_{A_1A_2}$ was considered as a unified quantity to describe both conventional orders and topological orders.
For conventional orders which are characterized by the spontaneous
symmetry breaking, it is found that the mutual information has the same expression as Eq.\ (\ref{RMIdisjoint}).
Here, we emphasize that this is not the case for the Renyi mutual information $I^{(n)}_{A_1A_2}$ with $n>1$. As shown in
Eq.\ (\ref{RMIdisjoint}), the Renyi mutual information depends on both the choice of ground state and the quantum dimensions $d_a$ which are absent in conventional orders. In short, the Renyi mutual information contains more information than the von Neumann mutual information.
On the other hand, if we focus on the Abelian Chern-Simons theories, Eq.\ (\ref{RMIdisjoint}) can be further simplified as
\begin{equation}
I_{A_1A_2}^{(n)}=\frac{1}{1-n}\ln \sum_a|\psi_a|^{2n},
\end{equation}
which can still be used as a unified quantity to describe both conventional orders and Abelian topological orders.

\section{Topological entanglement negativity}
\label{LREN}

In this section, we will study the entanglement negativity defined for
two spatial regions in Chern-Simons theories.
Note that both the mutual information and the entanglement negativity
are useful for understanding the entanglement property of a mixed state.
As will be seen later, however, compared to the mutual information, the entanglement negativity
may provide different information on the underlying theory.
At the technical level,
the calculations of the entanglement negativity
require a new layer of complexity
-- taking partial transpose of the reduced density matrix --,
as compared to the entanglement entropy or mutual information.

\subsection{Left-right entanglement negativity}

In this part, for illustration purpose,
we will calculate the entanglement negativity between the left-moving modes and the right-moving modes of
the general state in Eq.\ (\ref{bsNew0}), \textit{i.e.},
\begin{equation}\label{bsNewEN}
|\psi\rangle=\sum_a\psi_a|\mathfrak{h}_a\rangle\rangle.
\end{equation}

We start from the density matrix as follows
\begin{equation}
\begin{split}
\rho&=|\psi\rangle\langle\psi|
=\sum_{a,a'}\psi_a\psi^{\ast}_{a'}|\mathfrak{h}_a\rangle\rangle\langle\langle \mathfrak{h}_{a'}|\\
&=\sum_{a,a'}\psi_a\psi^{\ast}_{a'}\frac{1}{\sqrt{\mathfrak{n}_a}\sqrt{\mathfrak{n}_{a'}}}\\
&\quad \times \sum_{N,j}\sum_{N',j'}
e^{-\frac{4\pi\epsilon}{l}\left(h_a+N-\frac{c}{24}\right)}
e^{-\frac{4\pi\epsilon}{l}\left(h_{a'}+N'-\frac{c}{24}\right)}
\\
&\quad
\times |h_a,N;j\rangle\otimes \overline{|h_a,N;j}\rangle
\langle h_{a'},N';j'|\otimes  \langle\overline{h_{a'},N';j'}|.
\end{split}
\end{equation}
Next, without loss of generality, let us take
partial transposition over the right-moving modes. Then one can obtain
\begin{equation}\label{rhoTR}
\begin{split}
\rho^{T_R}
&=\sum_{a,a'}\psi_a\psi^{\ast}_{a'}\frac{1}{\sqrt{\mathfrak{n}_a}\sqrt{\mathfrak{n}_{a'}}}\\
&\quad \times \sum_{N,j}\sum_{N',j'}
e^{-\frac{4\pi\epsilon}{l}\left(h_a+N-\frac{c}{24}\right)}
e^{-\frac{4\pi\epsilon}{l}\left(h_{a'}+N'-\frac{c}{24}\right)}\\
&\quad
\times |h_a,N;j\rangle\otimes \overline{|h_{a'},N';j'}\rangle
\langle h_{a'},N';j'|\otimes  \langle\overline{h_{a},N;j}|,
\end{split}
\end{equation}
where $T_{R(L)}$ represents the partial transposition over the right(left)-moving modes.
To calculate the entanglement negativity $\mathcal{E}_{LR}$, we can use the
definitions either in Eq.\ (\ref{D1}) or in Eq.\ (\ref{D2}). In the main text of this work, we will use the definition in
Eq.\ (\ref{D2}). For the readers who are interested in the calculation of $\mathcal{E}_{LR}$
based on Eq.\ (\ref{D1}), one can find the explicit calculation in the Appendix.

Based on the expression of $\rho^{T_R}$
in Eq.\ (\ref{rhoTR}), one can get
\begin{align}\label{rhoNe}
\text{Tr}\left(\rho^{T_R}\right)^{n_e}
&=\Bigg[
\sum_{a}|\psi_a|^{n_e}\frac{1}{\left(\mathfrak{n}_a\right)^{n_e/2}}
\chi_{h_a}\left(e^{-\frac{4\pi n_e\epsilon}{l}}\right)
\Bigg]^2
\nonumber \\
&\to
\Bigg[
e^{\frac{\pi c l}{24\epsilon}\left(\frac{1}{n_e}-\frac{n_e}{4}\right)}\sum_a|\psi_a|^{n_e}(S_{a0})^{1-\frac{n_e}{2}}
\Bigg]^2.
\end{align}
where we take the thermodynamic limit in the second line.
Therefore, by using the definition in Eq.\ (\ref{D2}), one can immediately obtain the entanglement negativity between
the left-moving modes and the right-moving modes as follows
\begin{equation}\label{LREN0}
\begin{split}
\mathcal{E}_{LR}&=\lim_{n_e\to 1}\ln \text{Tr}\left(\rho^{T_R}\right)^{n_e}\\
&=\frac{3\pi c }{48}\cdot\frac{l}{\epsilon}-\ln \mathcal{D}+2\ln \left( \sum_a |\psi_a|\sqrt{d_a}
\right).
\end{split}
\end{equation}
By comparing with $S^{(n)}_L$ in Eq.\ (\ref{Srenyi}), it is found that $\mathcal{E}_{LR}$ equals to the $1/2$ Renyi entropy, i.e.,
\begin{equation}
\mathcal{E}_{LR}=S^{(1/2)}_L=S^{(1/2)}_R.
\end{equation}
This is actually a property of the entanglement negativity for a general pure state \cite{Calabrese2012b}.
Here we demonstrate it for
the left-right entanglement negativity through an explicit calculation. It is noted that for
$\psi_i=\delta_{ia}$, the universal
parts of the entanglement negativity are
\begin{equation}
\mathcal{E}_{LR}^{\text{\ top}}=-\ln\mathcal{D}+\ln d_a,
\end{equation}
which are the same as the universal parts of the Renyi/von Neumann entropy.

Before we end this part, the readers may be curious to ask what is the result of $\text{Tr}\left(\rho^{T_R}\right)^{n}$
 if we choose $n$ to be odd in Eq.\ (\ref{rhoNe}). After some simple algebra, one has
\begin{align}
\text{Tr}\left(\rho^{T_R}\right)^{n_o}
&=
\sum_a |\psi_a|^{2n_o}\frac{1}{\left(\mathfrak{n}_a\right)^{n_o}}\chi_{h_a}\left(e^{-\frac{8\pi\epsilon n_o}{l}}\right).
\end{align}
By comparing with Eq.\ (\ref{rhoLn}), it is found that
$
\text{Tr}\left(\rho^{T_R}\right)^{n_o}=\text{Tr}\rho_L^{n_0},
$
and therefore  $\lim_{n_0\to 1}\text{Tr}\left(\rho^{T_R}\right)^{n_o}=\text{Tr}\rho_L=1$, which is trivial.

\subsection{Bipartition of a torus}

For the bipartition of a torus in Fig.\ \ref{torus} (a) and (b),
$\mathcal{E}_{AB}$ can be immediately obtained by considering the property of the entanglement negativity for a pure state, i.e.,
$\mathcal{E}_{AB}=S_A^{(1/2)}=S_B^{(1/2)}$. Then the entanglement negativity $\mathcal{E}_{AB}$
corresponding to Fig.\ \ref{torus} (a) and (b) has the following form
\begin{align}\label{TorusBi1a}
\mathcal{E}_{AB}^{(a)}&=
(S_A^{(a)})^{(1/2)}
\nonumber \\
&=\frac{3\pi c}{48}\cdot\frac{l_1+l_2}{\epsilon}-2\ln \mathcal{D}+2\ln \left(\sum_a|\psi_a|d_a\right),
\nonumber \\
\mathcal{E}_{AB}^{(b)}&=(S_A^{(b)})^{(1/2)}
=\frac{3\pi c}{48}\cdot\frac{l_1+l_2}{\epsilon}-2\ln\mathcal{D}.
\end{align}

From the above analysis, one can find that for a pure state, the entanglement negativity cannot provide more information than the Renyi entropy.
As mentioned in the introduction, the entanglement negativity
becomes more useful for a mixed state.
In the following parts, we will mainly focus on the entanglement negativity
for different cases of mixed states.

\subsection{Two adjacent non-contractible regions on a torus with non-contractible $B$}

For two adjacent non-contractible regions on a torus with non-contractible $B$,
similar with the discussion on the mutual information,  we mainly focus on the two cases in Fig.\ \ref{MIC} (a) and (b).
In Fig.\ \ref{MIC} (a), the two adjacent regions $A_1$ and $A_2$ share a one-component $A_1A_2$ interface,
and in Fig.\ \ref{MIC} (b), the two adjacent regions share a two-component $A_1A_2$ interface.
In the following, we will study the entanglement negativity between $A_1$ and $A_2$ for these two cases separately.

\subsubsection{One component interface}

Let us start with
the entanglement negativity $\mathcal{E}_{A_1A_2}$
between two adjacent
non-contractible regions  $A_1$ and $A_2$
on a torus, as shown in Fig.\ \ref{MIC} (a).
Given the general ground state in Eq.\ (\ref{state001}),
the state at the interface (including the components $b_1$, $b_2$ and $b_3$) can be written as
\begin{equation}
|\psi\rangle=\sum_a\psi_a \bigotimes_{i=1,2,3}|\mathfrak{h}_a^{b_i}\rangle\rangle.
\end{equation}
Then it is straightforward to check that the reduced density matrix for $A=A_1\cup A_2$ has the expression
\begin{equation}
\begin{split}
\rho_{A_1\cup A_2}&=\text{Tr}_B{|\psi\rangle\langle \psi|}\\
&=\sum_a|\psi_a|^2
\bigotimes_{i=1,2,3}\rho_{A,a}^{b_i},
\end{split}
\end{equation}
where
\begin{align}\label{rho1}
\rho_{A,a}^{b_1}
&=\frac{1}{\mathfrak{n}_a^{b_1}}\sum_{N_1,j_1}e^{-\frac{8\pi\epsilon}{l_1}(h_a+N_1-\frac{c}{24})}|\overline{h_a^{b_1},N_1;j_1}\rangle
\langle \overline{h_a^{b_1},N_1;j_1}|,
\nonumber \\
\rho_{A,a}^{b_3}
&=\frac{1}{\mathfrak{n}_a^{b_3}}\sum_{N_3,j_3}e^{-\frac{8\pi\epsilon}{l_3}(h_a+N_3-\frac{c}{24})}|h_a^{b_3},N_3;j_3\rangle
\langle h_a^{b_3},N_3;j_3|,
\nonumber \\
\rho_{A,a}^{b_2}&=
\frac{1}{\mathfrak{n}_a^{b_2}}\sum_{N_2,j_2}\sum_{N_2',j_2'}
e^{-\frac{4\pi\epsilon}{l_2}(h_a+N_2-\frac{c}{24})}
e^{-\frac{4\pi\epsilon}{l_2}(h_a+N_2'-\frac{c}{24})}
\nonumber \\
&\quad
\times |h_a^{b_2},N_2,j_2\rangle|\overline{h_a^{b_2},N_2;j_2}\rangle
\langle h_a^{b_2},N_2';j_2'|
\langle\overline{ h_a^{b_2},N_2';j_2'}|.
\end{align}
By taking partial transposition over the subsystem $A_2$, one
obtains
\begin{equation}\label{rho12T}
\begin{split}
\rho_{A_1\cup A_2}^{T_2}
&=\sum_a|\psi_a|^2
\rho_{A,a}^{b_1}\otimes \left(\rho_{A,a}^{b_2}\right)^{T_2}\otimes \left(\rho_{A,a}^{b_3}\right)^T\\
&=\sum_a|\psi_a|^2
\rho_{A,a}^{b_1}\otimes \left(\rho_{A,a}^{b_2}\right)^{T_2}\otimes \rho_{A,a}^{b_3},\\
\end{split}
\end{equation}
where
\begin{equation}\label{rhob2T}
\begin{split}
&
\big(\rho_{A,a}^{b_2}\big)^{T_2}=
\nonumber \\
&\quad
\frac{1}{\mathfrak{n}_a^{b_2}}\sum_{N_2,j_2}\sum_{N_2',j_2'}
e^{-\frac{4\pi\epsilon}{l_2}(h_a+N_2-\frac{c}{24})}
e^{-\frac{4\pi\epsilon}{l_2}(h_a+N_2'-\frac{c}{24})}\\
&\quad
\times |h_a^{b_2},N_2,j_2\rangle|\overline{h_a^{b_2},N_2';j_2'}\rangle
\langle h_a^{b_2},N_2';j_2'|
\langle\overline{ h_a^{b_2},N_2;j_2}|,
\end{split}
\end{equation}
with $T_2$ representing the partial transposition over the subsystem $A_2$.
After some algebra, one obtains, by taking the thermodynamic limit,
\begin{align}\label{traceA1A2N}
&
\text{Tr}\big(\rho_{A_1\cup A_2}^{T_2}\big)^{n_e}
\nonumber \\
&=\sum_a |\psi_a|^{2n_e}
\frac{
\chi_{h_a}\big(e^{-\frac{8\pi n_e\epsilon}{l_1}}\big)}{(\mathfrak{n}_a^{b_1})^{n_e}}
\frac{
\chi_{h_a}\big(e^{-\frac{8\pi n_e\epsilon}{l_3}}\big)}{(\mathfrak{n}_a^{b_3})^{n_e}}
\nonumber \\
&\quad \times \frac{
\chi_{h_a}\big(e^{-\frac{4\pi n_e\epsilon}{l_2}}\big)
\chi_{h_a}\big(e^{-\frac{4\pi n_e\epsilon}{l_2}}\big) }{\left(\mathfrak{n_a}^{b_2}\right)^{n_e}}. \nonumber \\
&\to \sum_a |\psi_a|^{2n_e}
\left(\mathcal{S}_{a0}\right)^{2-2n_e}e^{\frac{\pi c(l_1+l_3)}{48\epsilon}\left(\frac{1}{n_e}-n_e\right)}
\nonumber \\
&\quad
\times \left(\mathcal{S}_{a0}\right)^{2-n_e}e^{\frac{\pi c l_2}{48\epsilon}\left(\frac{4}{n_e}-n_e\right)}.
\end{align}
Based on the definition (\ref{D2}),
one can immediately obtain the entanglement negativity
as
\begin{equation}\label{ENadjacentTorus}
\begin{split}
\mathcal{E}_{A_1A_2}
&=\lim_{n_e\to 1}\ln \text{Tr}\big(\rho_{A_1\cup A_2}^{T_2}\big)^{n_e}\\
&=\frac{3\pi c}{48}\cdot\frac{l_2}{\epsilon}-\ln \mathcal{D}+\ln \Big(\sum_a|\psi_a|^2d_a\Big).
\end{split}
\end{equation}
It is noted that the first term, which is the area-law term, is proportional to the length of the interface between
$A_1$ and $A_2$, but has nothing to do with the interface between $A_1(A_2)$ and $B$, as expected.
The second and third terms are related
only to the quantum dimensions and the choice of ground state, and therefore are
universal.
We call the second and third terms in Eq.\ (\ref{ENadjacentTorus}) `topological entanglement negativity'.
In particular, the third term is very useful since it can distinguish Abelian and non-Abelian theories.
For an Abelian Chern-Simons theory, we have $d_a=1$ for each topological sector $a$, and therefore  $\ln\left(\sum_a|\psi_a|^2d_a\right)
=\ln\left(\sum_a|\psi_a|^2\right)=0$.
For a non-Abelian Chern-Simons theory, however, we have $d_a\neq 1$ for at least one topological sector, and therefore $\ln\left(\sum_a|\psi_a|^2d_a\right)\neq 0$ for a general ground state. In practice, one can tune the ground state of a topological system, and observe if the topological entanglement negativity changes accordingly or not. This provides us a convenient way to distinguish an Abelian theory from a non-Abelian theory.

In Ref.\ \onlinecite{VidalEN}, the entanglement negativity for a toric code model was studied. For the case of two adjacent
non-contractible regions  as discussed in this part, they found that the entanglement negativity is independent of
the choice of ground state. This may be easily understood based on our result in Eq.\ (\ref{ENadjacentTorus})
considering that the toric code model is in an Abelian phase.

As a comparison, it is noted that the mutual information $I_{A_1A_2}$ for two adjacent non-contractible regions on a torus
depends on the choice of ground state for both Abelian and non-Abelian phases
[see Eqs.\ (\ref{RMIadjacent})-(\ref{vNMIadjacent})].
In other words, the mutual information of two adjacent
non-contractible regions on a torus
cannot distinguish an Abelian theory from a non-Abelian theory. From this point of view, the entanglement negativity is more useful in distinguishing
different topological phases.

\subsubsection{Two component interface}

Let us now consider the set up in Fig.\ \ref{MIC} (b),
where now the two adjacent non-contractible regions $A_1$ and $A_2$
share a two-component
$A_1A_2$ interface.
For the general ground state (\ref{state001}), the state at the interface
(including the components $b_1$, $b_2$, $b_3$ and $b_4$ ) has the expression
\begin{equation}
|\psi\rangle=\sum_a\psi_a \bigotimes_{i=1,2,3,4}|\mathfrak{h}_a^{b_i}\rangle\rangle.
\end{equation}
The reduced density matrix for $A_1\cup A_2$ can be expressed as
\begin{align}
\rho_{A_1\cup A_2}
&=\text{Tr}_{B}{|\psi\rangle\langle \psi|}
\nonumber\\
&=\sum_a|\psi_a|^2
\bigotimes_{i=1,2,3,4}\rho_{A,a}^{b_i},
\end{align}
where
\begin{equation}\label{rho1aA1B}
\begin{split}
\rho_{A,a}^{b_1}&=\frac{1}{\mathfrak{n}_a^{b_1}}\sum_{N_1,j_1}\sum_{N_1',j_1'}
e^{-\frac{4\pi\epsilon}{l_2}(h_a+N_1-\frac{c}{24})}
e^{-\frac{4\pi\epsilon}{l_2}(h_a+N_1'-\frac{c}{24})}\\
&\quad |\overline{h_a^{b_1},N_1,j_1}\rangle|h_a^{b_1},N_1;j_1\rangle
\langle \overline{h_a^{b_1},N_1';j_1'}|
\langle h_a^{b_1},N_1';j_1'|,
\end{split}
\end{equation}
\begin{equation}
\begin{split}
\rho_{A,a}^{b_2}&=\frac{1}{\mathfrak{n}_a^{b_2}}\sum_{N_2,j_2}\sum_{N_2',j_2'}
e^{-\frac{4\pi\epsilon}{l_2}(h_a+N_2-\frac{c}{24})}
e^{-\frac{4\pi\epsilon}{l_2}(h_a+N_2'-\frac{c}{24})}\\
&\quad |h_a^{b_2},N_2,j_2\rangle|\overline{h_a^{b_2},N_2;j_2}\rangle
\langle h_a^{b_2},N_2';j_2'|
\langle\overline{ h_a^{b_2},N_2';j_2'}|,
\end{split}
\end{equation}
\begin{equation}
\rho_{A,a}^{b_3}=\frac{1}{\mathfrak{n}_a^{b_3}}\sum_{N_3,j_3}e^{-\frac{8\pi\epsilon}{l_3}(h_a+N_3-\frac{c}{24})}|\overline{h_a^{b_3},N_3;j_3}\rangle
\langle \overline{h_a^{b_3},N_3;j_3}|,
\end{equation}
and
\begin{equation}
\rho_{A,a}^{b_4}=\frac{1}{\mathfrak{n}_a^{b_4}}\sum_{N_4,j_4}e^{-\frac{8\pi\epsilon}{l_4}(h_a+N_4-\frac{c}{24})}|h_a^{b_4},N_4;j_4\rangle
\langle h_a^{b_4},N_4;j_4|.
\end{equation}
Taking a partial transposition over region $A_2$, one can get
\begin{equation}\label{rhoA1TB}
\begin{split}
\rho_{A_1\cup A_2}^{T_2}
&=\sum_a|\psi_a|^2
\big(\rho_{A,a}^{b_1}\big)^{T_2}
\otimes
\big(\rho_{A,a}^{b_2}\big)^{T_2}\otimes \rho_{A,a}^{b_3}\otimes \rho_{A,a}^{b_4}.\\
\end{split}
\end{equation}
where
\begin{equation}
\begin{split}
\left(\rho_{A,a}^{b_1}\right)^{T_2}&=\frac{1}{\mathfrak{n}_a^{b_1}}\sum_{N_1,j_1}\sum_{N_1',j_1'}
e^{-\frac{4\pi\epsilon}{l_1}(h_a+N_1-\frac{c}{24})}
e^{-\frac{4\pi\epsilon}{l_1}(h_a+N_1'-\frac{c}{24})}\\
&\quad |\overline{h_a^{b_1},N_1,j_1}\rangle|h_a^{b_1},N_1';j_1'\rangle
\langle \overline{h_a^{b_1},N_1';j_1'}|
\langle h_a^{b_1},N_1;j_1|,
\end{split}
\end{equation}
and
\begin{equation}\label{rho2aTA1B}
\begin{split}
\left(\rho_{A,a}^{b_2}\right)^{T_2}&=\frac{1}{\mathfrak{n}_a^{b_2}}\sum_{N_2,j_2}\sum_{N_2',j_2'}
e^{-\frac{4\pi\epsilon}{l_2}(h_a+N_2-\frac{c}{24})}
e^{-\frac{4\pi\epsilon}{l_2}(h_a+N_2'-\frac{c}{24})}\\
&\quad |h_a^{b_2},N_2,j_2\rangle|\overline{h_a^{b_2},N_2';j_2'}\rangle
\langle h_a^{b_2},N_2';j_2'|
\langle\overline{ h_a^{b_2},N_2;j_2}|.
\end{split}
\end{equation}\label{traceA1BN}
Then one can obtain,
by taking the thermodynamic limit,
\begin{align}\label{traceA1BN}
&
\text{Tr}\big(\rho_{A_1\cup A_2}^{T_2}\big)^{n_e}
\nonumber \\
&=\sum_a |\psi_a|^{2n_e}
\prod_{i=3,4}
\frac{\chi_{h_a}\big(e^{-\frac{8\pi n_e\epsilon}{l_i}}\big)}{\left(\mathfrak{n}_a^{b_i}\right)^{n_e}}
\nonumber \\
&\quad
\times \prod_{i=1,2}\frac{
\chi_{h_a}\big(e^{-\frac{4\pi n_e\epsilon}{l_i}}\big)
\chi_{h_a}\big(e^{-\frac{4\pi n_e\epsilon}{l_i}}\big)
}{\left(\mathfrak{n_a}^{b_i}\right)^{n_e}}
\nonumber \\
&\to
\sum_a |\psi_a|^{2n_e}
\left(\mathcal{S}_{a0}\right)^{2-2n_e}e^{\frac{\pi c(l_3+l_4)}{48\epsilon}\left(\frac{1}{n_e}-n_e\right)}
\nonumber \\
&\quad
\quad
\times \left(\mathcal{S}_{a0}\right)^{4-2n_e}e^{\frac{\pi c( l_1+l_2)}{48\epsilon}\left(\frac{4}{n_e}-n_e\right)}.
\end{align}
Therefore, one can obtain the entanglement negativity between $A_1$ and $A_2$ as follows
\begin{equation}\label{ENadjacentTorusB}
\begin{split}
\mathcal{E}_{A_1A_2}
&=\lim_{n_e\to 1}\ln \text{Tr}\left(\rho_{A_1\cup A_2}^{T_1}\right)^{n_e}\\
&=\frac{3\pi c}{48}\cdot\frac{l_1+l_2}{\epsilon}-2\ln\mathcal{D}+\ln\Big(\sum_a|\psi_a|^2d_a^2\Big).
\end{split}
\end{equation}
Similar with the result of one component $A_1A_2$ interface in Eq.\ (\ref{ENadjacentTorus}), one can find that
$\mathcal{E}_{A_1A_2}$ is dependent(independent) of the choice of ground state for non-Abelian (Abelian) theories.

Therefore, the entanglement negativity of two adjacent non-contractible regions
for both configurations in Fig.\ \ref{MIC} (a) and (b) can serve as a quantity
to distinguish an Abelian theory from a non-Abelian theory.

\subsection{Two adjacent non-contractible regions on a torus with contractible $B$}

In this part, we study the entanglement negativity of two adjacent non-contractible regions $A_1$ and $A_2$
with a contractible region $B$, as shown in Fig.\ \ref{MIC} (c).
For the general ground state in Eq.\ (\ref{state001}), the state at the interface
(including the components $b_1$, $b_2$, $b_3$ and $b_4$ ) can be expressed as
\begin{equation}
|\psi\rangle=|\mathfrak{h}_I^{b_3}\rangle\rangle\otimes |\mathfrak{h}_I^{b_4}\rangle\rangle\otimes
\sum_a \psi_a |\mathfrak{h}_a^{b_1}\rangle\rangle\otimes |\mathfrak{h}_a^{b_2}\rangle\rangle.
\end{equation}
Then the reduced density matrix for $A=A_1\cup A_2$ can be obtained as follows
\begin{equation}
\begin{split}
\rho_{A_1\cup A_2}&=\rho_{A,I}^{b_3}\otimes \rho_{A,I}^{b_4}
\\
&\quad \otimes \sum_{aa'} \psi_a\psi_{a'}^{\ast}
|\mathfrak{h}_a^{b_1}\rangle\rangle\langle\langle \mathfrak{h}_{a'}^{b_1}|\otimes
|\mathfrak{h}_a^{b_2}\rangle\rangle\langle\langle \mathfrak{h}_{a'}^{b_2}|
\end{split}
\end{equation}
where
\begin{align}
\rho_{A,I}^{b_3}&=
\sum_{N,j}\frac{1}{\mathfrak{n}_I^{b_3}}e^{-\frac{8\pi \epsilon}{l_3}(h_I+N-\frac{c}{24})}|\overline{h_I^{b_3},N;j}\rangle\langle \overline{h_I^{b_3},N;j}|,
\nonumber \\
\rho_{A,I}^{b_4}&=
\sum_{N,j}\frac{1}{\mathfrak{n}_I^{b_4}}e^{-\frac{8\pi \epsilon}{l_4}(h_I+N-\frac{c}{24})}|h_I^{b_4},N;j\rangle\langle h_I^{b_4},N;j|.
\end{align}
The explicit expression of $\rho_{A_1\cup A_2}$ is as follows
\begin{widetext}
\begin{align}
\rho_{A_1\cup A_2}&=\rho_{A,I}^{b_3}\otimes \rho_{A,I}^{b_4}
\nonumber \\
&\quad \otimes\sum_{a,a'}\psi_a\psi^{\ast}_{a'}\frac{1}{\sqrt{\mathfrak{n}_a^{b_1}}\sqrt{\mathfrak{n}_{a'}^{b_1}}}
\frac{1}{\sqrt{\mathfrak{n}_a^{b_2}}\sqrt{\mathfrak{n}_{a'}^{b_2}}}
\nonumber \\
&\quad\quad
\times\Big(\sum_{N_1,j_1}\sum_{N_1',j_1'}
e^{-\frac{4\pi\epsilon}{l_1}\left(h_a+N_1-\frac{c}{24}\right)}
e^{-\frac{4\pi\epsilon}{l_1}\left(h_{a'}+N_1'-\frac{c}{24}\right)}
|h_a^{b_1},N_1;j_1\rangle \overline{|h_a^{b_1},N_1;j_1}\rangle
\langle h_{a'}^{b_1},N_1';j_1'| \langle\overline{h_{a'}^{b_1},N_1';j_1'}|\Big)
\nonumber \\
&\quad \otimes
\Big(\sum_{N_2,j_2}\sum_{N_2',j_2'}
e^{-\frac{4\pi\epsilon}{l_2}\left(h_a+N_2-\frac{c}{24}\right)}
e^{-\frac{4\pi\epsilon}{l_2}\left(h_{a'}+N_2'-\frac{c}{24}\right)}
|\overline{h_a^{b_2},N_2;j_2}\rangle |h_a^{b_2},N_2;j_2\rangle
\langle \overline{ h_{a'}^{b_2},N_2';j_2'}| \langle h_{a'}^{b_2},N_2';j_2'|\Big).
\end{align}
Next, we take partial transposition over $A_2$ on the reduced density matrix $\rho_{A_1\cup A_2}$. Then one can get
\begin{equation}\label{rhoA1A2ConB01}
\begin{split}
\rho_{A_1\cup A_2}^{T_2}&=\rho_{A,I}^{b_3}\otimes \rho_{A,I}^{b_4}\\
&\quad\otimes\sum_{a,a'}\psi_a\psi^{\ast}_{a'}\frac{1}{\sqrt{\mathfrak{n}_a^{b_1}}\sqrt{\mathfrak{n}_{a'}^{b_1}}}
\frac{1}{\sqrt{\mathfrak{n}_a^{b_2}}\sqrt{\mathfrak{n}_{a'}^{b_2}}}
\\
&\quad\quad
\times \Big(\sum_{N_1,j_1}\sum_{N_1',j_1'}
e^{-\frac{4\pi\epsilon}{l_1}\left(h_a+N_1-\frac{c}{24}\right)}
e^{-\frac{4\pi\epsilon}{l_1}\left(h_{a'}+N_1'-\frac{c}{24}\right)}
|h_a^{b_1},N_1;j_1\rangle \overline{|h_{a'}^{b_1},N_1';j_1'}\rangle
\langle h_{a'}^{b_1},N_1';j_1'| \langle\overline{h_{a}^{b_1},N_1;j_1}|\Big)\\
&\quad \otimes
\Big(\sum_{N_2,j_2}\sum_{N_2',j_2'}
e^{-\frac{4\pi\epsilon}{l_2}\left(h_a+N_2-\frac{c}{24}\right)}
e^{-\frac{4\pi\epsilon}{l_2}\left(h_{a'}+N_2'-\frac{c}{24}\right)}
|\overline{h_a^{b_2},N_2;j_2}\rangle |h_{a'}^{b_2},N_2';j_2'\rangle
\langle \overline{ h_{a'}^{b_2},N_2';j_2'}| \langle h_{a}^{b_2},N_2;j_2|\Big).\\
\end{split}
\end{equation}
After some tedious but straightforward algebra, one can get
\begin{align}
\text{Tr}\big( \rho_{A_1\cup A_2}^{T_2} \big)^{n_e}
&=
\frac{1}{\left(\mathfrak{n}_a^{b_3}\right)^{n_e}}
\chi_{h_I}\big(e^{-\frac{8\pi n \epsilon}{l_3}}\big)
\times
\frac{1}{\left(\mathfrak{n}_a^{b_4}\right)^{n_e}}
\chi_{h_I}\big(e^{-\frac{8\pi n \epsilon}{l_4}}\big)
\times
\Bigg[
\sum_{a}|\psi_a|^{n_e}
\frac{
\chi_{h_a}\big(e^{-\frac{4\pi n_e\epsilon}{l_1}}\big)}
{\big(\mathfrak{n}_a^{b_1}\big)^{n_e/2}}
\cdot
\frac{
\chi_{h_a}\big(e^{-\frac{4\pi n_e\epsilon}{l_2}}\big)}{\big(\mathfrak{n}_a^{b_2}\big)^{n_e/2}}
\Bigg]^2
\nonumber \\
&\to
e^{\frac{\pi c (l_3+l_4)}{48 \epsilon}\left(\frac{1}{n_e}-n_e\right)}\mathcal{S}_{00}^{2-2n_e}
\times\left[
e^{\frac{\pi c(l_1+l_2)}{24\epsilon}\left(\frac{1}{n_e}-\frac{n_e}{4}\right)}\sum_a|\psi_a|^{n_e}(\mathcal{S}_{a0})^{2-n_e}\right]^2.
\end{align}
\end{widetext}
Then the entanglement negativity between $A_1$ and $A_2$ can be expressed as
\begin{equation}\label{ENadjacentTorusConB}
\begin{split}
\mathcal{E}_{A_1A_2}
&=\lim_{n_e\to 1}\ln \text{Tr}\left(\rho_{A_1\cup A_2}^{T_1}\right)^{n_e}\\
&=\frac{3\pi c}{48}\cdot\frac{l_1+l_2}{\epsilon}-2\ln\mathcal{D}+2\ln\Big(\sum_a|\psi_a| d_a \Big),
\end{split}
\end{equation}
which is the same as the result in Eq.\ (\ref{TorusBi1a}) for a bipartited torus.
For this case, the entanglement negativity depends on the choice of ground state for both Abelian
and non-Abelian Chern-Simons theories.

\subsection{Two disjoint non-contractible regions on a torus}

In this part, we consider the entanglement negativity $\mathcal{E}_{A_1A_2}$ between two disjoint non-contractible regions
$A_1$
and $A_2$  on a torus, as shown in Fig.\ \ref{MIC} (d). For the general ground state in Eq.\ (\ref{state001}),
the state at the interface
(including the components $b_1$, $b_2$, $b_3$ and $b_4$ ) can be written as
\begin{equation}
|\psi\rangle=\sum_a\psi_a \bigotimes_{i=1,\cdots,4}|\mathfrak{h}_a^{b_i}\rangle\rangle,
\end{equation}
where $i=1,2$ correspond to the interface between $A_1$ and $B$, and $i=3,4$ correspond to the interface between $A_2$ and $B$. It is straightforward to check that
\begin{equation}
\rho_{A_1\cup A_2}=\sum_a|\psi_a|^2\rho_{A_1,a}\otimes\rho_{A_2,a},
\end{equation}
where
\begin{equation}
\begin{split}
\rho_{A_1,a}&=\frac{e^{-8\pi\epsilon/l_1}}{\mathfrak{n}_a^{b_1}}\sum_{N_1,j_1}|\overline{h_a^{b_1},N_1;j_1}\rangle
\langle \overline{h_a^{b_1},N_1;j_1}|\\
&\quad \otimes \frac{e^{-8\pi\epsilon/l_2}}{\mathfrak{n}_a^{b_2}}\sum_{N_1,j_1}|h_a^{b_2},N_1;j_1\rangle
\langle h_a^{b_2},N_1;j_1|,\\
\end{split}
\end{equation}
and
\begin{equation}
\begin{split}
\rho_{A_2,a}&=\frac{e^{-8\pi\epsilon/l_3}}{\mathfrak{n}_a^{b_3}}\sum_{N_3,j_3}|h_a^{b_3},N_3;j_3\rangle
\langle h_a^{b_3},N_3;j_3|\\
&\quad \otimes \frac{e^{-8\pi\epsilon/l_4}}{\mathfrak{n}_a^{b_4}}\sum_{N_4,j_4}|\overline{h_a^{b_4},N_4;j_4}\rangle
\langle\overline{ h_a^{b_4},N_4;j_4}|.\\
\end{split}
\end{equation}
In this case, the partial transposition of $\rho_{A_1\cup A_2}$ over $A_2$ can be expressed as
\begin{equation}
\begin{split}
\rho_{A_1\cup A_2}^{T_2}&=\sum_a|\psi_a|^2\rho_{A_1,a}\otimes\left( \rho_{A_2,a}\right)^{T}\\
&=\sum_a|\psi_a|^2\rho_{A_1,a}\otimes \rho_{A_2,a}\\
&=\rho_{A_1\cup A_2},
\end{split}
\end{equation}
based on which one obtains
$\text{Tr}\big(\rho_{A_1\cup A_2}^{T_2}\big)^{n_e}
=\text{Tr}\left(\rho_{A_1\cup A_2}\right)^{n_e}$ for two disjoint regions.
Then the entanglement negativity simply reads
\begin{equation}
\begin{split}
\mathcal{E}_{A_1A_2}
&=\lim_{n_e\to 1}\ln \text{Tr}\left(\rho_{A_1\cup A_2}^{T_2}\right)^{n_e}
=0.
\end{split}
\end{equation}
In Ref.\ \onlinecite{VidalEN}, the same conclusion was obtained based on the toric code model.
Here we demonstrate it for a general Chern-Simons field theory.

\section{Conclusions}
\label{con}

In this work, we develop an edge theory approach to study the topological entanglement entropy,
mutual information, and entanglement negativity in Chern-Simons theories.
Compared to the prior works, we propose a new regularized state to describe the spatial quantum entanglement
in Chern-Simons theories.
An advantage of our approach,
as compared to, \textit{e.g.}, the surgery method \cite{Dong},
is that
there is no need to consider the three dimensional spacetime manifold which may be quite complicated.
For all the cases studied by the replica and surgery method, our edge theory approach
reproduces the same results.

In addition, our edge theory approach is very flexible  to include various factors in the calculation of entanglement, including the choice of ground state, the fusion and braiding of Wilson lines and so on. In particular, through an
interference effect, we can detect the $R$-symbols and the monodromy of two quasipartilces/anyons in the entanglement entropy. We also generalize our edge theory approach to the calculation of entanglement entropy for a
manifold of genus $g$.

Furthermore, our edge theory approach is also applied to the calculation of topological mutual information
and entanglement
negativity in a mixed state. To our knowledge, this is the first calculation of the entanglement negativity for a general
Chern-Simons theory. It is found that the entanglement negativity between two adjacent non-contractible regions on a torus
provides a simple way to distinguish an Abelian Chern-Simons theory from a non-Abelian Chern-Simons theory.
To be concrete, for two adjacent non-contractible regions on a tripartited torus, the entanglement negativity is
independent of the choice of ground state for an Abelian Chern-Simons theory. On the other hand, for a non-Abelian
Chern-Simons theory, the entanglement negativity depends on the choice of ground state.
In the previous works,\cite{YiZhang2012,Sheng2014} 
to distinguish a non-Abelian phase
from an Abelian phase for a microscopic model, 
 one needs to tune the ground state to find out the MESs, based on which one can further obtain 
the quantum dimension corresponding to each anyon. With the method in our work, we only need to check whether the 
topological entanglement negativity is dependent on the choice of ground state or not, which is much easier in practice.

There are also some future problems we are interested in.
For example, in this paper we mainly focus on the quantum entanglement in Chern-Simons theories. It is interesting to
generalize our approach to non-chiral TQFTs. In addition, it is also interesting to apply the concept of
charged and shifted topological entanglement entropy that was proposed recently\cite{Shunji}
 to a general TQFT based on the edge theory approach developed in this work.

\textit{Note added}:
In a forthcoming paper, \cite{Wen_TEN} the entanglement negativity in Chern-Simons theories is studied based on 
the replica trick and sugery method. The results agree with the edge theory approach in this work for all the cases under
study.

\label{sum}

\acknowledgements

XW thanks Yingfei Gu and Jeffrey C. Y. Teo for helpful discussions, and
thanks Yanxiang Shi and Yingfei Gu for help with plotting.
XW and SR would like to thank the Kavli Institute for Theoretical Physics (KITP) at
Santa Barbara where this work was initiated.
XW would like to thank Prospects in Theoretical Physics 2015-Princeton Summer
School on Condensed Matter Physics, where part of this work was carried out.
This work was supported in part by the National Science Foundation grant
DMR-1455296 (XW and SR) at the University of Illinois,
and by Alfred P. Sloan foundation.

\appendix

\section{On modular tensor categories}

In this part, for the completeness of this work, we give  a short review of the modular tensor category (MTC) description of a
 (2+1)-dimensional topological quantum field theory. We will mainly review the properties of MTCs that are frequently used
 in this work. For more details and other interesting properties of MTCs, the readers may refer to Ref.\ \onlinecite{Kitaev1997,Kitaev2006b,BondersonT,Bernevig}.

The MTCs are also known as anyon models in physics. For an anyon model, one has a finite set $\mathcal{C}$ of superselection
 sectors which are called topological or anyonic charges. These anyons are usually labeled by $a,b,c,\cdots$, and they
 satisfy the so-called fusion algebra
\begin{equation}\label{FusionProduct}
a\otimes b=\bigoplus_{c\in\mathcal{C}}N_{ab}^c c,
\end{equation}
where the fusion coefficients $N_{ab}^c$ are non-negative integers, which denote different ways that the anyon charges
$a$ and $b$ fuse into $c$. Here we use the direct sum $\oplus$ to emphasize that different anyons lie in different Hilbert spaces. For each anyon model, there exists a trivial vacuum charge $I\in \mathcal{C}$, or the identity. Each charge $a$
has its own conjugate charge $\bar{a}\in \mathcal{C}$ so that $N_{a\bar{a}}^I=1$. For each fusion product in Eq.\ (\ref{FusionProduct}), we may assign a fusion vector space $V^{ab}_c$ which is spanned by the orthonormal set of basis
vectors $|a,b;c,\mu\rangle$, with $\mu=1,\cdots, N_{ab}^c$. If the fusion coefficients $N_{ab}^c$ are equal to
$0$ or $1$, we call the fusion rules \textit{multiplicity-free}.

The fusion rules in Eq.\ (\ref{FusionProduct}) are commutative and associative. For commutative, it means $a\otimes b=b\otimes a$, and therefore $N_{ab}^c=N_{ba}^c$. For associative, it means the results of $(a\otimes b)\otimes c$ and $a\otimes (b\otimes c)$ should be equivalent to each other. Then it is required that
\begin{equation}
\sum_{d,e}N_{ab}^dN_{dc}^e=\sum_{d,e}N_{ad}^eN_{bc}^d.
\end{equation}
Another quantity we frequently used in the main text is the quantum dimension $d_a$, which reflects the nontrivial
internal Hilbert space of the anyon $a$. It may be found by considering the dimension of the fusion space of $n$ anyons
$a$ with large $n$
\begin{equation}
\begin{split}
\text{dim}\left(\sum_{c_n}V_{a\cdots a}^{c_n}\right)&=\sum_{c_2,\cdots,c_n}N_{aa}^{c_2}N_{c_2a}^{c_3}\cdots
N_{c_{n-1}a}^{c_n}
\sim d_a^n.
\end{split}
\end{equation}
For arbitrary anyon models, one has $d_a\ge 1$. If the quantum dimensions of all the anyons in a TQFT are equal to $1$, then
the theory is Abelian. On the other hand, if there exist anyons with quantum dimensions $>1$, then the theory is non-Abelian.
The total quantum dimension of a TQFT is defined as
\begin{equation}
\mathcal{D}=\sqrt{\sum_a d_a^2}.
\end{equation}
With the quantum dimension introduced, the probability of fusing two anyons $a$ and $b$ into anyon $c$ can be expressed as
\begin{equation}
P_{ab\to c}=N_{ab}^c\frac{d_c}{d_ad_b}.
\end{equation}
The constraint $\sum_c P_{ab\to c}=1$ indicates that
\begin{equation}
d_ad_b=\sum_c N_{ab}^c d_c.
\end{equation}
Another useful concept in a TQFT is braiding. The effect of switching two anyons $a$ and $b$ adiabatically is described by the braiding operator $\mathcal{R}_{ab}$. It acts on the Hilbert space $V^{ab}_c$ as follows
\begin{equation}
\mathcal{R}_{ab}|a,b;c,\mu\rangle=\sum_{\nu}\big[
R^{ab}_c
\big]_{\mu\nu}|b,a;c,\nu\rangle,
\end{equation}
or diagrammatically,
\begin{eqnarray}
\begin{tikzpicture}[baseline={(current bounding box.center)}, scale=1.5]
\draw (10pt,0pt)..controls (10pt,5pt) and (0pt,5pt)..(0pt,10pt);
\draw[line width=5pt, draw=white] (0pt,0pt)..controls (0pt,5pt) and (10pt,5pt)..(10pt,10pt);
\draw (0pt,0pt)..controls (0pt,5pt) and (10pt,5pt)..(10pt,10pt);
\draw (0pt,0pt) arc (-180:0:5pt);
\draw (5pt,-5pt)--(5pt,-14pt);
\node at (10pt,-8pt){$\mu$};
\node at (0,15pt){$b$};
\node at (10pt,15pt){$a$};
\node at (5pt,-18pt){$c$};
\end{tikzpicture}=\sum_{\nu}\left[R^{ab}_c\right]_{\mu\nu}
\begin{tikzpicture}[baseline={(current bounding box.center)},scale=0.7]
\draw (0pt,-12pt)--(0pt,-28pt);
\draw (0pt,-12pt)--(15pt,20pt);
\draw (0pt,-12pt)--(-15pt,20pt);
\node [right] at (5pt,-12pt){$\nu$};
\node at (-15pt, 32pt) {$b$};
\node at (15pt, 30pt) {$a$};
\node at (0pt, -35pt) {$c$};
\end{tikzpicture}
\end{eqnarray}
where $R^{ab}_c$ are the so-called $R$-symbols, which are unitary matrices satisfying
\begin{equation}
\left[(R^{ab}_c)^{-1}\right]_{\mu\nu}=\left[
(R^{ab}_c)^{\dag}
\right]_{\mu\nu}=\left[
R^{ab}_c
\right]_{\nu\mu}^{\ast}.
\end{equation}
For a fusion multiplicity free theory, the $R$-symbol reduces to a phase.

Based on $R$-symbols, one can study the effect of double braiding of two anyons $a$ and $b$, which is governed by
 the monodromy equation, or ribbon property
\begin{equation}\label{monodromy001}
\sum_{\lambda}\left[R^{ab}_c\right]_{\mu\lambda}\left[R^{ba}_c\right]_{\lambda\nu}
=\frac{\theta_c}{\theta_a\theta_b}\delta_{\mu\nu},
\end{equation}
where $\theta_a$ is a root of unity called the topological spin of anyon $a$. It is related to the spin, or the scaling dimension $h_a$ in CFT as
\begin{equation}
\theta_a=e^{i2\pi h_a}.
\end{equation}
Alternatively, the topological spin $\theta_a$ can be expressed in terms of $R$-symbols as follows
\begin{equation}
\theta_a=\frac{1}{d_a}\sum_c d_c\text{Tr}_c\left[
R^{aa}_c
\right].
\end{equation}

Furthermore, given the $R$-symbols, one can also construct the modular $\mathcal{S}$ and $\mathcal{T}$ matrices as follows
\begin{equation}\label{SmatrixR}
\mathcal{S}_{ab}=\sum_cN_{ab}^c \text{Tr}\left[
R^{ab}_cR^{ba}_c
\right]d_c=\frac{1}{\mathcal{D}}\sum_cN_{ab}^c\frac{\theta_c}{\theta_a\theta_b}d_c,
\end{equation}
and
\begin{equation}\label{TmatrixR}
T_{ab}=\theta_a\delta_{ab}.
\end{equation}
In MTCs, the modular $\mathcal{S}$ and $\mathcal{T}$ matrices are unitary matrices satisfying $\mathcal{S}^{\dag}\mathcal{S}=\mathcal{S}\mathcal{S}^{\dag}=1$ and
$\mathcal{T}^{\dag}\mathcal{T}=\mathcal{T}\mathcal{T}^{\dag}=1$.
In addition, from Eq.\ (\ref{SmatrixR}), it is straightforward to check that
\begin{equation}\label{Qdimension}
d_a=\frac{\mathcal{S}_{a0}}{\mathcal{S}_{00}}=\frac{\mathcal{S}_{0a}}{\mathcal{S}_{00}}, \ \ \
\text{and}\ \ \ \mathcal{D}=\frac{1}{\mathcal{S}_{00}}.
\end{equation}
Other useful quantities such as the $F$-symbols will not be reviewed here, and one can refer to
Ref.\ \onlinecite{Kitaev1997,Kitaev2006b,BondersonT} for more details.

\subsection{Gauge freedom}
\label{gauge}

For any anyon models, there is a gauge freedom coming from the choice of bases in the fusion vector space $V^{ab}_c$.
We can always apply a unitary transformation in the vector space $V^{ab}_c$ without changing the theory. By using the notation where $\left[u_{c}^{ab}\right]_{\mu,\mu'}$ represents the unitary transformation of bases, i.e.,
\begin{equation}
|a,b;c,\mu\rangle=\sum_{\mu'}\left[u^{ab}_c\right]_{\mu\mu'}|a,b;c,\mu'\rangle,
\end{equation}
the $R$-symbols transform as
\begin{equation}
\left[R^{ab}_c\right]'_{\mu'\nu'}=\sum_{\mu,\nu}\left[(u^{ab}_c)^{-1}\right]_{\mu'\mu}\left[
R^{ab}_c
\right]_{\mu\nu}\left[u^{ba}_c\right]_{\nu\nu'}.
\end{equation}
For simplicity, let us consider the multiplicity-free case. Then the unitary transformations $u^{ab}_c$ are simply complex phases. In this case, the $R$-symbols transform as
\begin{equation}
\left[R^{ab}_c\right]'=\frac{u^{ba}_c}{u^{ab}_c}R^{ab}_c.
\end{equation}
It is found that the $R$-symbols are gauge dependent for $a\ne b$. For $a=b$, however, one always has
$\left[R^{aa}_c\right]'=R^{aa}_c$, which means $R^{aa}_c$ is a gauge invariant quantity.

The double braiding defined in Eq.\ (\ref{monodromy001}) transforms as
\begin{equation}
\begin{split}
\left[M^{ab}_c\right]':&=\left[R^{ab}_c\right]'\left[R^{ba}_c\right]'
=\frac{u^{ba}_c}{u^{ab}_c}R^{ab}_c\cdot\frac{u^{ab}_c}{u^{ba}_c}R^{ba}_c\\
&=R^{ab}_cR^{ba}_c=M^{ab}_c,
\end{split}
\end{equation}
which indicates that $M^{ab}_c$ is gauge invariant for arbitrary $a$ and $b$.
In a similar way, one can check that all the nontrivial $F$-symbols are gauge choice dependent\cite{BondersonT}.

\subsection{Topological data for $SU(2)_k$ theories}

In this part we give a brief review of the topological data of $SU(2)_k$ anyon theories \cite{BondersonT}.
The $SU(2)_k$ anyon theories are “$q$-deformed” versions of the usual $SU(2)$ for $q=e^{-2\pi i/(k+2)}$.
In other words, the integers in $SU(2)$ are replaced by the $q$-numbers
$[n]_q\equiv\frac{q^{n/2}-q^{-n/2}}{q^{1/2}-q^{-1/2}}$. These anyon theories  describe $SU(2)_k$ Chern-Simons theories, WZW CFTs, and the Jones polynomials of knot theory. The anyonic charges of a $SU(2)_k$ anyon theory is given by $\mathcal{C}=\left\{0,\frac{1}{2},\cdots, \frac{k}{2}\right\}$.

The fusion rules are given by a general version of the addition rules for a $SU(2)$ spin:
\begin{equation}\label{fusionSU2}
j_1\otimes j_2=\oplus_{j=|j_1-j_2|}^{\text{min}\{j_1+j_2,k-j_1-j_2\}} j,
\end{equation}
with $j\in \mathcal{C}$. The fusion rules can be alternatively written as
\begin{equation}
\begin{split}
j_1\otimes j_2&=\oplus_{j}N_{j_1j_2}^j j\\
&=|j_1-j_2|\oplus |j_1-j_2|+1\oplus\cdots \\
&\oplus \text{min}\{j_1+j_2,k-j_1-j_2\}.
\end{split}
\end{equation}
The $R$-symbols are given by the general formula
\begin{equation}
R_j^{j_1,j_2}=(-1)^{j-j_1-j_2}q^{\frac{1}{2}\left(j_1(j_1+1)+j_2(j_2+1)-j(j+1)\right)},
\end{equation}
based on which we can get the topological spins
\begin{equation}
\theta_j=e^{i2\pi\frac{j(j+1)}{k+2}}.
\end{equation}
In addition, based on the $R$-symbols, one can also obtain the modular $\mathcal{S}$ matrix and $\mathcal{T}$ matrix
according to Eqs.(\ref{SmatrixR}) and (\ref{TmatrixR}), respectively.
The quantum dimension for anyon $j$ has the expression
\begin{equation}
d_j=\frac{\sin\left(\frac{(2j+1)\pi}{k+2}\right)}{\sin\left(\frac{\pi}{k+2}\right)},
\end{equation}
and the total quantum dimension is
\begin{equation}
\mathcal{D}=\sqrt{\sum_i d_i^2}=\frac{\sqrt{\frac{k+2}{2}}}{\sin\left(\frac{\pi}{k+2}\right)}.
\end{equation}
For other topological data such as the $F$-moves (or $F$-symbols), one can refer to, \textit{e.g.}, Ref.\ \onlinecite{BondersonT}.

\section{Alternative calculations of entanglement negativity for different cases}

\subsection{Left-right entanglement negativity}

In the main text, we calculate the left-right entanglement negativity $\mathcal{E}_{LR}$
based on the definition in Eq.\ (\ref{D2}). In this part, we give an explicit calculation of $\mathcal{E}_{LR}$ based on
the definition in Eq.\ (\ref{D1}), i.e.,
\begin{equation}\
\mathcal{E}_{LR}=\ln\text{Tr}|\rho^{T_R}|.
\end{equation}
For the state in Eq.\ (\ref{bsNewEN}), $|\rho^{T_R}|$ can be evaluated as follows
\begin{equation}
\left|\rho^{T_R}\right|=\sqrt{ \left(\rho^{T_R}\right)^{\dag} \rho^{T_R}},
\end{equation}
where
\begin{equation}
\begin{split}
&\left(\rho^{T_R}\right)^{\dag} \rho^{T_R}\\
&=\sum_{aa'}|\psi_a|^2|\psi_{a'}|^2\frac{1}{\mathfrak{n}_a\mathfrak{n}_{a'}}
\sum_{N,j}\sum_{N',j'}
e^{-\frac{8\pi\epsilon}{l}\left(h_a+N-\frac{c}{24}\right)}\\
&\quad e^{-\frac{8\pi\epsilon}{l}\left(h_{a'}+N'-\frac{c}{24}\right)}
|h_{a'},N';j'\rangle\otimes \overline{|h_{a},N;j}\rangle\\
&\quad \langle h_{a'},N';j'|\otimes  \langle\overline{h_{a},N;j}|,
\end{split}
\end{equation}
which is of the diagonal form. Then one can get
\begin{equation}
\begin{split}
\left|\rho^{T_R}\right|&=\sum_{aa'}\sum_{N,j}\sum_{N',j'}|\psi_a||\psi_{a'}|\frac{1}{\sqrt{\mathfrak{n}_a
\mathfrak{n}_{a'}}}
e^{-\frac{4\pi\epsilon}{l}\left(h_a+N-\frac{c}{24}\right)}\\
&\quad e^{-\frac{4\pi\epsilon}{l}\left(h_{a'}+N'-\frac{c}{24}\right)}
|h_{a'},N';j'\rangle\otimes \overline{|h_{a},N;j}\rangle\\
&\quad \langle h_{a'},N';j'|\otimes  \langle\overline{|h_{a},N;j}|.
\end{split}
\end{equation}
Then the left-right entanglement negativity may be expressed as
\begin{align}
\mathcal{E}_{LR}
&=\ln\text{Tr}|\rho^{T_R}|
\nonumber \\
&=
2\ln \left\{\sum_a|\psi_a|\frac{\chi_{h_a}\left(e^{-\frac{4\pi\epsilon}{l}}\right)}{\sqrt{\mathfrak{n}_a}}\right\}
\nonumber \\
&\to
2\ln \left(\sum_a|\psi_a|\mathcal{S}_{a0}^{1/2} e^{\frac{3\pi l}{4\epsilon} \frac{c}{24}}
\right)
\nonumber \\
&=\frac{3\pi c}{48}\cdot\frac{l}{\epsilon}-\ln \mathcal{D}
+2\ln \left(\sum_a|\psi_a|\sqrt{d_a}\right),
\end{align}
where we recall that $\mathfrak{n}_a$ is expressed in Eq.\ (\ref{na1}),
and take the thermodynamic limit.
This is exactly the same as the result in Eq.\ (\ref{LREN0}).

\subsection{Entanglement negativity of two non-contractible regions on a torus}

In this part, we calculate the entanglement negativity of two non-contractible regions on a torus [see Fig.\ \ref{MIC}] based on the definition of entanglement negativity in Eq.\ (\ref{D1}).
Following the structure in the main text, we study these cases
one by one, as follows.

\begin{center}
\emph{\small{
(a) Two adjacent non-contractible regions with non-contractible $B$:
one component interface}
}
\end{center}

As shown in  Fig.\ \ref{MIC} (a), we study the
entanglement negativity between $A_1$ and $A_2$ on a torus with a one-component $A_1A_2$ interface.
We may start from the partially transposed reduced density matrix $\rho_{A_1\cup A_2}^{T_2}$
in Eq.\ (\ref{rho12T}), \textit{i.e.},
\begin{equation}\label{rho12Ta}
\begin{split}
\rho_{A_1\cup A_2}^{T_2}
&=\sum_a|\psi_a|^2
\rho_{A,a}^{b_1}\otimes \left(\rho_{A,a}^{b_2}\right)^{T_2}\otimes \rho_{A,a}^{b_3},\\
\end{split}
\end{equation}
where $\rho_{A,a}^{b_1}$, $ (\rho_{A,a}^{b_2})^{T_2}$ and $\rho_{A,a}^{b_3}$ are defined in Eqs.\ (\ref{rho1})$\sim$(\ref{rhob2T}).
Next, let us calculate $|\rho_{A_1\cup A_2}^{T_2}|$ as follows
\begin{equation}
\left|\rho_{A_1\cup A_2}^{T_2}\right|=\sqrt{ \left(\rho_{A_1\cup A_2}^{T_2}\right)^{\dag} \rho_{A_1\cup A_2}^{T_2}},
\end{equation}
where
\begin{equation}
\begin{split}
&\left(\rho_{A_1\cup A_2}^{T_2}\right)^{\dag} \rho_{A_1\cup A_2}^{T_2}\\
&=\sum_{a}|\psi_a|^4\left(\rho_{A,a}^{b_1}\right)^2\otimes
\left[\left(\rho_{A,a}^{b_2}\right)^{T_2}\right]^{\dag}\left(\rho_{A,a}^{b_2}\right)^{T_2}\otimes \left(\rho_{A,a}^{b_3}\right)^2.
\end{split}
\end{equation}
In particular
\begin{equation}
\begin{split}
&\left[\left(\rho_{A,a}^{b_2}\right)^{T_2}\right]^{\dag}\left(\rho_{A,a}^{b_2}\right)^{T_2}\\
&=\frac{1}{(\mathfrak{n}_a^{b_2})^2}\sum_{N_2,j_2}\sum_{N_2',j_2'}e^{-\frac{8\pi}{l_2}(h_a+N_2-\frac{c}{24})}
e^{-\frac{8\pi}{l_2}(h_a+N_2'-\frac{c}{24})}\\
&\quad |h_a^{b_2},N_2';j_2'\rangle|\overline{h_a^{b_2},N_2;j_2}\rangle\langle h_a^{b_2},N_2';j_2'|\langle \overline{h_a^{b_2},N_2;j_2}|.
\end{split}
\end{equation}
Then one can get
\begin{equation}
\begin{split}
&\left|\rho_{A_1\cup A_2}^{T_2}\right|\\
&=
\sum_a|\psi_a|^2 \rho_{A,a}^{b_1} \otimes \rho_{A,a}^{b_3}\\
&\quad \otimes \frac{1}{\mathfrak{n}_a^{b_2}}
\sum_{N_2,j_2}\sum_{N_2',j_2'}e^{-\frac{4\pi}{l_2}(h_a+N_2-\frac{c}{24})}
e^{-\frac{4\pi}{l_2}(h_a+N_2'-\frac{c}{24})}\\
&\quad |h_a^{b_2},N_2';j_2'\rangle|\overline{h_a^{b_2},N_2;j_2}\rangle\langle h_a^{b_2},N_2';j_2'|\langle \overline{h_a^{b_2},N_2;j_2}|.
\end{split}
\end{equation}
Then, by using the definition in Eq.\ (\ref{D1}), one can obtain the entanglement negativity as follows
\begin{align}
\mathcal{E}_{A_1A_2}&=
\ln \text{Tr}\left|\rho_{A_1\cup A_2}^{T_2}\right|
\nonumber \\
&=\ln\left\{ \sum_a|\psi_a|^2 \frac{\left[\chi_{h_a}\left(e^{-\frac{4\pi\epsilon}{l_2}}\right)\right]^2}
{\chi_{h_a}\left(e^{-\frac{8\pi\epsilon}{l_2}}\right)}
\right\}
\nonumber\\
&\to
\frac{3\pi}{48}\cdot \frac{l_2}{\epsilon}+\ln \left(\sum_a|\psi_a|^2\mathcal{S}_{a0}\right)
\nonumber \\
&=\frac{3\pi}{48}\cdot \frac{l_2}{\epsilon}-\ln \mathcal{D}+\ln \left(\sum_a|\psi_a|^2 d_a\right)
\end{align}
which is exactly the same as Eq.\ (\ref{ENadjacentTorus}).

\begin{center}
\emph{\small{
(b) Two adjacent non-contractible regions with non-contractible $B$:
two component interface}
}
\end{center}

As shown in  Fig.\ \ref{MIC} (b), we study the
entanglement negativity between $A_1$ and $A_2$ on a torus with a two-component $A_1A_2$ interface.
We may start from the partially transposed reduced density matrix $\rho_{A_1\cup A_2}^{T_2}$ in
Eq.\ (\ref{rhoA1TB}) directly, \textit{i.e.},
\begin{equation}\label{rhoA1B002}
\begin{split}
\rho_{A_1\cup A_2}^{T_2}
&=\sum_a|\psi_a|^2
\left(\rho_{A,a}^{b_1}\right)^{T_2}\otimes \left(\rho_{A,a}^{b_2}\right)^{T_2}\otimes \rho_{A,a}^{b_3}\otimes \rho_{A,a}^{b_4},\\
\end{split}
\end{equation}
where the definition of $(\rho_{A,a}^{b_1})^{T_2}$, $(\rho_{A,a}^{b_2})^{T_2}$, $\rho_{A,a}^{b_3}$
and $\rho_{A,a}^{b_4}$ can be found in Eqs.(\ref{rho1aA1B})$\sim$(\ref{rho2aTA1B}). Based on $\rho_{A_1\cup A_2}^{T_2}$ in Eq.\ (\ref{rhoA1B002}), one can get
\begin{equation}
\begin{split}
&\left(\rho_{A_1\cup A_2}^{T_2}\right)^{\dag} \rho_{A_1\cup A_2}^{T_2}\\
&=\sum_{a}|\psi_a|^4\left[\left(\rho_{A,a}^{b_1}\right)^{T_2}\right]^{\dag}\left(\rho_{A,a}^{b_1}\right)^{T_2}
\otimes
\left[\left(\rho_{A,a}^{b_2}\right)^{T_2}\right]^{\dag}\left(\rho_{A,a}^{b_2}\right)^{T_2}\\
&\quad \otimes\left(\rho_{A,a}^{b_3}\right)^{2}\otimes \left(\rho_{A,a}^{b_4}\right)^2.
\end{split}
\end{equation}
In particular, one has
\begin{equation}
\begin{split}
&\left[\left(\rho_{A,a}^{b_1}\right)^{T_2}\right]^{\dag}\left(\rho_{A,a}^{b_1}\right)^{T_2}\\
&=\frac{1}{(\mathfrak{n}_a^{b_1})^2}\sum_{N_1,j_1}\sum_{N_1',j_1'}e^{-\frac{8\pi}{l_1}(h_a+N_1-\frac{c}{24})}
e^{-\frac{8\pi}{l_1}(h_a+N_1'-\frac{c}{24})}\\
&\quad |h_a^{b_1},N_1';j_1'\rangle|\overline{h_a^{b_1},N_1;j_1}\rangle\langle h_a^{b_1},N_1';j_1'|\langle \overline{h_a^{b_1},N_1;j_1}|.
\end{split}
\end{equation}
and
\begin{equation}
\begin{split}
&\left[\left(\rho_{A,a}^{b_2}\right)^{T_2}\right]^{\dag}\left(\rho_{A,a}^{b_2}\right)^{T_2}\\
&=\frac{1}{(\mathfrak{n}_a^{b_2})^2}\sum_{N_2,j_2}\sum_{N_2',j_2'}e^{-\frac{8\pi}{l_2}(h_a+N_2-\frac{c}{24})}
e^{-\frac{8\pi}{l_2}(h_a+N_2'-\frac{c}{24})}\\
&\quad |h_a^{b_2},N_2';j_2'\rangle|\overline{h_a^{b_2},N_2;j_2}\rangle\langle h_a^{b_2},N_2';j_2'|\langle \overline{h_a^{b_2},N_2;j_2}|.
\end{split}
\end{equation}
It is noted that now $\big(\rho_{A,a}^{b_3}\big)^2$, $\big(\rho_{A,a}^{b_4}\big)^2$, $\big[(\rho_{A,a}^{b_1})^{T_2}\big]^{\dag}(\rho_{A,a}^{b_1})^{T_2}$, and
$\big[(\rho_{A,a}^{b_2})^{T_2}\big]^{\dag}(\rho_{A,a}^{b_2})^{T_2}$,
are all of the diagonal form. Then one can easily check that
\begin{equation}
\begin{split}
\left|\rho_{A_1\cup A_2}^{T_2}\right|
&=\sqrt{ \left(\rho_{A_1\cup A_2}^{T_2}\right)^{\dag} \rho_{A_1\cup A_2}^{T_2}}\\
&=\sum_a|\psi_a|^2 \rho_{A,a}^{b_3}\otimes \rho_{A,a}^{b_4}\\
&\quad \otimes
\frac{1}{\mathfrak{n}_a^{b_1}}
\sum_{N_2,j_2}\sum_{N_2',j_2'}e^{-\frac{4\pi}{l_1}(h_a+N_1-\frac{c}{24})}
e^{-\frac{4\pi}{l_1}(h_a+N_1'-\frac{c}{24})}\\
&\quad |\overline{h_a^{b_1},N_1';j_1'}\rangle|h_a^{b_1},N_1;j_1\rangle\langle \overline{h_a^{b_1},N_1';j_1'}|\langle h_a^{b_1},N_1;j_1|\\
&\quad \otimes
\frac{1}{\mathfrak{n}_a^{b_2}}
\sum_{N_2,j_2}\sum_{N_2',j_2'}e^{-\frac{4\pi}{l_2}(h_a+N_2-\frac{c}{24})}
e^{-\frac{4\pi}{l_2}(h_a+N_2'-\frac{c}{24})}\\
&\quad |h_a^{b_2},N_2';j_2'\rangle|\overline{h_a^{b_2},N_2;j_2}\rangle\langle h_a^{b_2},N_2';j_2'|\langle \overline{h_a^{b_2},N_2;j_2}|.
\end{split}
\end{equation}
Then, one can obtain the entanglement negativity between $A_1$ and $A_2$ as follows
\begin{align}
\mathcal{E}_{A_1A_2}
&=
\ln \text{Tr}\left|\rho_{A_1\cup A_2}^{T_2}\right|
\nonumber \\
&=\ln\left\{ \sum_a|\psi_a|^2 \frac{\left[\chi_{h_a}\left(e^{-\frac{4\pi\epsilon}{l_1}}\right)\right]^2
\left[\chi_{h_a}\left(e^{-\frac{4\pi\epsilon}{l_2}}\right)\right]^2
}{\chi_{h_a}\left(e^{-\frac{8\pi\epsilon}{l_1}}\right)\chi_{h_a}\left(e^{-\frac{8\pi\epsilon}{l_2}}\right)}
\right\}
\nonumber \\
&\to
\frac{3\pi c}{48}\cdot\frac{l_1+l_2}{\epsilon}-2\ln\mathcal{D}+\ln\Big(\sum_a|\psi_a|^2d_a^2\Big).
\end{align}
which agrees with the result in Eq.\ (\ref{ENadjacentTorusB}).

\begin{center}
\emph{\small{
(c)
Two adjacent non-contractible regions on a torus with contractible $B$}
}
\end{center}

As shown in Fig.\ \ref{MIC} (c), we study the entanglement negativity between two adjacent
non-contractible regions $A_1$ and $A_2$
with a contractible region $B$.
We may start from the partially transposed reduced density matrix $\rho_{A_1\cup A_2}^{T_2}$
in Eq.\ (\ref{rhoA1A2ConB01}), based on which we can get
\begin{equation}
\begin{split}
&\left(\rho_{A_1\cup A_2}^{T_2}\right)^{\dag} \rho_{A_1\cup A_2}^{T_2}\\
&=\left(\rho_{A,I}^{b_3}\right)^2 \otimes \left(\rho_{A,I}^{b_4}\right)^2
\otimes \sum_{aa'}|\psi_a|^2|\psi_{a'}|^2\frac{1}{\mathfrak{n}_a^{b_1}\mathfrak{n}_{a'}^{b_1}}
\frac{1}{\mathfrak{n}_a^{b_2}\mathfrak{n}_{a'}^{b_2}}\\
&\quad\sum_{N_1,j_1}\sum_{N_1',j_1'}e^{-\frac{8\pi\epsilon}{l_1}(h_a+N_1-\frac{c}{24})}
e^{-\frac{8\pi\epsilon}{l_1}(h_{a'}+N_1'-\frac{c}{24})}\\
&\quad
|h_{a'}^{b_1},N_1';j_1'\rangle |\overline{h_a^{b_1},N_1,j_1}\rangle\langle h_{a'}^{b_1},N_1';j_1'|\langle
\overline{h_a^{b_1},N_1,j_1}|\\
&\quad \otimes
\sum_{N_2,j_2}\sum_{N_2',j_2'}e^{-\frac{8\pi\epsilon}{l_2}(h_a+N_2-\frac{c}{24})}
e^{-\frac{8\pi\epsilon}{l_2}(h_{a'}+N_2'-\frac{c}{24})}\\
&
\quad |\overline{h_{a'}^{b_2},N_2';j_2'}\rangle |h_a^{b_2},N_2,j_2\rangle\langle\overline{ h_{a'}^{b_2},N_2';j_2'}|\langle
h_a^{b_2},N_2,j_2|,\\
\end{split}
\end{equation}
which is of the diagonal form.
Then one can get
\begin{equation}
\begin{split}
\left|\rho_{A_1\cup A_2}^{T_2}\right|&=\sqrt{ \left(\rho_{A_1\cup A_2}^{T_2}\right)^{\dag} \rho_{A_1 \cup A_2}^{T_2}}\\
&=\rho_{A,I}^{b_3}\otimes \rho_{A,I}^{b_4}
\otimes \sum_{aa'}|\psi_a||\psi_{a'}|\frac{1}{\sqrt{\mathfrak{n}_a^{b_1}\mathfrak{n}_{a'}^{b_1}}}
\frac{1}{\sqrt{\mathfrak{n}_a^{b_2}\mathfrak{n}_{a'}^{b_2}}}\\
&\quad\sum_{N_1,j_1}\sum_{N_1',j_1'}e^{-\frac{4\pi\epsilon}{l_1}(h_a+N_1-\frac{c}{24})}
e^{-\frac{4\pi\epsilon}{l_1}(h_{a'}+N_1'-\frac{c}{24})}\\
&\quad
|h_{a'}^{b_1},N_1';j_1'\rangle |\overline{h_a^{b_1},N_1,j_1}\rangle\langle h_{a'}^{b_1},N_1';j_1'|\langle
\overline{h_a^{b_1},N_1,j_1}|\\
&\quad\otimes
\sum_{N_2,j_2}\sum_{N_2',j_2'}e^{-\frac{4\pi\epsilon}{l_2}(h_a+N_2-\frac{c}{24})}
e^{-\frac{4\pi\epsilon}{l_2}(h_{a'}+N_2'-\frac{c}{24})}\\
&\quad
|\overline{h_{a'}^{b_2},N_2';j_2'}\rangle |h_a^{b_2},N_2,j_2\rangle\langle\overline{ h_{a'}^{b_2},N_2';j_2'}|\langle
h_a^{b_2},N_2,j_2|.\\
\end{split}
\end{equation}
Then, the entanglement negativity between $A_1$ and $A_2$ can be obtained as follows
\begin{align}
\mathcal{E}_{A_1A_2}&=
\ln \text{Tr}\left|\rho_{A_1\cup A_2}^{T_2}\right|\\
&=
2\ln\Bigg\{
\sum_a |\psi_a| \frac{\chi_{h_a}\left(e^{-\frac{4\pi\epsilon}{l_1}}\right)}{\sqrt{\mathfrak{n}_a^{b_1}}}
\cdot
\frac{\chi_{h_a}\left(e^{-\frac{4\pi\epsilon}{l_2}}\right)}{\sqrt{\mathfrak{n}_a^{b_2}}}
\Bigg\}
\nonumber \\
&\to
2\ln \left(
\sum_a|\psi_a|\mathcal{S}_{a0}e^{\frac{3\pi(l_1+l_2)}{4\epsilon}\cdot\frac{c}{24}}\right)
\nonumber \\
\end{align}
which is the same as Eq.\ (\ref{ENadjacentTorusConB}).

\bibliography{TEE}

\end{document}